\documentclass[fleqn,usenatbib]{mnras}

\usepackage{newtxtext,newtxmath,url}

\usepackage[T1]{fontenc}

\DeclareRobustCommand{\VAN}[3]{#2}
\let\VANthebibliography\thebibliography
\def\thebibliography{\DeclareRobustCommand{\VAN}[3]{##3}\VANthebibliography}

\usepackage{graphicx}	
\usepackage{amsmath}	

\title[For modified gravity, it's the LITTLE THINGS that matter]{For modified gravity, it's the LITTLE THINGS that matter.}

\author[G. Beck]{
Geoff Beck,\thanks{E-mail: geoffrey.beck@wits.ac.za}
\\
School of Physics and Centre for Astrophysics, University of the Witwatersrand, 1 Jan Smuts Avenue, Johannesburg, WITS-2050, Gauteng, South Africa.\\
}

\date{Accepted XXX. Received YYY; in original form ZZZ}

\pubyear{2026}

\begin{document}
\label{firstpage}
\pagerange{\pageref{firstpage}--\pageref{lastpage}}
\maketitle

\begin{abstract}
Dwarf galaxies have long been recognised as important testing grounds for models of dark matter. For instance, it is here where the cusp-core problem is most apparent. In this work we select two dwarf galaxy samples: LITTLE THINGS and dwarf galaxies in SPARC. We use these to examine whether there are preferences for MOND or dark matter halos in these objects. Notably, our analysis employs the latest developments in Hamiltonian Monte Carlo sampling methodology and robust model comparison via ELPD differences. Our findings suggest a $>4\sigma$ preference for cored halo models over MOND. However, this relies on significant preferences from 7 out of 19 SPARC galaxies and 11 of 18 from LITTLE THINGS (few of which are overwhelming). It is notable that only a single galaxy prefers MOND over a cored halo. Thus, this evidence is suggestive, but does not conclusively decide against MOND. We also test for evidence of a MOND external field effect, and find weak evidence against its presence. Despite these statistical preferences, most SPARC galaxies remain compatible with a universal MOND scale. In LITTLE THINGS, a free MOND model is preferred to a universal value at $\sim 8\sigma$, but this is of doubtful physical significance. For MOG, the story is different, here we find $\gtrsim 8\sigma$ preferences for all halos (or MOND) against universal MOG models with significant exclusions in individual galaxies across both samples. Thus, a proposed universal rotation curve model derived from MOG is quite strongly disfavoured.    
\end{abstract}

\begin{keywords}
dark matter -- galaxies: dwarf 
\end{keywords}



\section{Introduction}

Dark Matter (DM) is a hypothetical variety of particle(s) that would account for numerous astrophysical and cosmological mass deficits in gravitational phenomena. It is largely compatible with observational data, but suffers from several persistent difficulties. Namely, the cusp-core problem~\citep{2021Galax...9..123D}, the baryonic Tully-Fisher relation~\citep{2016ApJ...816L..14L}, an inability to exclude modified gravity theories, and, most seriously, the lack of a discovered candidate particle. In light of these issues, there exist many models of modified gravity that can dispense with DM in various scenarios. 

MOND~\citep{mond1983}, or MOdified Newtonian Dynamics, replaces Newton's law of universal gravitation with one that differs under conditions of extremely low acceleration (an alternative formulation modifies inertia instead~\citep{mond-inertia-1994}). In this limit, gravitational acceleration tends to a constant small value, rather than continuing to zero. This has long performed well in explaining the flat rotation curves of galaxies (see the reviews \citet{mond-review-2012,mond-rev-2015} for more details of MOND's successes and challenges). However, it faces certain difficulties: the first being that it is purely phenomenological and not relativistically covariant. However, it remains a useful proxy for a modification of Newtonian gravity, as its foundational issues are mitigated by formulations like AQUAL~\citep{aqual1984} and QUMOND~\citep{qumond2010}. More recent challenges to this model have arisen from several areas. Firstly, the MOND formulation that is most effective in galaxy rotation curves experiences substantial difficulties in the solar system~\citep{mond-issues-1}. MOND also faces increasing constraints from wide binary systems~\citep{mond-issues-2}, kinetic Sunyaev-Zeldovich probes of inter-galaxy gravitation~\citep{mond-issues-3}, and increasing evidence from diffuse galaxies that DM is a separate substance~\citep{2026arXiv260315860K}. Finally, a recent claim has also been made, in \citet{khelashvili2024}, that SPARC~\citep{lelli_sparc_2016} galaxies generally prefer DM halos to MOND.

The modified gravity model known as MOG (or scalar-vector-tensor gravity) is an attempt to modify gravity in a self-consistent and relativistically covariant manner. MOG is argued to have many successes in explaining astrophysical and cosmological observations. Particularly, MOG does not appear to stumble on the bullet cluster~\citep{mog-bullet}. See \citet{moffat_fundamental_2009} and references therein for further details. MOG operates by adding additional fields that couple to matter similarly to gravity. In the weak field limit, this results in a Yukawa-like, long-range boost to the gravitational field strength (relative to the Newtonian expectation). It is this long range effect that can account for the flattening of rotation curves outside the major concentrations of visible matter in galaxies. Previous studies have suggested tensions~\citep{little-things-mond-mog} between universal MOG effects and rotation curves in the LITTLE THINGS~\citep{hunter_little_2012} dwarf galaxies, as well as dwarf galaxy velocity dispersions~\citep{mog-velocity-disp}. Although the latter may not be an unambiguous problem~\citep{mog-velocity-disp-reply}. On the other hand there are cosmological issues raised against MOG in \citet{mog-issues-cosmology}. 

In this work, we will make use of dwarf galaxies from the SPARC~\citep{lelli_sparc_2016} and LITTLE THINGS~\citep{hunter_little_2012} samples to test DM, MOND, and MOG. In particular, we will explore multiple halo geometries as well as both fixed universal modified gravity parameters and those assigned flat priors to determine whether universal models are statistically compatible with individual galaxy fits. We compare the fits made in each galaxy with each model on the basis of leave-one-out cross-validation, via the Pareto-stabilised importance-sampling implemented in the \texttt{ArviZ} package~\citep{Martin_2026}. The choice of dwarf galaxies is made to better examine cusped vs cored DM halo profiles, to retain a more homogeneous sample than the full SPARC catalogue, and for comparison to LITTLE THINGS.

In section~\ref{sec:samples} we detail our selection criteria for dwarf galaxies from the studied catalogues. In \ref{sec:fitting}, we list fitting functions for all of our models. We then proceed to document our MCMC approach and priors in section~\ref{sec:mcmc}. Our results are displayed in section~\ref{sec:results} and compared to the literature in \ref{sec:comp}. Finally, we draw conclusions in section~\ref{sec:conc}.

\section{Dwarf galaxy samples}
\label{sec:samples}
The dwarf galaxies we study are selected from the SPARC~\citep{lelli_sparc_2016} and the LITTLE THINGS~\citep{hunter_little_2012} samples. For the former we select only galaxies in the first two quality levels, having more than 8 points in their rotation curve data, and measured stellar masses of $< 10^9$ M$\odot$ when the mass-to-light ratio is normalised to unity. For the latter sample, we select only galaxies with measured stellar contributions to the rotation curve~\citep{oh_high-resolution_2015} and discard some based on their dynamical state. In particular, we neglect WLM, due to dynamical asymmetry and ram pressure stripping~\citep{wlm2022,wlm2026}, and IC1613 due to the presence of a bar\footnote{\url{https://ned.ipac.caltech.edu/byname?objname=IC1613}}. Note that we extracted the rotation curve data for LITTLE THINGS from the online source~\citep{oh_high-resolution_2015} and used \texttt{LabPlot} to digitise the baryonic contributions from the same work.    

We display our full considered sample in Table~\ref{tab:sample}. In some cases the LITTLE THINGS galaxies are assigned an inclination error of $0^\circ$, we replace this with $0.1^\circ$ (the lowest error otherwise quoted). We supplement LITTLE THINGS distances from \citet{oh_high-resolution_2015} with consistent errors determined from the NASA/IPAC Extragalactic Database\footnote{\url{https://ned.ipac.caltech.edu/}}, as this was used to determine the distances originally. Note that for galaxies in LITTLE THINGS that do not have a consistent error estimate in NED, we substitute an arbitrary 10\% relative error for the sake of a uniform fitting methodology.  

\begin{table}
    \caption{Input parameters for the dwarf galaxy sample. The upper table is SPARC~\citep{lelli_sparc_2016} and the lower is LITTLE THINGS~\citep{oh_high-resolution_2015}. Note that $M_*$ is determined when the mass-to-light ratio is unity.}
    \label{tab:sample}
    \resizebox{0.99\hsize}{!}{\begin{tabular}{|l|c|c|c|c|}
        \hline
        Galaxy & $d_L$ [Mpc] & $\theta$ & $M_{HI}$ [$10^7$ M$_\odot$] & $M_*$ [$10^7$ M$_\odot$] \\
        \hline
        F583-1 & 35.4 $\pm$ 8.85 & 63.0$^\circ$ $\pm$ 5.0$^\circ$ & 212.60 & 41.69 \\
        NGC3741 & 3.21 $\pm$ 0.17 & 70.0$^\circ$ $\pm$ 4.0$^\circ$ & 18.20 & 1.56 \\
        NGC4214 & 2.87 $\pm$ 0.14 & 15.0$^\circ$ $\pm$ 10.0$^\circ$ & 48.60 & 76.38 \\
        UGC04499 & 10.5 $\pm$ 2.3 & 50.0$^\circ$ $\pm$ 3.0$^\circ$ & 110.00 & 69.18 \\
        UGC05716 & 21.3 $\pm$ 5.3 & 54.0$^\circ$ $\pm$ 10.0$^\circ$ & 109.40 & 21.33 \\
        UGC05721 & 12.55 $\pm$ 1.96 & 61.0$^\circ$ $\pm$ 5.0$^\circ$ & 56.20 & 24.89 \\
        UGC05918 & 7.66 $\pm$ 2.3 & 46.0$^\circ$ $\pm$ 5.0$^\circ$ & 29.70 & 6.75 \\
        UGC06446 & 20.8 $\pm$ 4.79 & 51.0$^\circ$ $\pm$ 3.0$^\circ$ & 137.90 & 42.56 \\
        UGC06667 & 18.0 $\pm$ 2.5 & 89.0$^\circ$ $\pm$ 1.0$^\circ$ & 80.90 & 60.26 \\
        UGC07399 & 8.43 $\pm$ 2.53 & 55.0$^\circ$ $\pm$ 3.0$^\circ$ & 74.50 & 53.21 \\
        UGC07603 & 6.7 $\pm$ 1.54 & 78.0$^\circ$ $\pm$ 3.0$^\circ$ & 25.80 & 18.28 \\
        UGC08286 & 6.5 $\pm$ 0.21 & 90.0$^\circ$ $\pm$ 3.0$^\circ$ & 64.20 & 97.72 \\
        UGC08490 & 4.65 $\pm$ 0.53 & 50.0$^\circ$ $\pm$ 3.0$^\circ$ & 72.00 & 55.59 \\
        UGC08550 & 10.5 $\pm$ 2.96 & 90.0$^\circ$ $\pm$ 3.0$^\circ$ & 28.80 & 24.10 \\
        UGC08837 & 7.21 $\pm$ 0.36 & 80.0$^\circ$ $\pm$ 5.0$^\circ$ & 32.00 & 33.65 \\
        UGC11820 & 18.1 $\pm$ 5.43 & 45.0$^\circ$ $\pm$ 10.0$^\circ$ & 197.70 & 36.81 \\
        DDO161 & 6.03 $\pm$ 0.25 & 70.0$^\circ$ $\pm$ 10.0$^\circ$ & 137.80 & 25.64 \\
        NGC2915 & 4.06 $\pm$ 0.2 & 56.0$^\circ$ $\pm$ 4.0$^\circ$ & 50.80 & 37.33 \\
        UGCA442 & 4.35 $\pm$ 0.22 & 64.0$^\circ$ $\pm$ 7.0$^\circ$ & 26.30 & 6.73 \\
        \hline
        CVnldwA & 3.6 $\pm$ 0.08 & 66.5$^\circ$ $\pm$ 5.2$^\circ$ & 2.91 & 1.58 \\
        DDO101 & 6.4 $\pm$ 0.64 & 51.0$^\circ$ $\pm$ 7.1$^\circ$ & 3.48 & 11.82 \\
        DDO126 & 4.9 $\pm$ 0.55 & 65.0$^\circ$ $\pm$ 0.0$^\circ$ & 16.36 & 7.57 \\
        DDO133 & 3.5 $\pm$ 0.35 & 43.4$^\circ$ $\pm$ 0.1$^\circ$ & 12.85 & 7.49 \\
        DDO154 & 3.7 $\pm$ 0.37 & 68.2$^\circ$ $\pm$ 3.1$^\circ$ & 35.27 & 4.23 \\
        DDO168 & 4.3 $\pm$ 0.49 & 46.5$^\circ$ $\pm$ 0.1$^\circ$ & 25.94 & 15.09 \\
        DDO210 & 0.9 $\pm$ 0.04 & 66.7$^\circ$ $\pm$ 0.1$^\circ$ & 0.14 & 0.09 \\
        DDO216 & 1.1 $\pm$ 0.05 & 63.7$^\circ$ $\pm$ 4.6$^\circ$ & 0.49 & 2.86 \\
        DDO50 & 3.4 $\pm$ 0.05 & 49.7$^\circ$ $\pm$ 6.0$^\circ$ & 132.52 & 37.65 \\
        DDO52 & 10.3 $\pm$ 1.0 & 43.0$^\circ$ $\pm$ 0.0$^\circ$ & 33.43 & 20.00 \\
        DDO53 & 3.6 $\pm$ 0.05 & 27.0$^\circ$ $\pm$ 0.0$^\circ$ & 7.00 & 2.59 \\
        DDO70 & 1.3 $\pm$ 0.02 & 50.0$^\circ$ $\pm$ 0.0$^\circ$ & 3.80 & 3.65 \\
        DDO87 & 7.7 $\pm$ 0.77 & 55.5$^\circ$ $\pm$ 4.8$^\circ$ & 29.12 & 15.85 \\
        HARO36 & 9.3 $\pm$ 0.93 & 70.0$^\circ$ $\pm$ 0.0$^\circ$ & 11.16 & 16.60 \\
        NGC1569 & 3.4 $\pm$ 0.19 & 69.1$^\circ$ $\pm$ 0.1$^\circ$ & 20.24 & 42.22 \\
        NGC2366 & 3.4 $\pm$ 0.16 & 63.0$^\circ$ $\pm$ 0.8$^\circ$ & 108.24 & 36.03 \\
        NGC3738 & 4.9 $\pm$ 0.56 & 22.6$^\circ$ $\pm$ 0.1$^\circ$ & 12.58 & 33.73 \\
        UGC8508 & 2.6 $\pm$ 0.17 & 82.5$^\circ$ $\pm$ 0.1$^\circ$ & 1.19 & 0.61 \\
        \hline
    \end{tabular}}
\end{table}

\section{Fitting rotation curves}
\label{sec:fitting}
To fit the rotation curves of our samples we define the baryonic contribution $v_b$ via the same convention as \citet{lelli_sparc_2016}, or
\begin{equation}
    v_b^2 = \left(v_g \vert v_g \vert + v_s^2 Y_d \right)\frac{d_L}{d_{L,\mathrm{obs}}} \; ,
\end{equation}
where $v_g$ and $v_s$ are the gas and stellar disk velocity contributions sourced from \citet{oh_high-resolution_2015} or \citet{lelli_sparc_2016}, $Y_d$ is the stellar disk mass-to-light ratio (to be fitted), $d_L$ is the distance (nuisance parameter), and $d_{L,\mathrm{obs}}$ is the distance assumed in determining $v_g$ and $v_s$. 

This is used to produce a total model prediction $v_\mathrm{model}$ and compared to the observed values $v_\mathrm{obs} \frac{\sin(\theta_\mathrm{obs})}{\sin(\theta)}$, where $\theta_\mathrm{obs}$ is the originally assumed inclination and $\theta$ is the inclination treated as a nuisance parameter. We will now proceed to detail how we determine $v_\mathrm{model}$ in each case studied.

\subsection{Dark matter halos}
The contribution from DM is added in quadrature to that of baryons, i.e.
\begin{equation}
    v^2_\mathrm{model} = v^2_\mathrm{DM} + v_b^2 \; .
\end{equation}

We consider six halo profiles in our fitting: NFW~\cite{nfw1996}, DC14~\citep{cintio_mass-dependent_2014}, L13~\citep{li_comprehensive_2020}, Burkert~\cite{burkert1995}, isothermal (or pseudo-isothermal), and Einasto~\cite{einasto1968}. All of these, barring NFW and Einasto, are intrinsically cored profiles. Meaning they feature a central region of constant density. Note that the Einasto halo can produce core-like behaviour when the Einasto index $\alpha_e \gtrsim 0.4$. NFW, on the other hand, has a cusped density profile such that $\rho \to \infty$ as $r \to 0$.

In order to perform our fitting we use a universal form for the DM contributions specified in terms of the fitting parameters $v_{200}$ and $c_{200}$. These being the circular velocity at $r_{200}$ (the radius within which the average halo density exceeds the cosmological background by a factor of 200) and halo concentration, defined as $\frac{r_{200}}{r_s}$, respectively. Here $r_s$ is the halo characteristic length scale. We follow \citet{li_comprehensive_2020} with $r_s = \frac{v_{200}}{10 c_{200} H_0}$, where $H_0$ is the Hubble constant (we assume $H_0 = 67.4$ km s$^{-1}$ Mpc$^{-1}$ following \citet{planck2020}). Our DM contributions are then written as
\begin{equation}
    v^2_\mathrm{DM}(x) = v_{200}^2 \frac{c_{200}}{x} \frac{f(x)}{f(c_{200})} \; ,
\end{equation}
where $x = \frac{r}{r_s}$, $r$ is the radial coordinate, and $f$ is the dimensionless, spherical mass function for the DM halo. For our chosen halo profiles we have~\citep{li_comprehensive_2020,cintio_mass-dependent_2014}
\begin{align}
    f_\mathrm{NFW} (x) & = \ln(1+x) - \frac{x}{1+x} \; , \\
    f_\mathrm{Burkert} (x) & = \frac{1}{2}\ln(1+x^2) + \ln(1+x) - \arctan(x) \; , \\
    f_\mathrm{Iso} (x) & = x - \arctan(x) \; , \\
    f_\mathrm{L13} (x) & = \ln(1+x) + \frac{2}{1+x} - \frac{1}{2(1+x)^2} - \frac{3}{2} \; , \\
    f_\mathrm{DC14} (x,\alpha_{14},\beta_{14},\gamma_{14}) & = \beta\left(\frac{3-\gamma_{14}}{\alpha_{14}},\frac{3-\beta_{14}}{\alpha_{14}}+1,\frac{x^{\alpha_{14}}}{1+x^{\alpha_{14}}}\right)\nonumber\\ &{} + \beta\left(\frac{3-\gamma_{14}}{\alpha_{14}}+1,\frac{3-\beta_{14}}{\alpha_{14}},\frac{x^{\alpha_{14}}}{1+x^{\alpha_{14}}}\right) \; ,  \\
    f_\mathrm{Ein} (x,\alpha_e) & = \Gamma\left(\frac{3}{\alpha_e},\frac{2 x^{\alpha_e}}{\alpha_e}\right) \; ,
\end{align}
where $\beta$ and $\Gamma$ are the incomplete $\beta$ and $\Gamma$ functions. Note that the DC14 coefficients, e.g. $\alpha_{14}$, are not free parameters. Instead, they specified by \citet{cintio_mass-dependent_2014} as
\begin{align}
    \alpha_{14} & = 2.94 - \log_{10}\left(10^{-1.08(X+2.33)} + 10^{2.29(X+2.33)}\right) \; , \\
    \beta_{14} & = 4.23 + 1.34 X + 0.26 X^2 \; , \\
    \gamma_{14} & = -0.06 + \log_{10}\left(10^{-0.68(X+2.56)} + 10^{X+2.56}\right) \; ,
\end{align}
where $X = \log_{10}\left(\frac{M_{*}}{M_{200}}\right)$.

\subsection{MOND}
For MOND, the acceleration $a$ of a test particle (mass $m$), that would otherwise be subject to a Newtonian gravitational force $F_N$, is defined as~\citep{mond1983}
\begin{equation}
    F_N = m \mu\left(\frac{a}{a_0}\right) a \; , \label{eq:mond}
\end{equation}
where $a_0$ is the MOND threshold, and $\mu(x)$ is such that $\mu \to 1$ when $x \gg 1$ but $\mu \to x$ when $x \ll 1$. This results in Newtonian mechanics at large accelerations (relative to $a_0$) and modified behaviour below the threshold $a_0$. In particular, the low acceleration gravitational field scales as the square-root of the source mass and linearly, rather than quadratically, with distance. There are two ways to interpret Eq.~(\ref{eq:mond}), either as a general modification to Newton's second law or as a modification only to the universal gravitation law. The former of these is described in \citet{mond-inertia-1994} but is constrained by second law tests, e.g. \citet{2007PhRvL..98o0801G}. 

For a test particle undergoing circular motion due to gravity we have
\begin{equation}
    v^2_\mathrm{model} = v_b^2 + \frac{v_b^2}{2}\left(\sqrt{1+\frac{4 a_0 r}{v_b^2}}-1\right) \; , \label{eq:plain-mond}
\end{equation}
where $v_b$ is the Newtonian velocity contribution of baryonic matter. This stems from our use of the ``simple'' interpolation function
\begin{equation}
    \mu(x) = \frac{x}{1+x} \; .
\end{equation} 

This can be extended to include the external field effect, i.e. the fact that MOND necessitates a nonlinear contribution from the gravitational environment of a galaxy. Our rotational velocity becomes approximately~\citep{sep-zeta}
\begin{equation}
    v^2_\mathrm{model} \approx v_b^2\left[\frac{1}{2} - \frac{A_\zeta}{y} + \sqrt{\left(\frac{1}{2} - \frac{A_\zeta}{y}\right)^2 + \frac{1+y_\zeta}{y}}\right] \; , \label{eq:zeta-mond}
\end{equation}
where $y = \frac{v_b^2}{r a_0}$. We define the external field gravitational acceleration as 
\begin{equation}
y_\zeta = \frac{\zeta \times 1.2 \times 10^{-10} \ \mathrm{m \ s}^{-2}}{a_0} \; ,
\end{equation} 
such that
\begin{equation}
    A_\zeta = \frac{y_\zeta (1 + 0.5y_\zeta)}{1+y_\zeta} \; .
\end{equation} 
This formulation reduces to Eq.~(\ref{eq:plain-mond}) when $\zeta \to 0$ and the Newtonian external field contributions can be computed as approximations to the MOND values for $\zeta$~\citep{sep-zeta}. Our choice of $\zeta$ normalisation accommodates the data in \citet{sep-zeta} that uses a fixed MOND scale of $1.2 \times 10^{-10}$ m s$^{-2}$.

\subsection{MOG}
For a test particle near a point mass $M$ in this scenario \citet{moffat_fundamental_2009} found 
\begin{equation}
    \ddot{r} = - \frac{G M}{r^2} \left[1 + \alpha_M - \alpha_M (1+\mu_M r) \exp(-\mu_M r)\right] \; , \label{eq:mog}
\end{equation}
where $\mu_M = \frac{D}{\sqrt{M}}$, and $\alpha_M \sim \frac{19 M}{(\sqrt{M} + E)^2}$. The free parameters of the model are thus $D$ and $E$ which are found to be universal from rotation curves fits in \citet{moffat_fundamental_2009} with values $D = 6250$ M$^{1/2}_\odot$ kpc$^{-1}$ and $E = 25000$ M$^{1/2}_\odot$. MOG results in a correction to Newtonian gravity that scales as $r^{-1}$ when $r \ll \mu_M^{-1}$, in the limit $r > \mu_m^{-1}$ we have merely a mass-dependent rescaling of the field strength.   

For rotation curve fitting we follow the prescription in \citet{brownstein2006} for dealing with an extended mass distribution, and combine it with Eq.~(\ref{eq:mog}) to yield
\begin{equation}
    v^2_\mathrm{model} = v_b^2 \left[1 + \alpha_M - \alpha_M (1+\mu_M r) \exp(-\mu_M r)\right] \; ,
\end{equation}
where $\mu_M = \frac{D}{\sqrt{M_\mathrm{tot}}}$, and $\alpha_M \sim \frac{19 M}{(\sqrt{M_\mathrm{tot}} + E)^2}$. Here $M_\mathrm{tot} = M_d Y_d + M_\mathrm{gas}$, where $M_d$ is the disk stellar mass when $Y_d = 1$. 

\section{MCMC methodology}
\label{sec:mcmc}
Our MCMC approach uses \texttt{pymc}\footnote{\url{https://www.pymc.io} v6.0.0}~\citep{pymc2023}. We make use of the \texttt{nutpie}\footnote{\url{https://pymc-devs.github.io/nutpie/} v0.16.10}~\citep{nutpie2026} NUTS sampler with 6 sample chains tuned for 1000 steps each and producing 3000 samples each. This is sufficient to provide excellent convergence for all parameters in the vast majority of cases (chains are fully available in the online data\footnote{DOI 10.5281/zenodo.20397291}). Avoiding excessive divergences in the chains requires we set the target acceptance rate at $> 0.9$ ($0.97$ for MOG). Reliable sampling is ensured by setting the maximum tree depth at $30$ for MOND, MOG, DC14, and Einasto. In select cases we need to run more chains and/or samples per chain for reliability.

Our likelihood takes the form
\begin{equation}
    \mathcal{L} \propto \exp\left[\sum_i \frac{(v_\mathrm{obs}(r_i) - v_\mathrm{model}(r_i,\Theta))^2}{2 \delta v_\mathrm{obs}^2 (r_i)}\right] \; , \label{eq:likelihood}
\end{equation}
where $v_\mathrm{obs}(r_i)$ is the observed rotation curve velocity at sampled radius $r_i$, $v_\mathrm{model}(r_i)$ is the model predicted velocity as a function of the parameter set $\Theta$, and $\delta v_\mathrm{obs} (r_i)$ is the rotation curve error. 

We will now proceed to detail the prior distributions used for each potential parameter in $\Theta$. We note that wide uniform priors are not, in general, uninformative. However, \citet{desmond-rar-2023} found MOND fits to the radial acceleration relation do not appear to be strongly sensitive to the choice of prior. For our DM models our priors are all physically motivated (saving $c_{200}$ for Burkert and isothermal halos). This only leaves MOG in danger of being  unduly influenced by the choice of prior. In practice, we find only a handful of problematic galaxies in SPARC. Here the width of the uniform prior on the MOG parameter $D$ can result in a bi-modal posterior if $D_{max} > 10^5$ M$_\odot^{1/2}$. This bi-modality is physically interesting (more on this later). However, it causes various sampling warnings in \texttt{pymc}. Thus, we check its results by also considering lognormal priors, with 1 dex width, for both $D$ and $E$.

\subsection{Common priors}
In Table~\ref{tab:common-priors} we detail the priors used in all fitting attempts.
\begin{table}
    \caption{Common priors for all fitting.}
    \label{tab:common-priors}
    \begin{tabular}{|l|l|l|}
        \hline
        Quantity & Prior function & Notes \\
        \hline
        $Y_d$ & Lognormal & $\mu = \ln(0.5)$, $\sigma = 0.23$ \\
        $d_L$ & Normal & Nuisance\\
        $\theta$ & Normal & Nuisance\\
        \hline
    \end{tabular}
\end{table}
Our prior from $Y_d$ matches that used in \citet{li_comprehensive_2020}, sourced from \citet{schombert_2018}. Parameters labelled ``Nuisance'' have normal distributions with the mean as the measured value and the standard deviation being the error. One should note that \texttt{pymc} implements lognormal distributions in terms of the natural log, so our standard deviations are quoted for the natural logarithm unless specified.   

\subsection{DM Priors}
We detail the additional DM priors in Table~\ref{tab:dm-priors}. 
\begin{table*}
    \caption{Priors for DM halo fitting.}
    \label{tab:dm-priors}
    \begin{tabular}{|l|l|l|}
        \hline
        Quantity & Prior function & Notes \\
        \hline 
        $v_{200}$ & Uniform on [5,100] km s$^{-1}$ & \\
        $c_{200}$ & Lognormal & $\mu = \ln(c_{200,\mathrm{dm14}} (M_{200}))$, $\sigma = 0.25$ \\
        $c_{200}$ &  Uniform on [0,1000] & For isothermal and Burkert only \\
        $M_*(M_{200})$ & Lognormal & $\mu = \ln(M_*(M_{200}))$, $\sigma = 0.35$ \\
        $\alpha_e$ & Lognormal & $\mu = \ln(\alpha_{e,\mathrm{dm14}})$, $\sigma = 0.37$\\
        \hline
    \end{tabular}
\end{table*}
For the $c_{200}$ prior we use
\begin{equation}
    c_{200,\mathrm{dm14}} (M_{200}) = a - b\log_{10}\left(\frac{M_{200} h}{10^{12} \ \mathrm{M}_\odot}\right) \; ,
\end{equation} 
where $h = \frac{H_0}{100 \ \mathrm{km/(Mpc s)}}$. Where $a = 0.905$ and $b = 0.101$, except for the Einasto case where we have $0.977$ and $0.13$ following \citet{dutton2014}. Additionally, for the Einasto halo we have~\citep{dutton2014}
\begin{equation}
    \alpha_{e,\mathrm{dm14}} = 0.155 + 0.0095 \nu^2 \; ,
\end{equation}
where $\log_{10}(\nu) = -0.11 + 0.146 m + 0.0138 m^2 + 0.00123 m^3$, and $m = \frac{M_{200}h}{10^{12} \ \mathrm{M}_\odot}$. 
Finally, for $M_*(M_{200})$ we make use of the relation from \citet{moster_2013}
\begin{equation}
    \frac{M_*}{M_{200}} = 0.0702 \left[\left(\frac{M_{200}}{M_1}\right)^{-1.376} + \left(\frac{M_{200}}{M_1}\right)^{0.608}\right]^{-1} \; ,
\end{equation}
where $M_1 = 10^{11.59}$ M$_\odot$. Note that $M_{200} = \frac{v_{200}^3}{10 H_0 G}$, but $M_*$ depends on $Y_d$. Thus, this ``$M_*(M_{200})$ prior'' acts like a constraint on $v_{200}$ and $Y_d$. Note that this relation does not differ from the more recent one in \citet{2018AstL...44....8K} on the mass scales relevant to our work.

\subsection{MOND priors}
MOND requires two parameters stipulated in Table~\ref{tab:mond-priors}. Note that we normalise $a_0$ to $10^{-13}$ km s$^{-2}$ for convenience throughout this paper. We will compare a case where $a_0$ is free (``MOND Free'') to one where it is fixed at $a_0 = 1.2\times 10^{-13}$ km s$^{-2}$ (``MOND Universal''). The $\zeta$ parameter is only considered on the SPARC sample, it is set to $0$ for LITTLE THINGS. For SPARC, we source the $\zeta$ prior parameters from the estimations in \citet{sep-zeta}.
\begin{table}
    \caption{Priors for MOND fitting.}
    \label{tab:mond-priors}
    \begin{tabular}{|l|l|l|}
        \hline
        Quantity & Prior function & Notes \\
        \hline
        $a_0$ & Uniform on $[0,10^2]$ & Units $10^{-13}$ km s$^{-2}$  \\
        $\zeta$ & Normal & SPARC from \citet{sep-zeta} \\
        \hline
    \end{tabular}
\end{table}

\subsection{MOG priors}
We place our MOG priors in Table~\ref{tab:mog-priors}.
\begin{table}
    \caption{Priors for MOG fitting.}
    \label{tab:mog-priors}
    \begin{tabular}{|l|l|l|}
        \hline
        Quantity & Prior function & Notes \\
        \hline
        $D$ & Uniform on [$0$,$10^8$] M$^{1/2}_\odot$ kpc$^{-1}$ & \\
        $D$ & Lognormal & $\mu = \ln\left(6250\right)$, $\sigma = 2.3$ \\
        $E$ & Uniform on [$0$,$10^7$] M$^{1/2}_\odot$ & \\
        $E$ & Lognormal & $\mu = \ln\left(25000\right)$, $\sigma = 2.3$ \\
        \hline
    \end{tabular}
\end{table}
We will consider two main scenarios here: one where $D$ and $E$ are free (``MOG Free'') and one where they are fixed to the values found in \citet{moffat_fundamental_2009} (``MOG Universal''). A substantial number of galaxies displayed bi-modal posteriors in $D$. This made obtaining consistent results in \texttt{pymc} difficult. To overcome this, we implemented more samples and chains, 6000 and 24 respectively, where appropriate. We also considered a lognormal prior centred on the universal value with a 1 dex width. This was chosen to limit only very large or small values of $D$ and $E$.

\subsection{Model comparison} \label{sec:model-comp}
In order to compare how each rotation curve model performs we will determine the expected log predictive density (ELPD) via the Pareto-stabilised leave-one-out method implemented in \texttt{ArviZ}\footnote{This work was done with \texttt{ArviZ} v1.1.0}~\citep{Martin_2026} via the \texttt{compare} method. To determine the significance of model discrimination we will compare the $\Delta \mathrm{ELPD}$ between models and divide it by the associated standard error. This will roughly equate to a frequentist-type statistical confidence interval. 

For comparisons across the whole galaxy sample we will sum $\Delta$ELPD values for each galaxy, while adding the standard errors for $\Delta$ELPD in quadrature. 

An obvious risk in the use of this model comparison technique is that many dwarf galaxies have few data points in their rotation curves. This can have a serious impact on the reliability of ELPD variances~\citep{Bayesian2017}. Indeed, several galaxy-model combinations have data points with Pareto-$k$ values above $0.7$ (it is at most 1 per combination). Fortunately, only a single data point in UGCA442, 2 in UGC11820, and 3 in UGC08286 have $k \geq 1$ with free MOG/MOND models. The vast majority of all data points in all galaxies and models have $k < 0.7$. To maintain reliability, we ensure our standard error and ELPD values are stable in each model fitting, i.e. they do not vary substantially with different MCMC random seeds. This can require that we customise the sampling parameters for troublesome cases. We also check all our model comparisons with both LOO and WAIC methods, in all presented cases these give reasonable agreement. Finally, all significantly preferred models are also strongly favoured by the Bayesian bootstrap model weight (also computed in \texttt{ArviZ}).

\section{Results}
\label{sec:results}
\begin{figure}
    \resizebox{0.99\hsize}{!}{\includegraphics{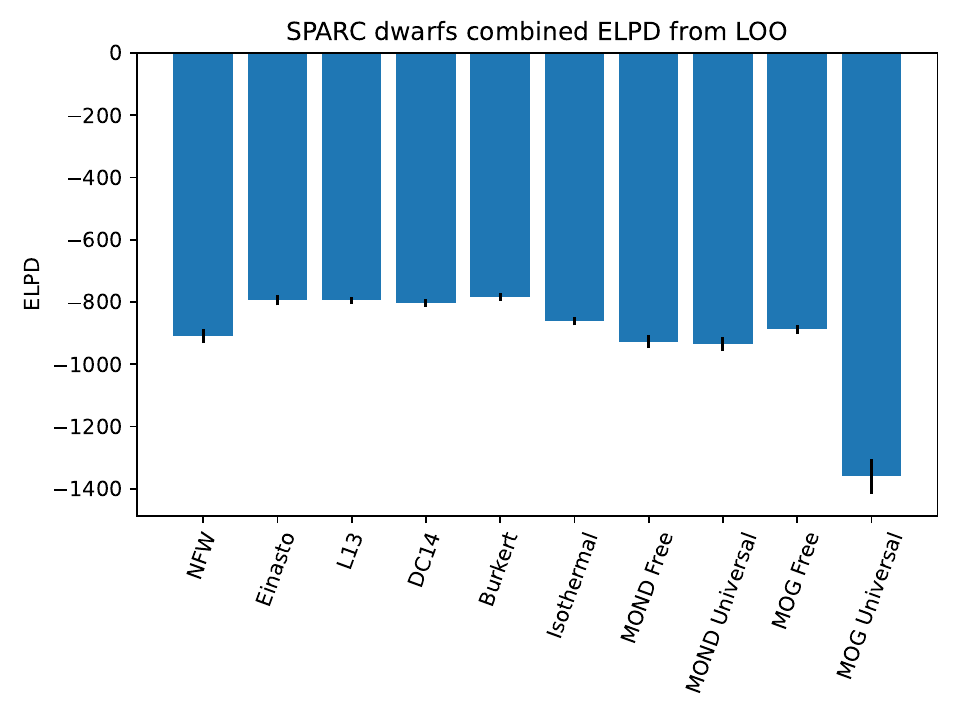}}
    \resizebox{0.99\hsize}{!}{\includegraphics{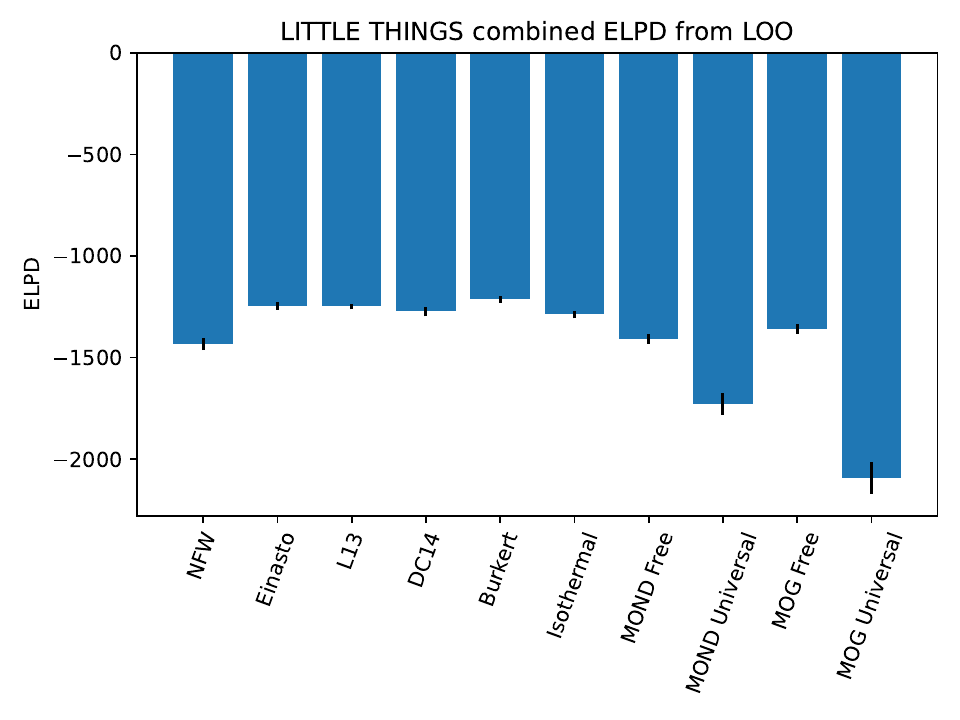}}
    \caption{Combined leave-one-out ELPD for SPARC (upper) and LITTLE THINGS (lower) samples. Burkert values are in black, MOND in magneta, and MOG in green. The error bars show the standard error of the ELPD. The dotted lines show the standard error on $\Delta$ELPD. Note that more negative values indicate a worse model.}
    \label{fig:ts}
\end{figure}

We start by presenting the main result of the paper: our cumulative ELPD scores for the LITTLE THINGS and SPARC samples in Fig.~\ref{fig:ts}. Here we see that cored DM halos are generally preferred over all other models. These plots themselves are not used to produce final statistical preferences. Instead, we iterate over each galaxy and sum the $\Delta$ELPD values for the two models being compared, while adding the standard error on $\Delta$ELPD in quadrature. 

In the SPARC sample (Table~\ref{tab:prefs}) the $\sigma$ value represents the cumulative $\Delta$ELPD divided by the cumulative standard error. A positive value indicates the model in the row is favoured, while negative favours the column. Here we find that Einasto, isothermal, L13, DC14, and Burkert DM halos are significantly preferred over MOND and MOG. However, there is no significant preference for or against NFW, except when compared to the universal MOG model. For Burkert, Einasto, L13, and DC14 this preference is very significant indeed, exceeding $6\sigma$ against MOND. It is less significant against MOG ($\sim 4\sigma$) until we require the MOG parameters take universal values, in which case it becomes $\gtrsim 8 \sigma$. In terms of intra-halo preferences, isothermal and NFW are significantly disfavoured relative to other models. 

There is no significant difference between free and universal MOND models, in agreement with previous literature results~\citep{desmond-rar-2023,rar_lelli,rar_li}. However, the universal MOG rotation curve values from~\citep{moffat_fundamental_2009} are very significantly disfavoured over a free model. This demonstrates that, without the intervention of strong modelling systematics, dwarf galaxies are not compatible with the proposed universal model of MOG. 

\begin{table*}
    \caption{Significance of SPARC total ELPD preference between DM halos and modified gravity. A positive value favours the model in the row, negative favours that in the column. These are computed via the $\Delta \mathrm{ELPD}$ values from \texttt{ArviZ}.}
    \label{tab:prefs}
    \begin{tabular}{|l|c|c|c|c|c|c|c|c|c|c|c|c|}
        \hline
        & NFW & Burkert & Einasto & L13 & DC14 & Isothermal & MOND & MOND U & MOG & MOG U \\
        \hline
        NFW &  -  & $-5.8\sigma$ & $-8.1\sigma$ & $-6.6\sigma$ & $-5.0\sigma$ & $-3.2\sigma$ & $0.5\sigma$ & $1.5\sigma$ & $-1.8\sigma$ & $7.7\sigma$\\
        Burkert & $5.8\sigma$ &  -  & $0.6\sigma$ & $1.7\sigma$ & $1.7\sigma$ & $4.9\sigma$ & $6.8\sigma$ & $7.0\sigma$ & $4.1\sigma$ & $10.2\sigma$\\
        Einasto & $8.1\sigma$ & $-0.6\sigma$ &  -  & $0.1\sigma$ & $0.1\sigma$ & $5.2\sigma$ & $6.6\sigma$ & $6.8\sigma$ & $3.6\sigma$ & $9.9\sigma$\\
        L13 & $6.6\sigma$ & $-1.7\sigma$ & $-0.1\sigma$ &  -  & $0.7\sigma$ & $5.8\sigma$ & $7.2\sigma$ & $7.5\sigma$ & $4.2\sigma$ & $9.9\sigma$\\
        DC14 & $5.0\sigma$ & $-1.7\sigma$ & $-0.1\sigma$ & $-0.7\sigma$ &  -  & $3.5\sigma$ & $6.4\sigma$ & $6.8\sigma$ & $3.3\sigma$ & $9.8\sigma$\\
        Isothermal & $3.2\sigma$ & $-4.9\sigma$ & $-5.2\sigma$ & $-5.8\sigma$ & $-3.5\sigma$ &  -  & $4.0\sigma$ & $5.5\sigma$ & $-0.1\sigma$ & $9.6\sigma$\\
        MOND & $-0.5\sigma$ & $-6.8\sigma$ & $-6.6\sigma$ & $-7.2\sigma$ & $-6.4\sigma$ & $-4.0\sigma$ &  -  & $1.2\sigma$ & $-3.6\sigma$ & $9.3\sigma$\\
        MOND U & $-1.5\sigma$ & $-7.0\sigma$ & $-6.8\sigma$ & $-7.5\sigma$ & $-6.8\sigma$ & $-5.5\sigma$ & $-1.2\sigma$ &  -  & $-3.4\sigma$ & $9.1\sigma$\\
        MOG & $1.8\sigma$ & $-4.1\sigma$ & $-3.6\sigma$ & $-4.2\sigma$ & $-3.3\sigma$ & $0.1\sigma$ & $3.6\sigma$ & $3.4\sigma$ &  -  & $9.8\sigma$\\
        MOG U & $-7.7\sigma$ & $-10.2\sigma$ & $-9.9\sigma$ & $-9.9\sigma$ & $-9.8\sigma$ & $-9.6\sigma$ & $-9.3\sigma$ & $-9.1\sigma$ & $-9.8\sigma$ &  - \\
        \hline
    \end{tabular}
\end{table*}

In Table~\ref{tab:prefs-lt} we display the sample preferences for the LITTLE THINGS galaxies. Once again, the Burkert, Einasto, and L13 halos are the strongest performers. Interestingly, the preference for cored over cuspy halos is substantially stronger than in the SPARC galaxies. However, this data set shows a significant tension with a universal MOND acceleration scale. Once again, a universal MOG model is strongly rejected. 

\begin{table*}
    \caption{Significance of LITTLE THINGS total ELPD preference between DM halos and modified gravity. A positive value favours the model in the row, negative favours that in the column. These are computed via the $\Delta \mathrm{ELPD}$ values from \texttt{ArviZ}.}
    \label{tab:prefs-lt}
    \begin{tabular}{|l|c|c|c|c|c|c|c|c|c|c|c|c|}
        \hline
         & NFW & Burkert & Einasto & L13 & DC14 & Isothermal & MOND & MOND U & MOG & MOG U \\
         \hline
        NFW &  -  & $-6.8\sigma$ & $-8.2\sigma$ & $-6.8\sigma$ & $-4.4\sigma$ & $-5.9\sigma$ & $-0.8\sigma$ & $6.3\sigma$ & $-2.1\sigma$ & $8.7\sigma$\\
        Burkert & $6.8\sigma$ &  -  & $1.9\sigma$ & $3.4\sigma$ & $5.7\sigma$ & $3.6\sigma$ & $8.0\sigma$ & $9.8\sigma$ & $7.0\sigma$ & $11.4\sigma$\\
        Einasto & $8.2\sigma$ & $-1.9\sigma$ &  -  & $0.2\sigma$ & $1.4\sigma$ & $2.3\sigma$ & $7.4\sigma$ & $10.0\sigma$ & $5.1\sigma$ & $11.4\sigma$\\
        L13 & $6.8\sigma$ & $-3.4\sigma$ & $-0.2\sigma$ &  -  & $1.4\sigma$ & $3.1\sigma$ & $9.0\sigma$ & $10.4\sigma$ & $6.4\sigma$ & $11.8\sigma$\\
        DC14 & $4.4\sigma$ & $-5.7\sigma$ & $-1.4\sigma$ & $-1.4\sigma$ &  -  & $0.5\sigma$ & $4.8\sigma$ & $8.4\sigma$ & $3.7\sigma$ & $10.5\sigma$\\
        Isothermal & $5.9\sigma$ & $-3.6\sigma$ & $-2.3\sigma$ & $-3.1\sigma$ & $-0.5\sigma$ &  -  & $6.7\sigma$ & $10.2\sigma$ & $3.7\sigma$ & $11.7\sigma$\\
        MOND & $0.8\sigma$ & $-8.0\sigma$ & $-7.4\sigma$ & $-9.0\sigma$ & $-4.8\sigma$ & $-6.7\sigma$ &  -  & $7.7\sigma$ & $-2.5\sigma$ & $10.0\sigma$\\
        MOND U & $-6.3\sigma$ & $-9.8\sigma$ & $-10.0\sigma$ & $-10.4\sigma$ & $-8.4\sigma$ & $-10.2\sigma$ & $-7.7\sigma$ &  -  & $-8.0\sigma$ & $7.5\sigma$\\
        MOG Free, D high & $2.1\sigma$ & $-7.0\sigma$ & $-5.1\sigma$ & $-6.4\sigma$ & $-3.7\sigma$ & $-3.7\sigma$ & $2.5\sigma$ & $8.0\sigma$ &  -  & $10.7\sigma$\\
        MOG U & $-8.7\sigma$ & $-11.4\sigma$ & $-11.4\sigma$ & $-11.8\sigma$ & $-10.5\sigma$ & $-11.7\sigma$ & $-10.0\sigma$ & $-7.5\sigma$ & $-10.7\sigma$ &  - \\
        \hline
    \end{tabular}
\end{table*}

\begin{figure}
    \resizebox{0.99\hsize}{!}{\includegraphics{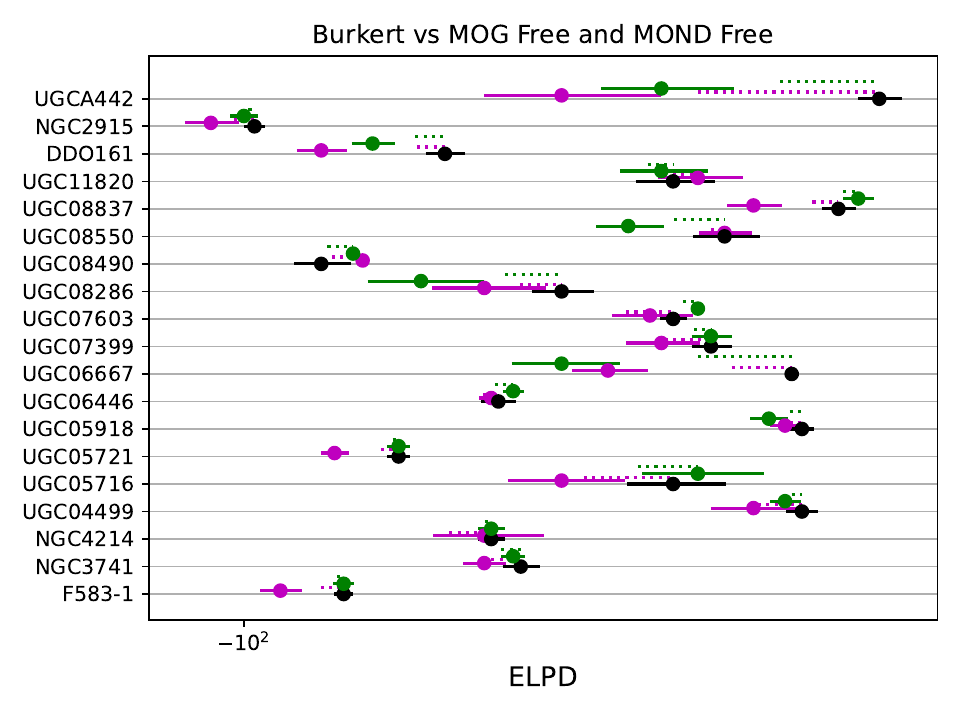}}
    \resizebox{0.99\hsize}{!}{\includegraphics{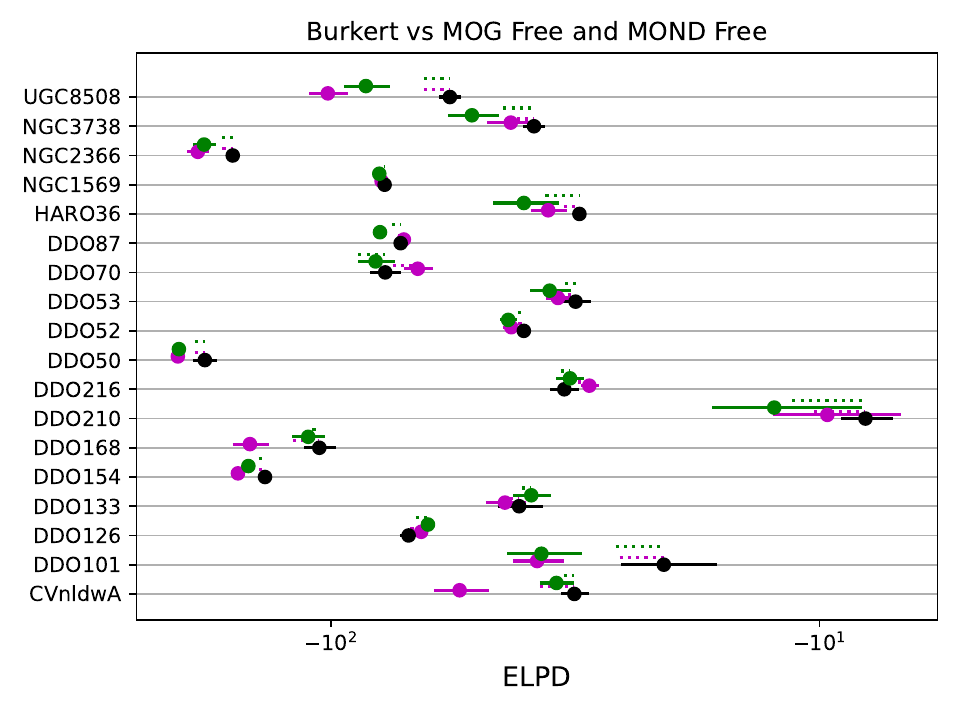}}
    \caption{Per galaxy ELPD for SPARC (upper) and LITTLE THINGS (lower) samples. Burkert values are in black, MOND in magneta, and MOG in green. The error bars show the standard error of the ELPD. The dotted lines show the standard error on $\Delta$ELPD. Note that more negative values indicate a worse model.}
    \label{fig:ts-g}
\end{figure}

To discuss individual galaxy preferences in the SPARC sample we will use the Burkert and NFW halos as benchmarks to compare against modified gravity. Our Burkert results are displayed in Fig.~\ref{fig:ts-g} (upper panel). We find that 8 galaxies significantly prefer Burkert halos over MOND with free $a_0$, and none prefer MOND. Two galaxies, UGC05721 and DDO161 achieve $>4.3\sigma$ preferences for cored halos. For the free MOG case, there are 5 galaxies significantly preferring Burkert and none that prefer MOG with any significance. The best-performing MOG galaxy is UGC08490. This has the rotation curve most lacking in features, when compared to others with many data points in the flat velocity region. In the case of NFW we find 6 galaxies preferring NFW to MOND at between $2.5$ and $3.8\sigma$, and 5 galaxies preferring MOND at $2.0$ to $3.3\sigma$. We find 5 galaxies prefer NFW to MOG with NGC4214 achieving an $8\sigma$ preference, while 4 galaxies have opposite preferences ranging from $2$ to $3.8\sigma$.

For LITTLE THINGS (Fig.~\ref{fig:ts-g} lower panel), we find that 11 galaxies significantly prefer Burkert to free MOND, with only DDO216 breaking the other way. The most significant preference for Burkert here achieves $5.9\sigma$ in UGC08508. For the free MOG model, 9 galaxies prefer Burkert and none prefer MOG. The strongest Burkert preference is DDO87 with $4.8\sigma$. There are 3 galaxies that prefer NFW to MOND here, with DDO101 reaching $4.3\sigma$. In contrast, 5 galaxies significantly prefer MOND, with DDO87 reaching $5\sigma$. NFW is preferred to MOG by DDO101 ($2.9\sigma$) and DDO216 ($2.5\sigma$).

\begin{figure}
    \resizebox{0.99\hsize}{!}{\includegraphics{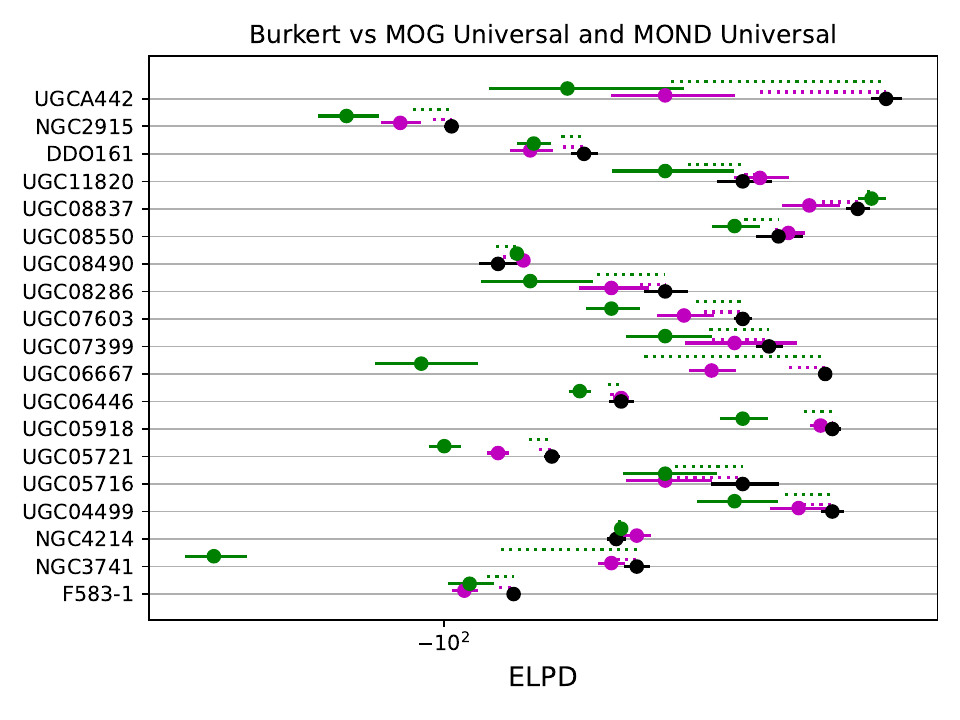}}
    \resizebox{0.99\hsize}{!}{\includegraphics{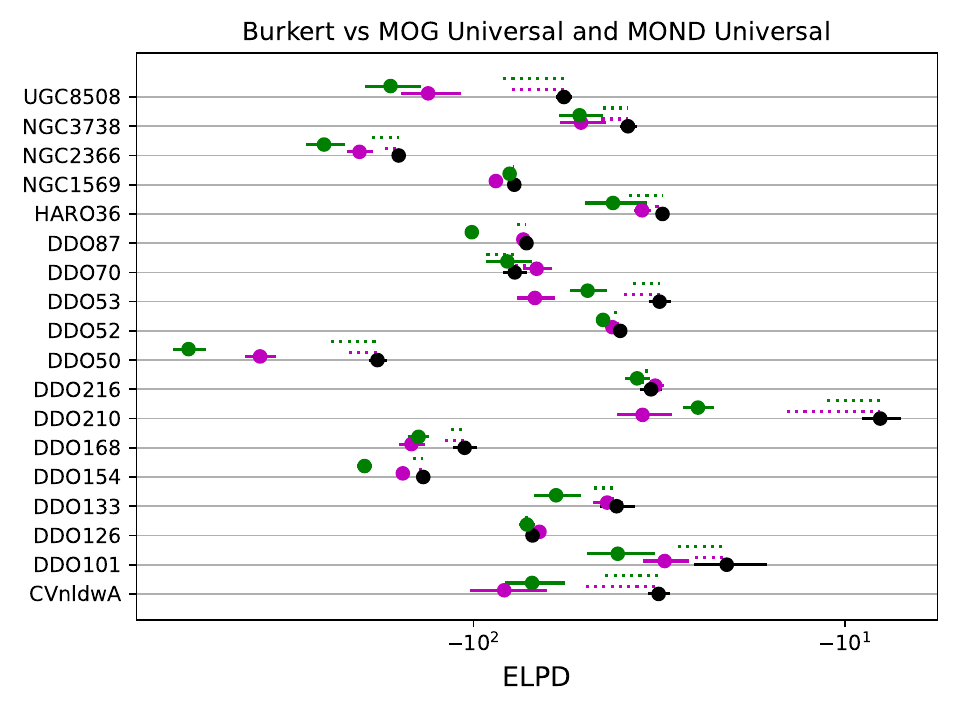}}
    \caption{Per galaxy ELPD for SPARC (upper) and LITTLE THINGS (lower) samples. Burkert values are in black, MOND in magneta, and MOG in green. The error bars show the standard error of the ELPD. The dotted lines show the standard error on $\Delta$ELPD. Note that more negative values indicate a worse model.}
    \label{fig:ts-g-u}
\end{figure}

For the universal MOND model in SPARC (Fig.~\ref{fig:ts-g-u} upper panel) there are now 6 galaxies preferring Burkert, with UGC05918 and UGC8837 preferences becoming insignificant. The largest preference for Burkert is $4.7\sigma$ in UGC05721. Additionally, NGC4214 prefers universal MOND over a Burkert halo at $3.7\sigma$. For the universal MOG model, 11 galaxies significantly prefer Burkert and only 1 galaxy favours MOG at $3.6\sigma$ (UGC08837). Both NGC3741 and UGC05721 achieve $>5\sigma$ preferences for Burkert halos over universal MOG. There are 6 galaxies that prefer NFW to universal MOND in this sample with UGC08490 reaching $5.1\sigma$, while 2 have preferences for universal MOND. A total of 9 galaxies prefer NFW to universal MOG with NGC3741 achieving $6.1\sigma$, only DDO16 and UGC08837 prefer universal MOG at $2.3\sigma$ and $3.8\sigma$ respectively. 

For the universal form of MOND in LITTLE THINGS (Fig.~\ref{fig:ts-g-u} lower panel), the Burkert halo preferences are largely unchanged from the free case. However, no galaxies now prefer MOND, and UCG08508's Burkert preference is reduced to $3.5\sigma$. Additionally, there are 5 galaxies with $>4\sigma$ preference for Burkert over MOND. In the universal MOG model we find 15 galaxies prefer Burkert, with DDO50 having a $6.7\sigma$ preference and 3 other galaxies $>5\sigma$. For the NFW halo, 7 galaxies prefer it to universal MOND, at levels ranging from $2.3$ to $5.3\sigma$. There are two galaxies preferring universal MOND at $2.2$ and $4.4\sigma$. There are 6 galaxies preferring NFW to universal MOG, with 3 of them passing the $5\sigma$ level and the highest being $7.3\sigma$ in DDO210. No galaxies significantly prefer universal MOG to NFW.

These results suggest a reasonably strong sample-level preference against modified gravity models, with universal forms being very significantly disfavoured against non-cuspy DM halos. However, a repeat of this analysis with future telescope data~\citep{bigsparc} would be necessary before one draws firm conclusions about the status of modified gravity. This is especially true for the LITTLE THINGS galaxies, whose rotation curves are noted to contain significant non-circular effects~\citep{lelli_sparc_2016}. At the level of individual galaxies, the universal MOG model seems unviable in these samples, with 6 galaxies disfavouring it at $>5\sigma$. Whereas, only 2 galaxies in LITTLE THINGS have a $5\sigma$ preference against universal MOND. Although this is highly suggestive against these modified gravity models, one would require a very close examination of the individual galaxies that seem to reject modified gravity before one could claim to rule such models out. To see if anything further can be gleaned from the individual galaxies we will examine how well the physically motivated priors are recovered by the models.

\subsection{SPARC sample}
We begin by assessing the $c_{200}$ fits. First, the two cases we do not expect to the reproduce the \citet{dutton2014} concentration relation: isothermal and Burkert. We see that isothermal requires far larger concentrations and thus far smaller cores. Surprisingly, Burkert does not stray overly far from \citet{dutton2014} values. For the other halo models we find that DC14 recovers the prior best with only UGC05721 deviating at $2.6\sigma$, in agreement with \citet{li_comprehensive_2020}. The Einasto halo reproduces the prior fairly well, but has a tendency towards lower concentrations, with significant deviations in  DDO161 ($2\sigma$), UGC08837 ($2.9\sigma$), and UGC11820 ($2.7\sigma$). This is similar to NFW, where we have DDO161 ($2.8\sigma$), UGC05721 ($2.0\sigma$), UGC08837 ($2.9\sigma$), and UGC11820 ($2.2\sigma$). We see that L13 has considerable scatter above the prior relation with 7 galaxies deviating at $>2.9\sigma$. 
\begin{figure*}
    \resizebox{0.49\hsize}{!}{\includegraphics{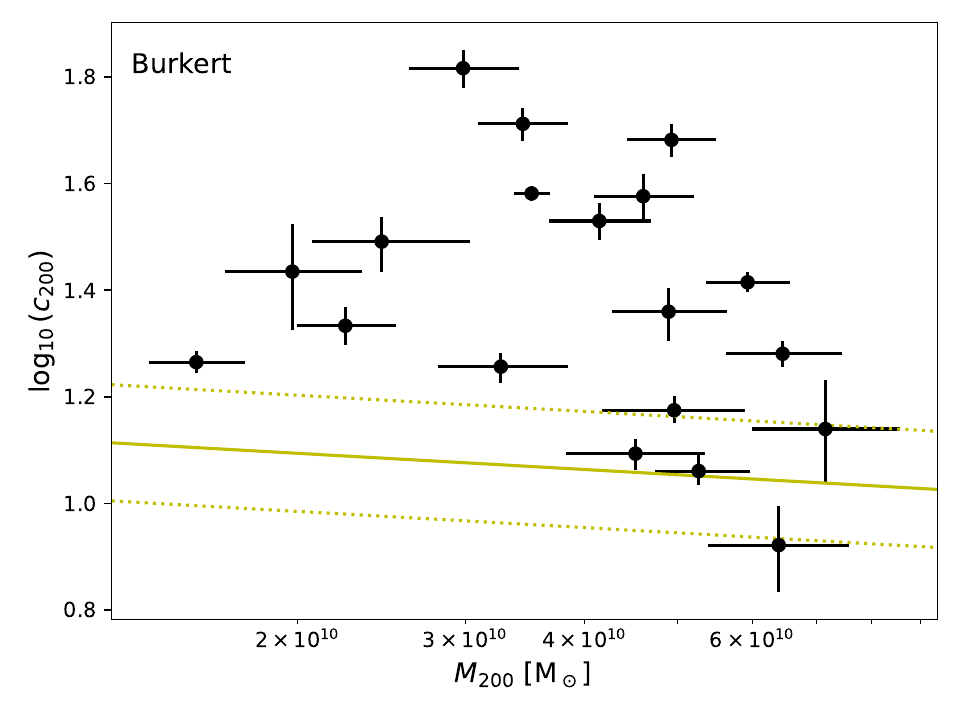}}
    \resizebox{0.49\hsize}{!}{\includegraphics{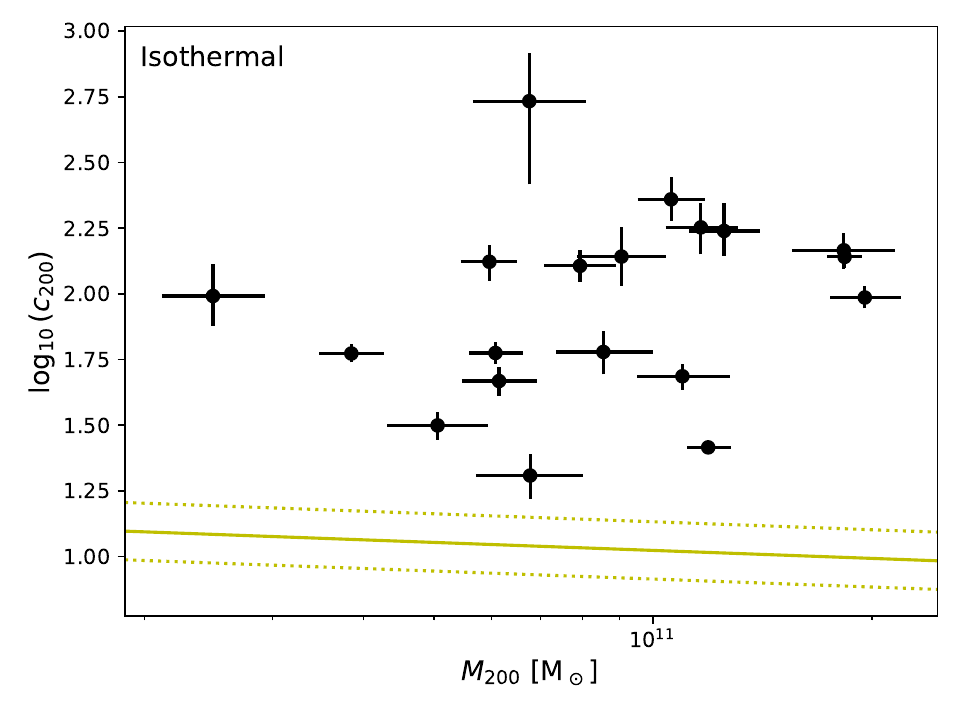}}

    \resizebox{0.49\hsize}{!}{\includegraphics{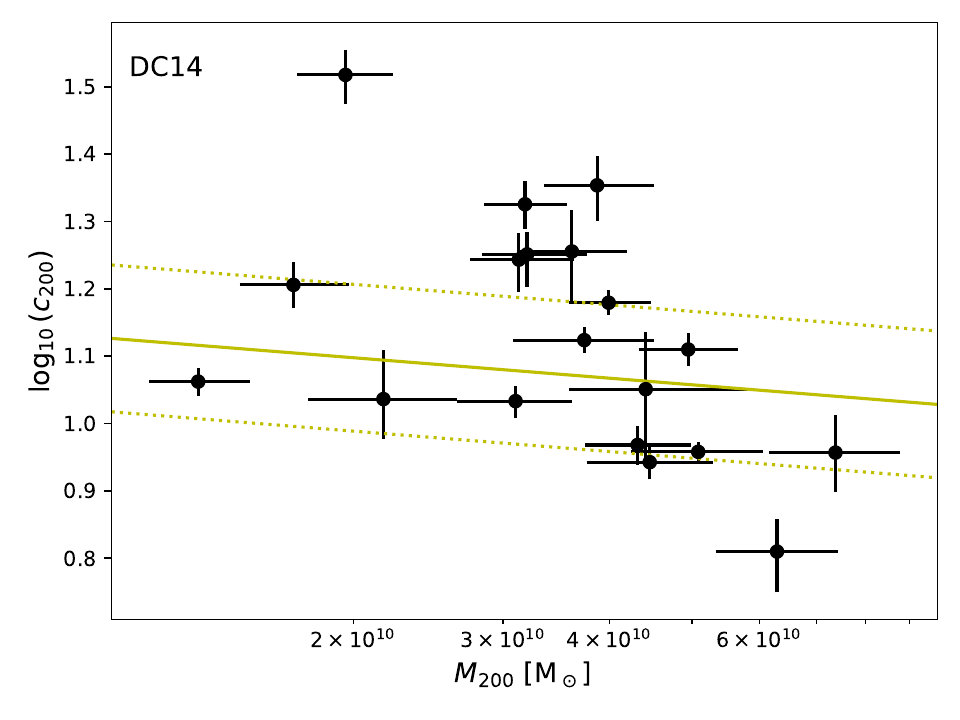}}
    \resizebox{0.49\hsize}{!}{\includegraphics{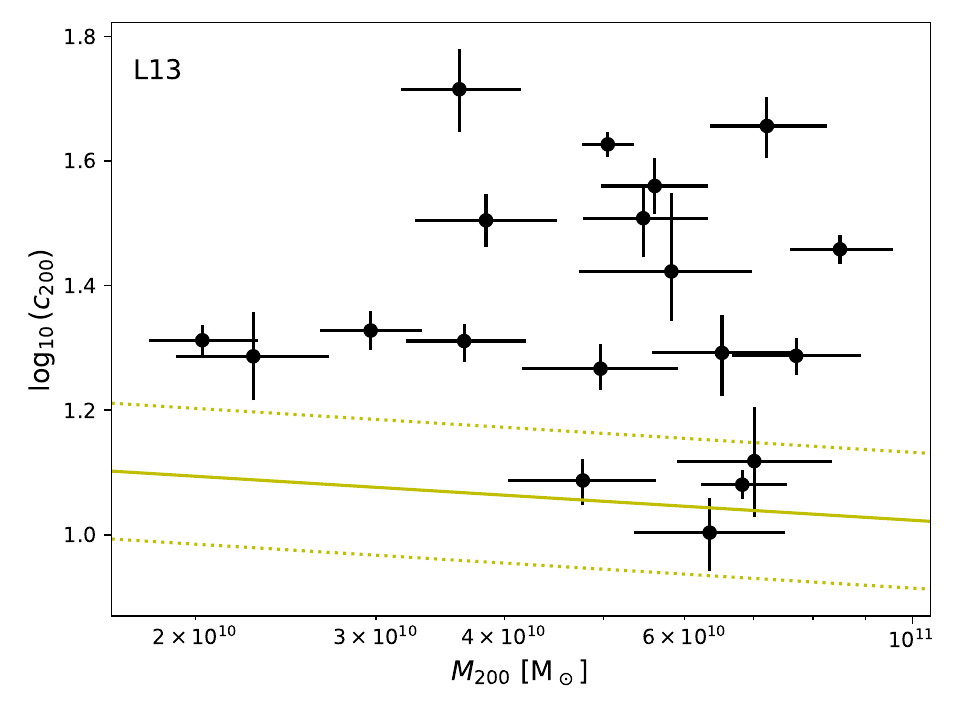}}

    \resizebox{0.49\hsize}{!}{\includegraphics{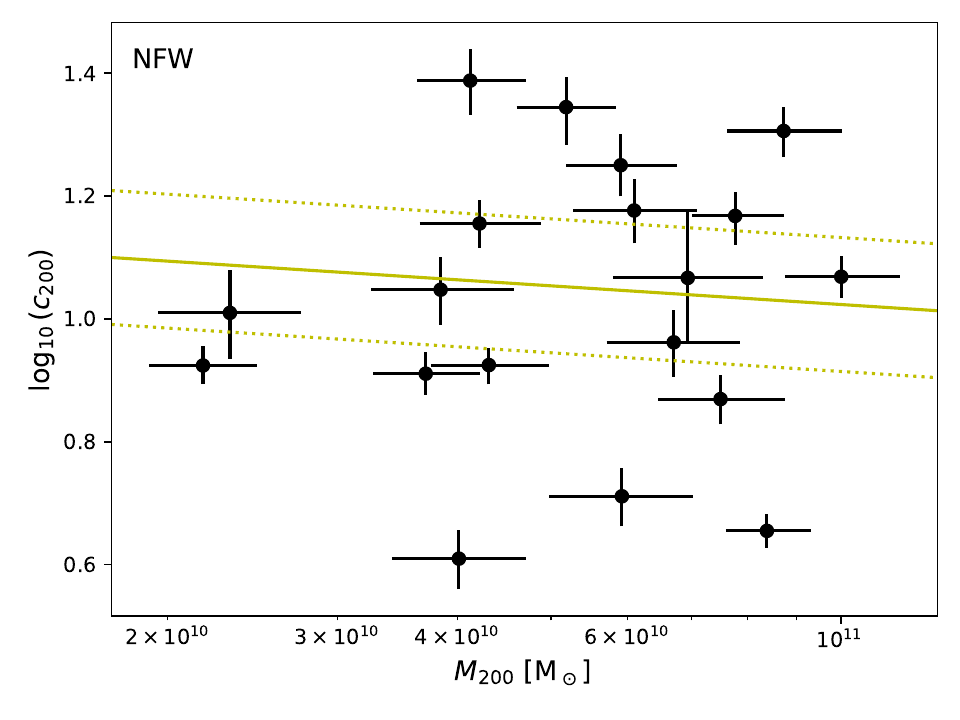}}
    \resizebox{0.49\hsize}{!}{\includegraphics{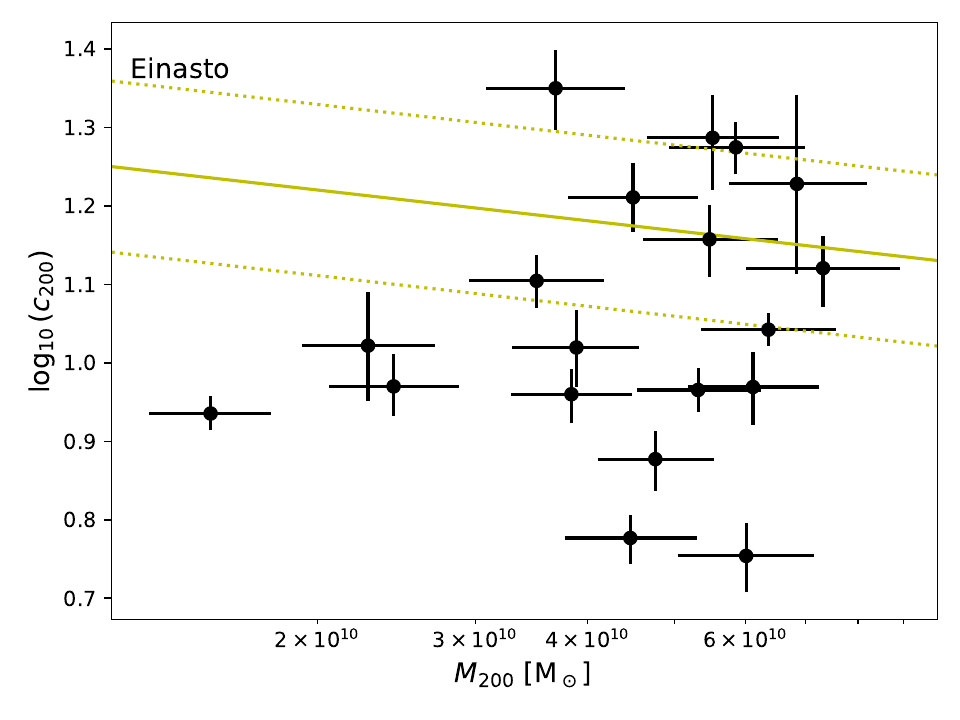}}
    \caption{$c_{200}$ vs $M_{200}$ for the various halo models fitted to the SPARC dwarf sample. The yellow lines reflect the concentration model from \citet{dutton2014}. Note that this prior was not applied to the Burkert or isothermal halos.}
    \label{fig:sparc-c200}
\end{figure*}

As a novel test, we attempt to re-fit the relation for $c_{200}$. We focus on the cored halos: L13, Burkert, and isothermal. We are thus fitting 
\begin{equation}
    c_{200} (M_{200}) = a - b\log_{10}\left(\frac{M_{200} h}{10^{12} \ \mathrm{M}_\odot}\right) \; ,
\end{equation} 
to find $a$ and $b$ (we use lognormal priors, centred on the \citet{dutton2014} values, with a width of $0.3$ dex). The aim is to explore whether the cored halos are consistent a modified $a$ normalisation. This might be expected because cored halos do not have $r_{-2} = r_s$ in general, where $r_{-2}$ is the point where $\frac{d \ln \rho}{d \ln r} = -2$. Importantly, \citet{dutton2014} derive their $c_{200}$ from NFW halos which have $r_{-2} = r_s$. To give it an objective definition, $c_{200}$ is more properly defined by $r_{-2}$ than $r_s$. Thus, we might reasonably expect a rescaling of $c_{200}$ by $r_{-2}/r_s$. Interestingly, we find $c_{200,\mathrm{L13}}$ is between a factor of $1.81$ and $2.37$ times larger than that from \citet{dutton2014} at $1\sigma$. This suggests compatibility with the fact that L13 has $r_{-2} = 2 r_s$. Similarly, Burkert requires a multiplicative factor of between $1.28$ and $1.65$, very compatible with $r_{-2} \approx 1.52 r_s$. For L13 there is no significant adjustment of $b$ (and none expected since it is from a generalised NFW family). With Burkert, $b$ approximately doubles, indicating a far stronger mass dependence. These fits also reduce the significance of deviations drastically, only 4 galaxies now lie at $\sim 2\sigma$ with both profiles. 

To achieve a similar result in isothermal, we require a multiplicative factor of $\sim 7$ on $c_{200}$ and reducing $b$ by a factor of 10. However, there is no physically motivated factor to consider here as $r_{-2} \to \infty$, and two galaxies remain $\gtrsim 3\sigma$ away among a total of 8 at $> 2\sigma$. Additionally, the required lack of mass dependence is alarming. Einasto and NFW both end up requiring negligible mass dependence as well. For Burkert and L13, these $c_{200}$ relationships are clearly worth exploring in the full SPARC sample. If this result is robust in larger samples it provides a fascinating consistency check, and a means to leverage priors derived for NFW halos in other contexts.

For the mass relation from \citet{moster_2013}, we find that Einasto and L13 halos reproduce this the best, followed by NFW and Burkert. This can be quantified in terms of the number of significant deviations from the trend. Einasto and L13 have no significant deviations, Burkert has only UGC08286 at $2.0\sigma$, and NFW has DDO161 at $2.9\sigma$. DC14 produces some substantially smaller $M_{200}$ values than expected, with 4 galaxies exceeding a $2\sigma$ deviation. Isothermal generally requires larger halo masses, with 9 galaxies deviating at $> 2.2\sigma$. 
\begin{figure*}
    \resizebox{0.49\hsize}{!}{\includegraphics{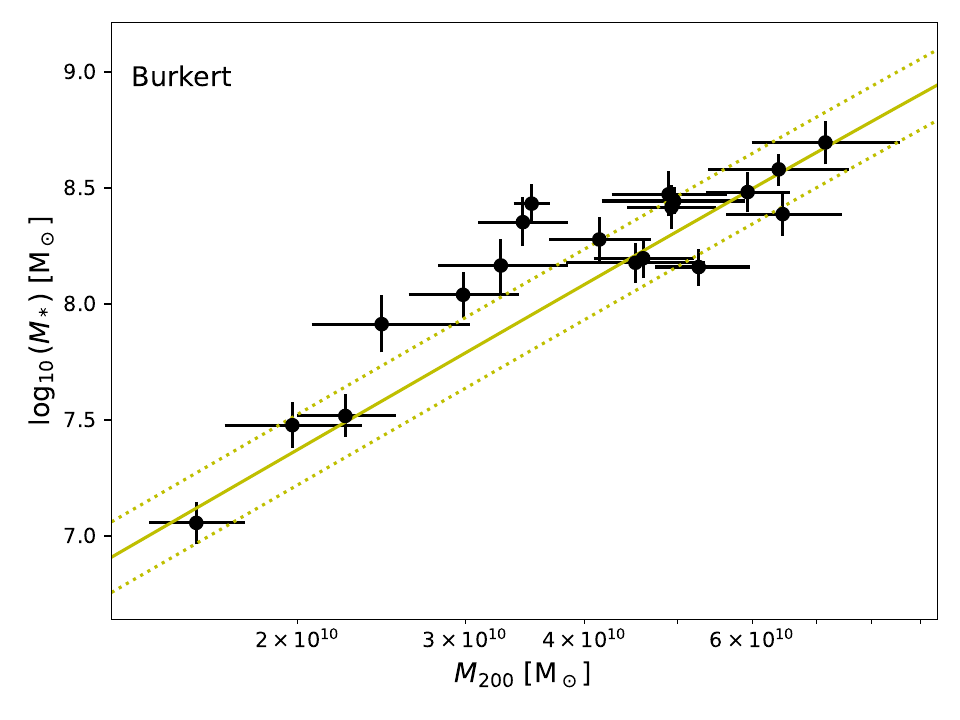}}
    \resizebox{0.49\hsize}{!}{\includegraphics{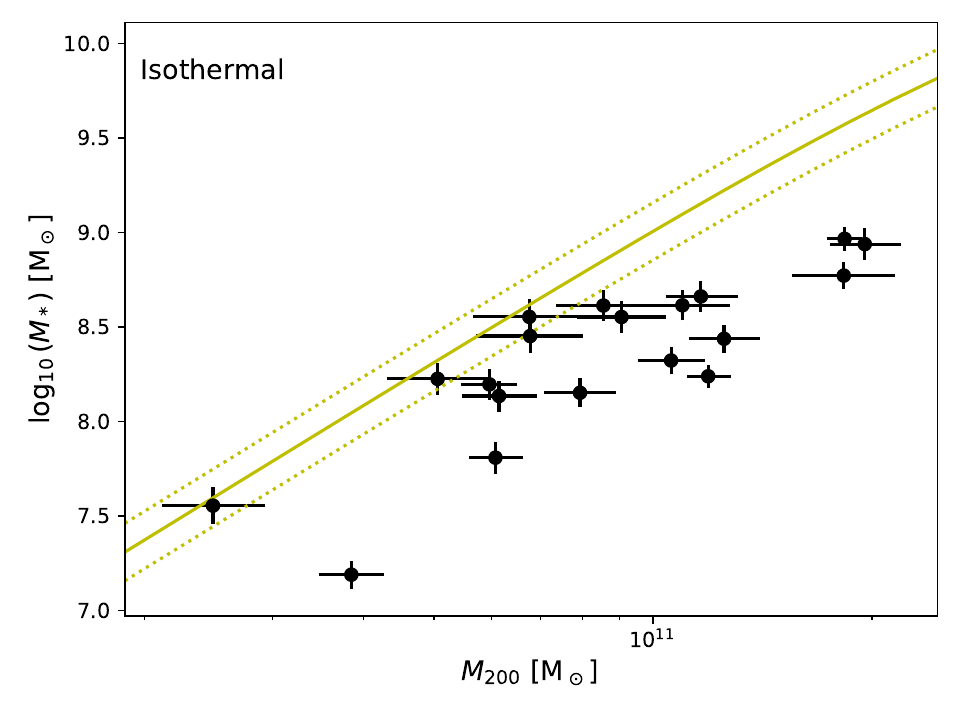}}

    \resizebox{0.49\hsize}{!}{\includegraphics{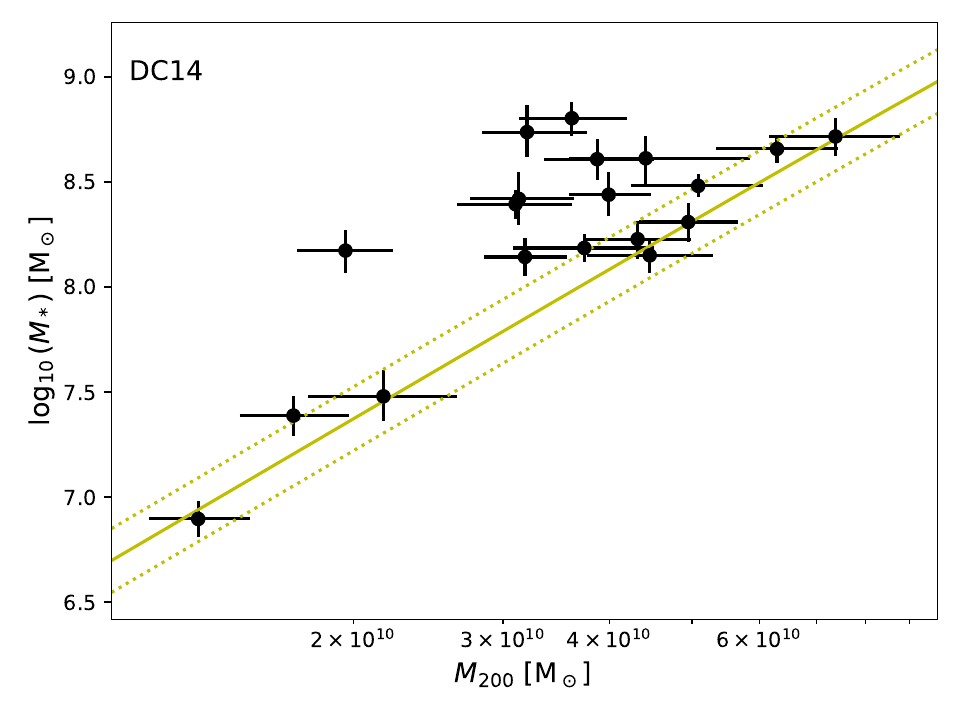}}
    \resizebox{0.49\hsize}{!}{\includegraphics{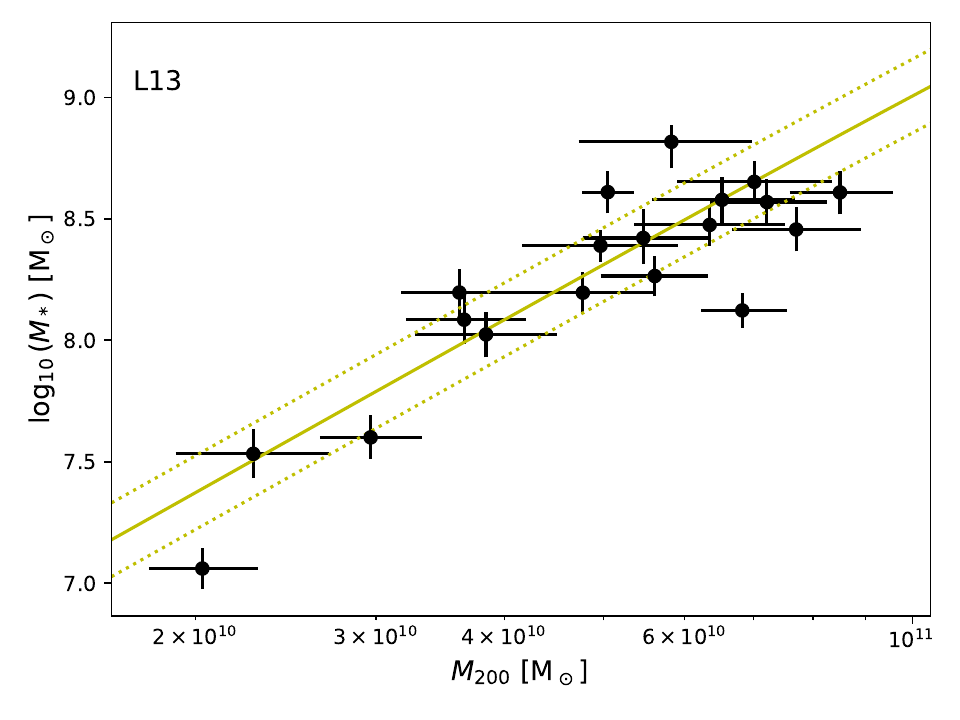}}

    \resizebox{0.49\hsize}{!}{\includegraphics{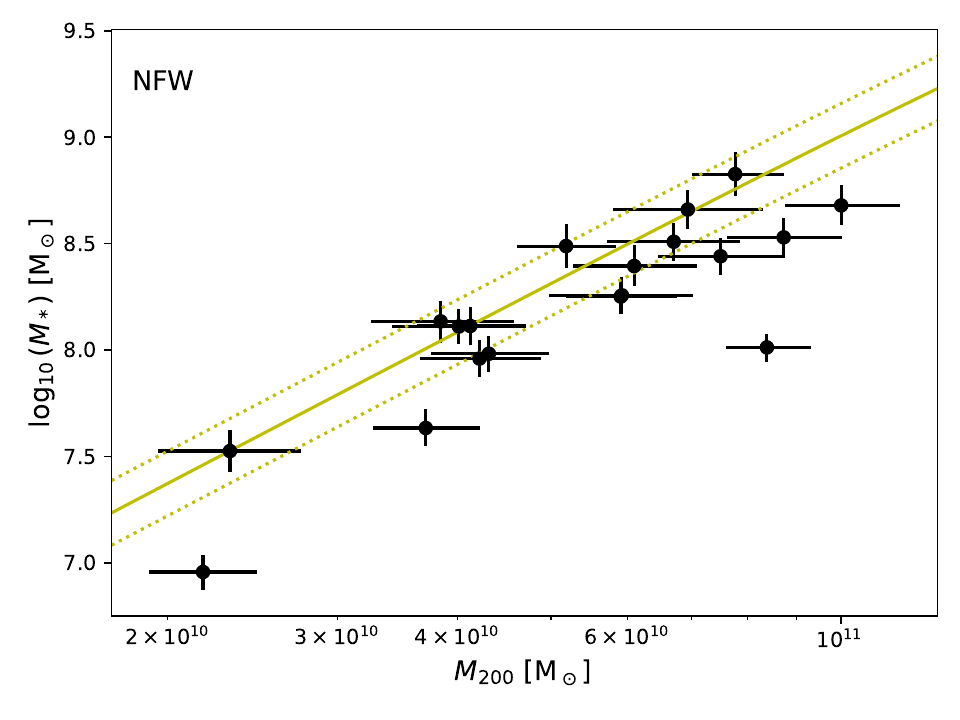}}
    \resizebox{0.49\hsize}{!}{\includegraphics{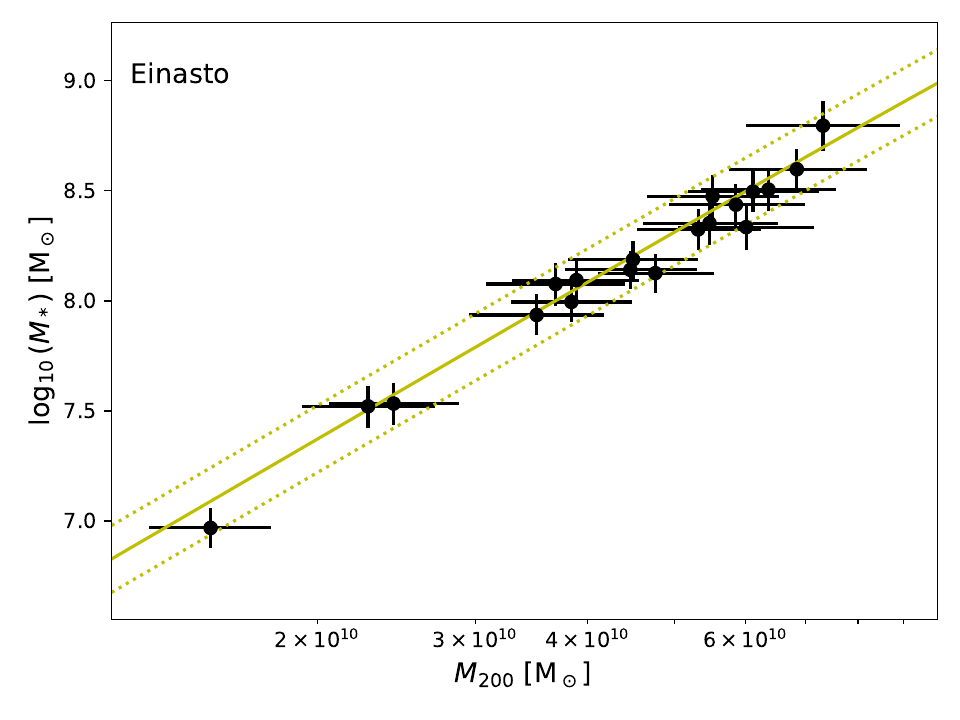}}
    \caption{$M_*$ vs $M_{200}$ for the various halo models fitted to the SPARC dwarf sample. The yellow lines prior from \citet{moster_2013}.}
    \label{fig:sparc-m200}
\end{figure*}

Next for the DM halos we have the mass-to-light ratios. Once again the Einasto halo recovers the prior the best, perhaps to be expected given its extra free parameter. Interestingly, NFW is the next best. Burkert has good agreement with the prior, with only two values outside the $1\sigma$ confidence interval. The L13 and DC14 halos have 3 and 4 weakly outlying values respectively. The isothermal halo tends to require larger mass-to-light ratios in general. However, there are no significant deviations from the prior for any halo. 
\begin{figure*}
    \resizebox{0.49\hsize}{!}{\includegraphics{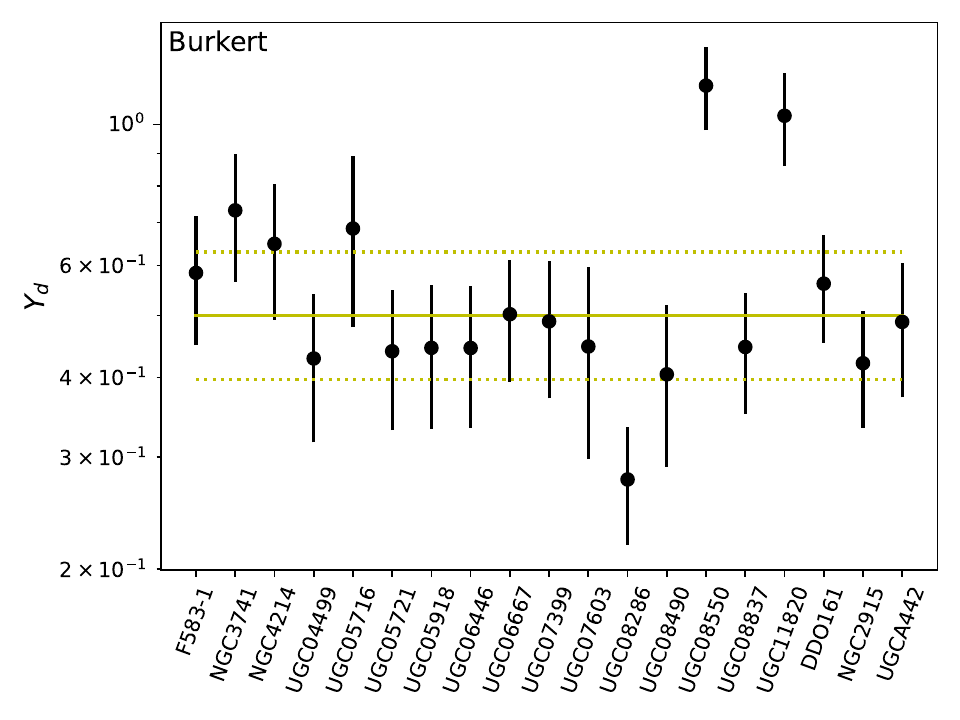}}
    \resizebox{0.49\hsize}{!}{\includegraphics{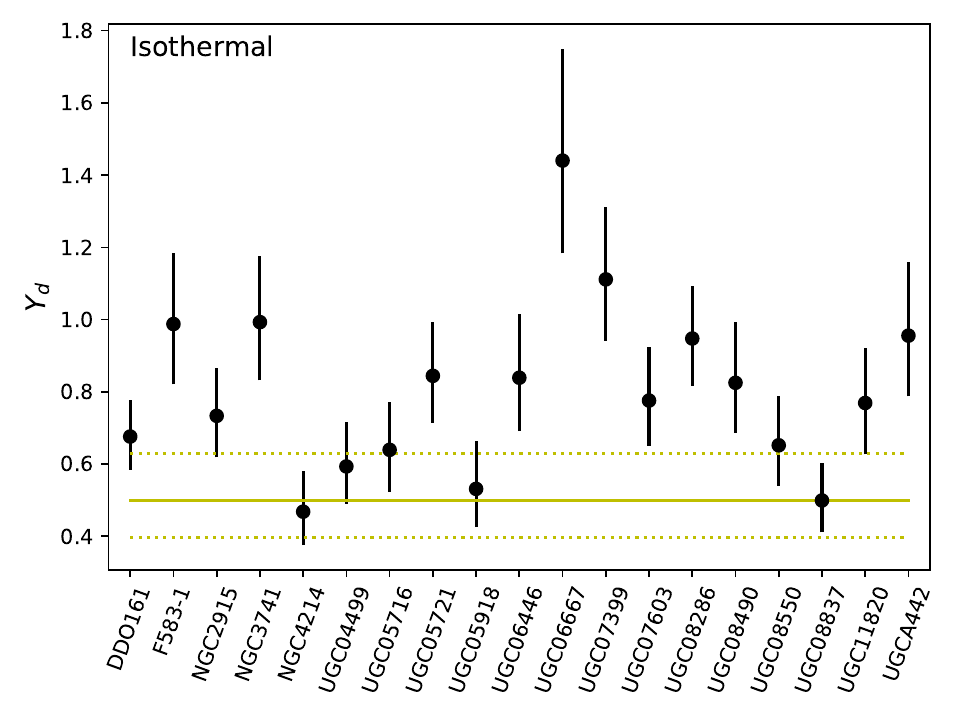}}

    \resizebox{0.49\hsize}{!}{\includegraphics{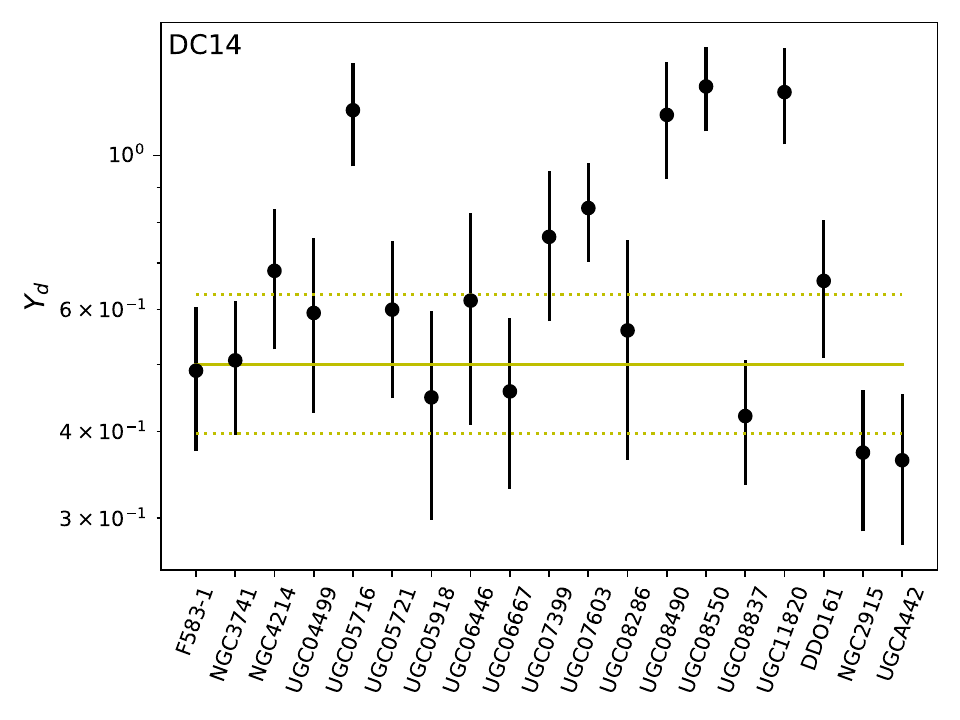}}
    \resizebox{0.49\hsize}{!}{\includegraphics{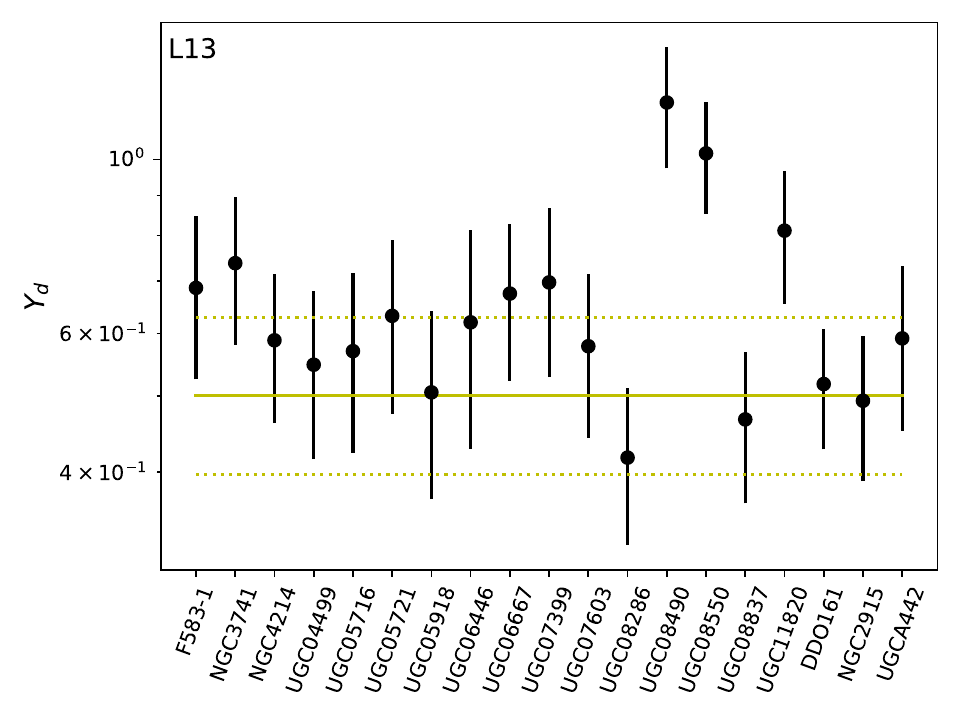}}

    \resizebox{0.49\hsize}{!}{\includegraphics{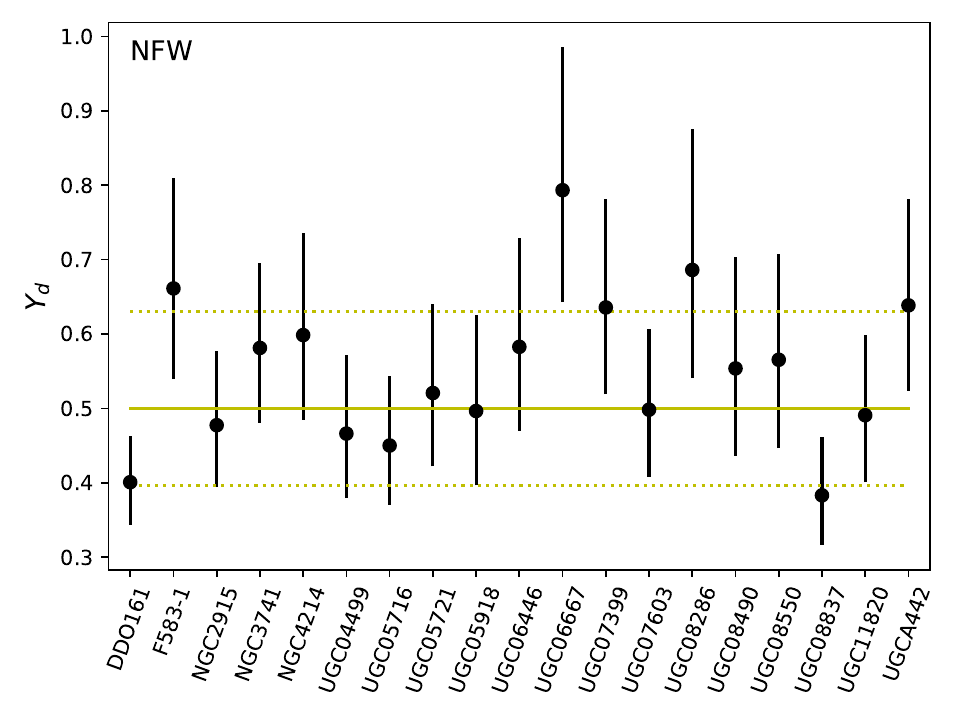}}
    \resizebox{0.49\hsize}{!}{\includegraphics{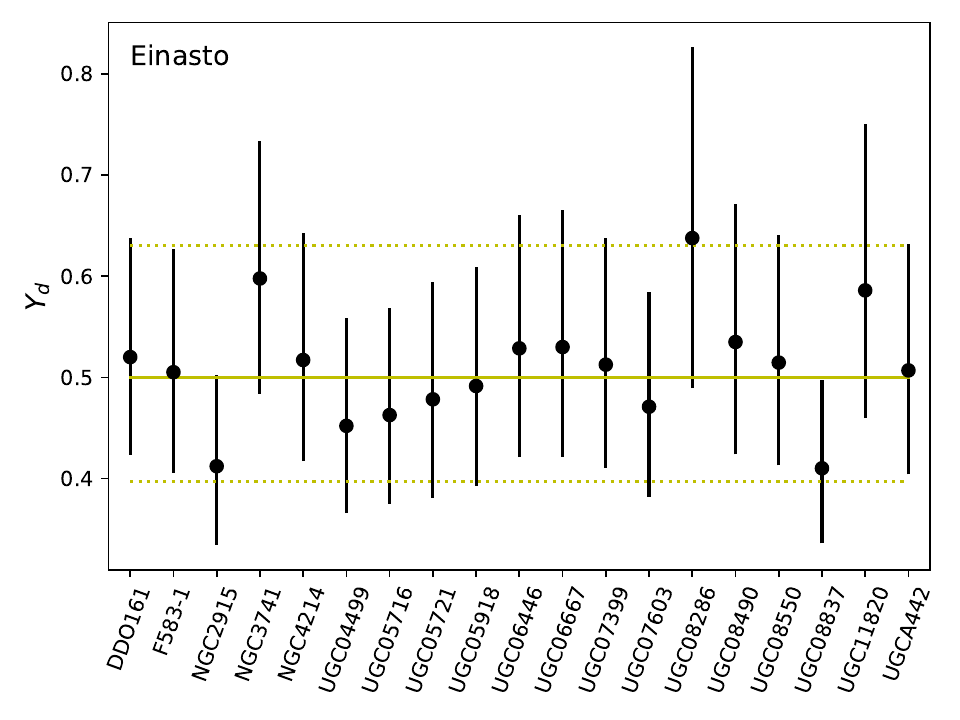}}
    \caption{$Y_d$ fitted for the various halo models fitted to the SPARC dwarf sample. The yellow lines reflect the prior from \citet{schombert_2018}.}
    \label{fig:sparc-yd}
\end{figure*}

Finally, we can report on whether any of the models require significant adjustments of the nuisance parameters. We do so in Table~\ref{tab:dm-nuisance}. Here we can see that only the isothermal and DC14 halos commonly require significant deviations in inclination. 
\begin{table}
    \caption{Average relative deviation in distance and inclination with halo model in SPARC.}
    \label{tab:dm-nuisance}
    \begin{tabular}{|l|l|l|}
        \hline
        Halo & $\bar{\sigma}_d$ & $\bar{\sigma}_\theta$ \\
        \hline
        Burkert & 0.48 & 0.83 \\
        DC14 & 0.58 & 1.10 \\
        L13 & 0.64 & 0.72 \\
        Einasto & 0.14 & 0.69 \\
        isothermal & 0.69 & 1.51 \\
        NFW & 0.34 & 0.89 \\
        \hline
    \end{tabular}
\end{table}

\begin{figure*}
    \resizebox{0.49\hsize}{!}{\includegraphics{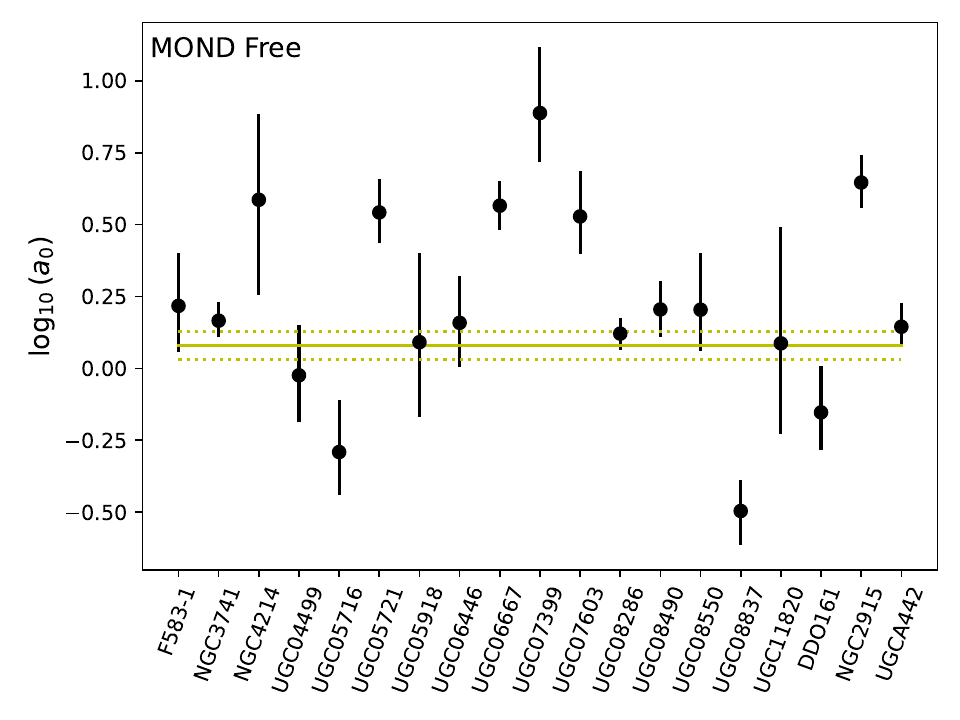}}
    \resizebox{0.49\hsize}{!}{\includegraphics{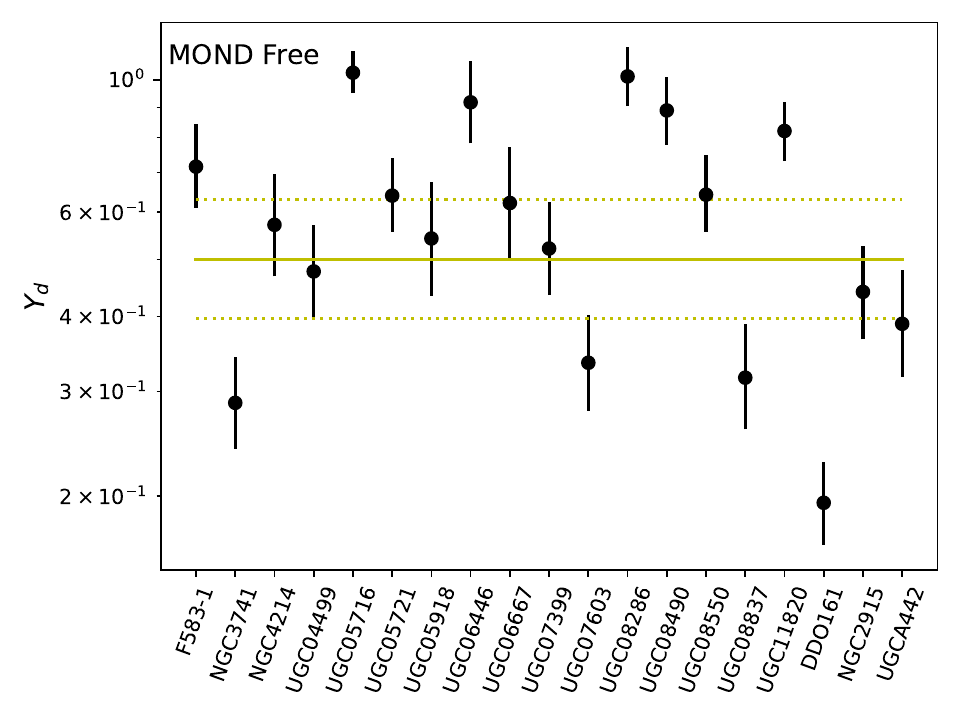}}

    \resizebox{0.49\hsize}{!}{\includegraphics{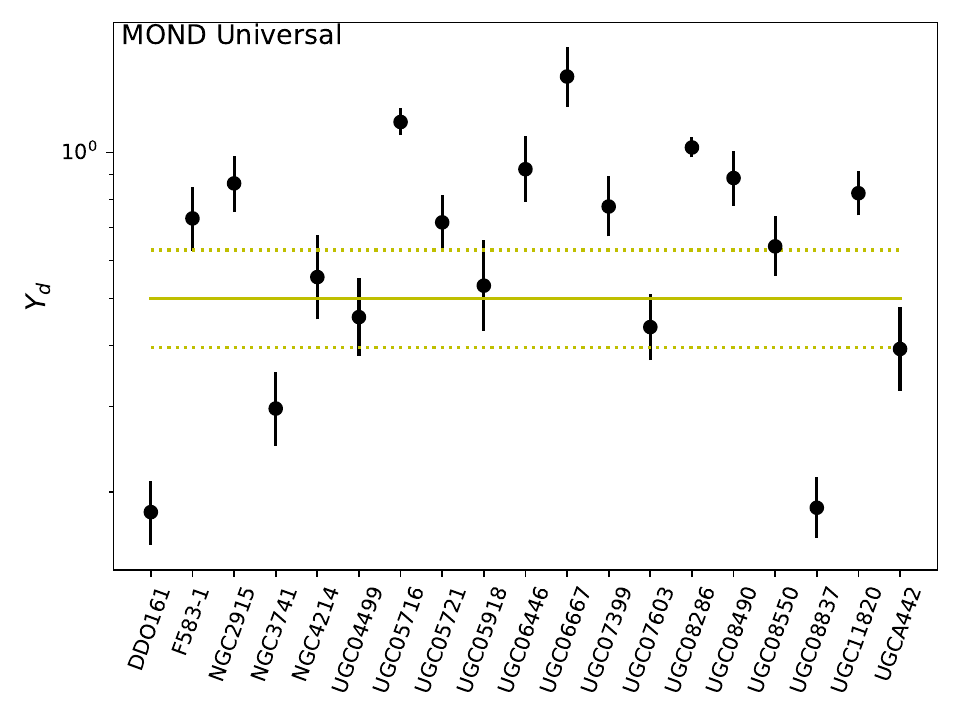}}
    \caption{MOND model values of $a_0$ and $Y_d$ for the various galaxies in the SPARC dwarf sample. For $Y_d$ the yellow lines reflect the prior from \citet{schombert_2018}. Whereas, for $a_0$ it reflects $1.2\times 10^{-13}$ km s$^{-2}$ with a 12\% tolerance from \citet{desmond-rar-2023}.}
    \label{fig:sparc-mond-prior}
\end{figure*}

For the MOND fits to SPARC we present the priors (with $\zeta = 0$) in Fig.~\ref{fig:sparc-mond-prior}. We see that $Y_d$ has 2 significant deviations around the prior these are only just $> 2\sigma$. Additionally, 6 galaxies deviate from the universal $a_0$, while including the statistical and systematic errors from \citet{desmond-rar-2023}. Notably, 5 of these achieve $>3\sigma$. The mean value of $a_0 = 1.66$ with a relative error factor of $1.3$.

When $a_0$ is fixed, the $Y_d$ values must generally become larger, with 4 significant deviations around the prior, 3 of these are $\sim 3\sigma$ (notable given the width of the prior).

\begin{figure*}
    \resizebox{0.49\hsize}{!}{\includegraphics{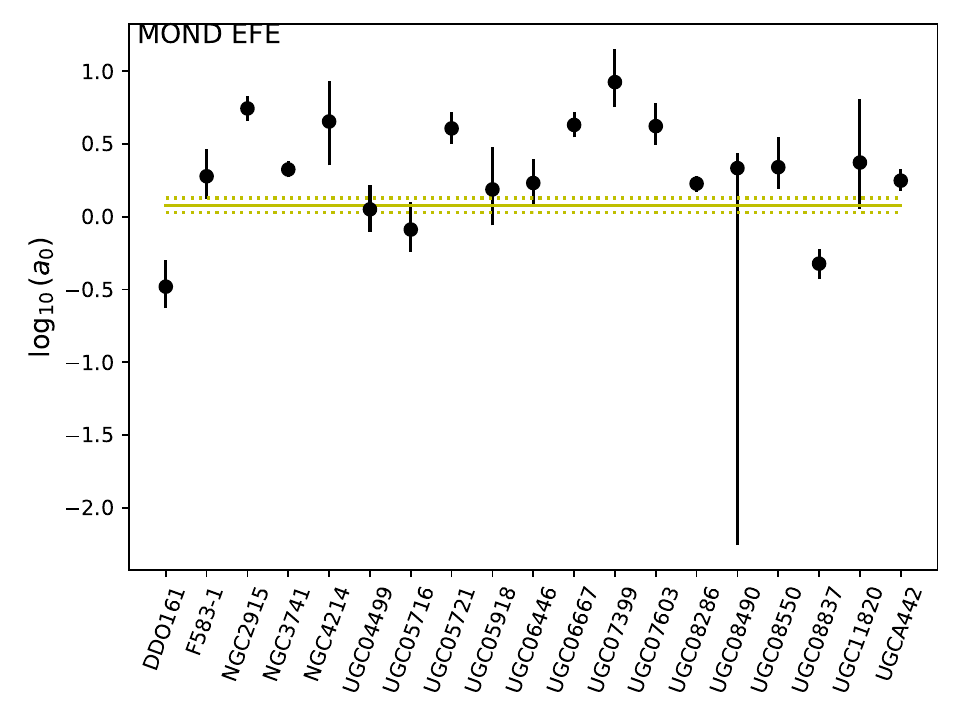}}
    \resizebox{0.49\hsize}{!}{\includegraphics{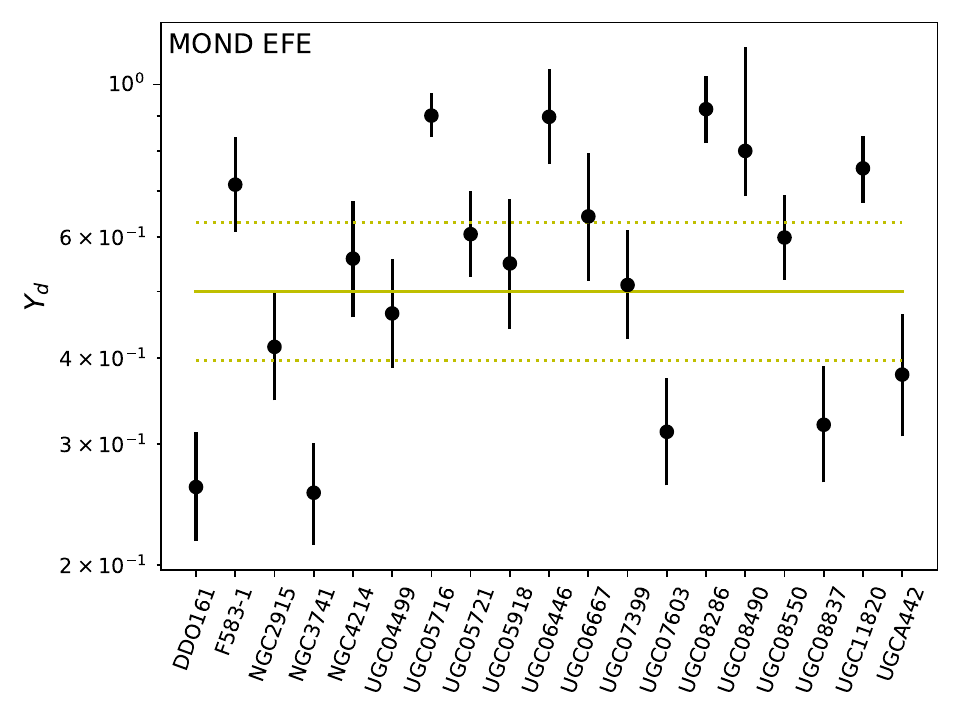}}

    \resizebox{0.49\hsize}{!}{\includegraphics{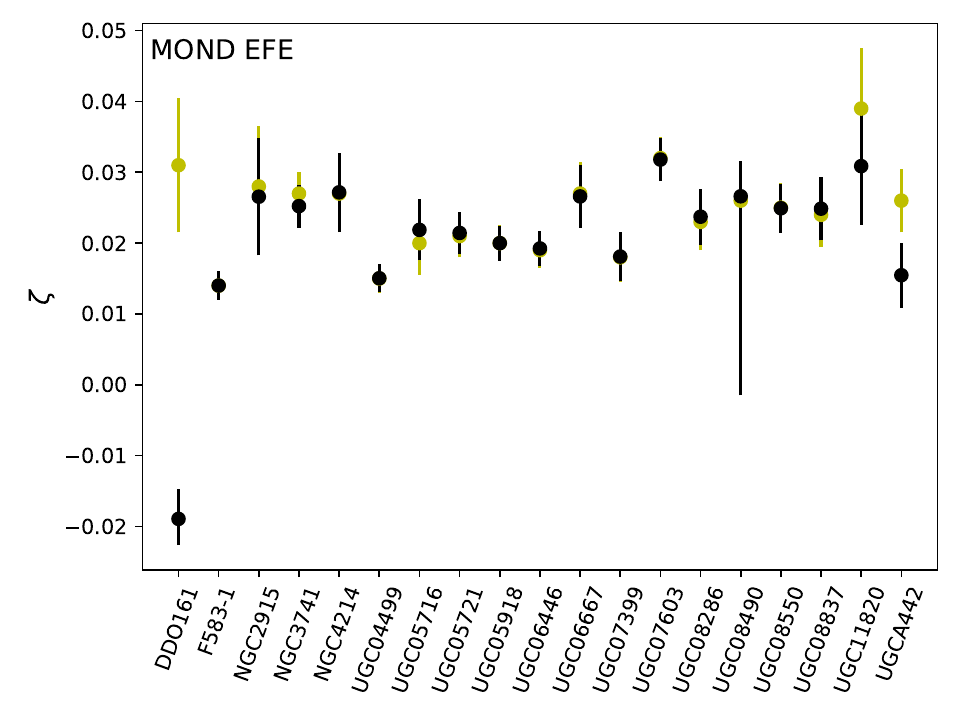}}
    \resizebox{0.49\hsize}{!}{\includegraphics{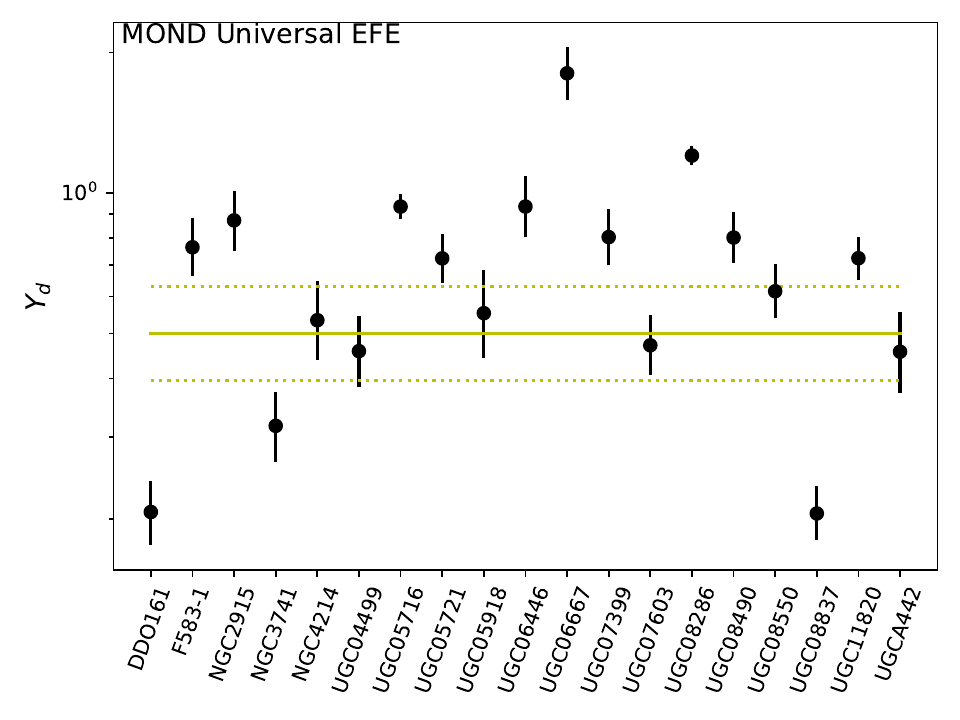}}

    \resizebox{0.49\hsize}{!}{\includegraphics{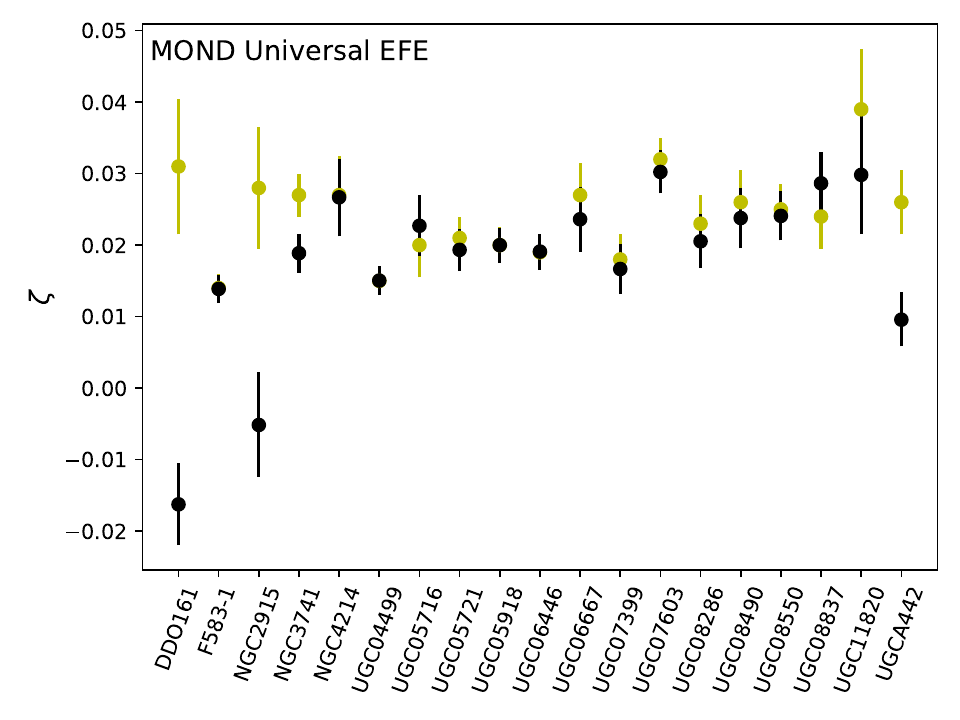}}
    \caption{MOND model values of $a_0$, $\zeta$, and $Y_d$ for the various galaxies in the SPARC dwarf sample. For $Y_d$ the yellow lines reflect the prior from \citet{schombert_2018}. Whereas, for $a_0$ it reflects $1.2\times 10^{-13}$ km s$^{-2}$ with a 12\% tolerance from \citet{desmond-rar-2023}. Finally, for $\zeta$ the yellow points reflect the values found in \citet{sep-zeta}.}
    \label{fig:sparc-mond-prior-zeta}
\end{figure*}

Figure~\ref{fig:sparc-mond-prior-zeta} displays our consideration of the MOND external field effect. In this case, our statistical question is ``can our MOND model accommodate the $\zeta$ estimates from local structure~\citep{sep-zeta} while $a_0$ is a free parameter?'' For $Y_d$, the deviations are reduced, with only UGC05716 at $2\sigma$. However, we see a slight increase in deviations from the universal $a_0$ value from $\zeta = 0$. There are 8 galaxies with significant deviations, with NGC2915 achieving $4.8\sigma$. Lastly, DDO161 can conform to the universal $a_0$, but then requires a $3.6\sigma$ divergence from the value found for $\zeta$~\citep{sep-zeta} (to the extent of a sign reversal even).

When we fix $a_0$ and test $\zeta$, we find a general increase in $Y_d$ and a larger divergence in $\zeta$ from the \citet{sep-zeta} values. The only significant change in $\zeta$ is seen in NGC2915. For universal $a_0$, $\zeta \neq 0$ offers no advantage over the $\zeta = 0$ case in any galaxy.

When $a_0$ is free we have no significant sample-level preference for or against an external field effect. There are 5 galaxies with significant preferences for $\zeta \neq 0$ with significances $2.6\sigma$, $2.7\sigma$, $3.1\sigma$, $3.8\sigma$, and $4.2\sigma$. However, NGC3741 prefers a zero external field effect at $4.6\sigma$ and UGCA442 achieves a $2.9\sigma$ preference. 

When we fix $a_0$ there is sample-level preference for $\zeta=0$ at $2.4\sigma$. There are now 4 galaxies with significant preferences both for and against the effect. The largest preference for $\zeta=0$ remains NGC3741 at $4.3\sigma$. The largest preference for $\zeta \neq 0$ is DDO161 at $3.6\sigma$.

Given that we are testing in dwarf galaxies, we should expect the external field effect to be most apparent in our sample. However, we have found only weak evidence against the effect, contrary to the results of \citet{sep-zeta}. A similar conclusion was drawn for the larger SPARC sample in \citet{2022MNRAS.517..130P,2025Parti...8...65S}.

\begin{figure*}
    \resizebox{0.49\hsize}{!}{\includegraphics{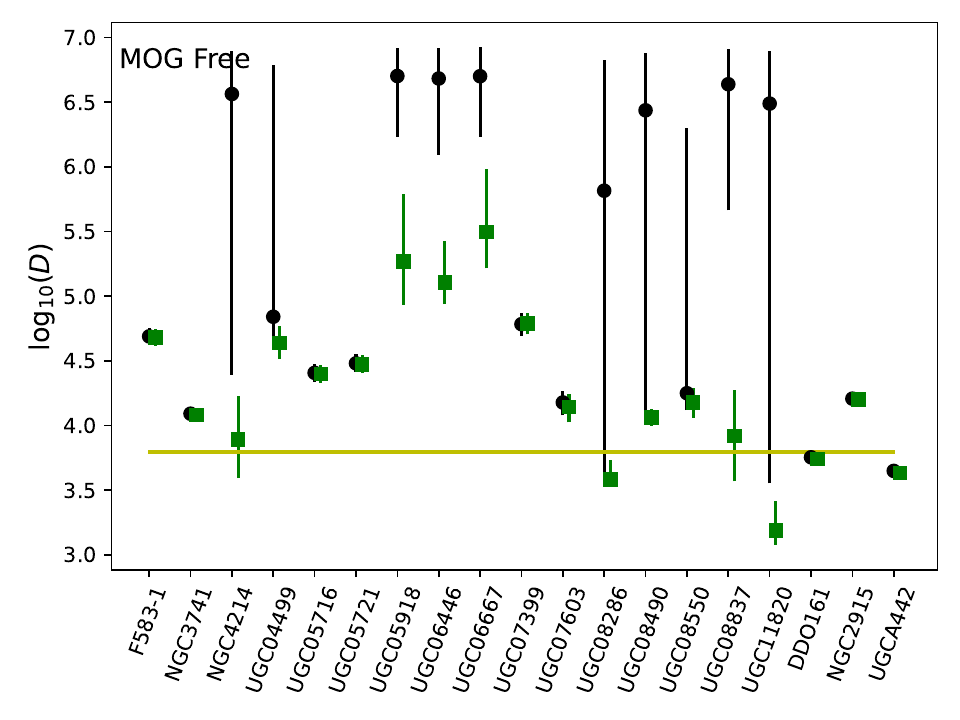}}
    \resizebox{0.49\hsize}{!}{\includegraphics{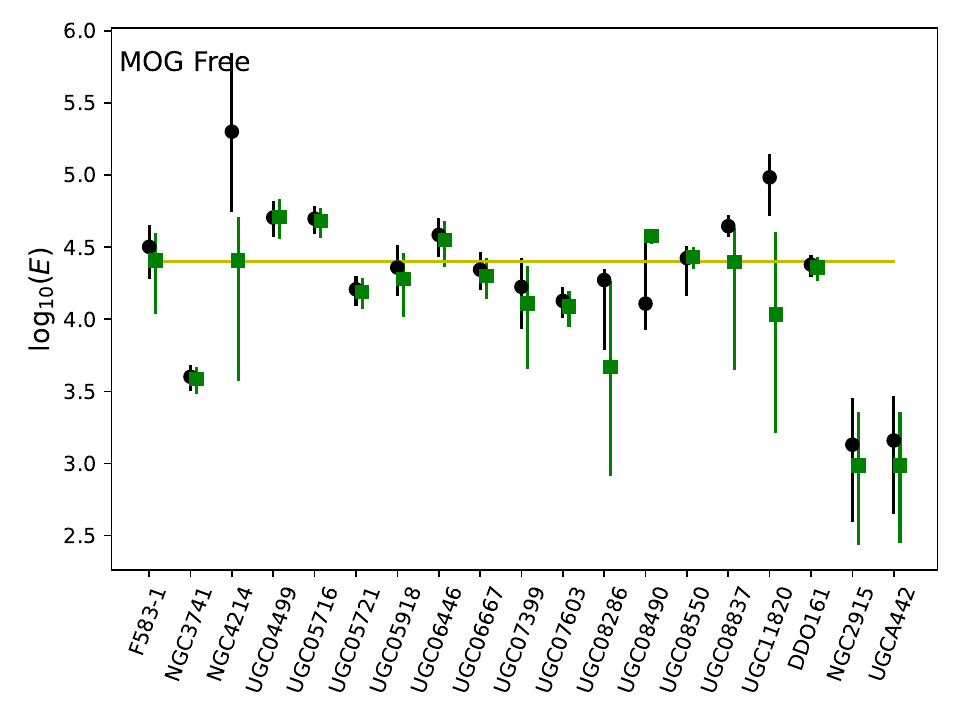}}

    \resizebox{0.49\hsize}{!}{\includegraphics{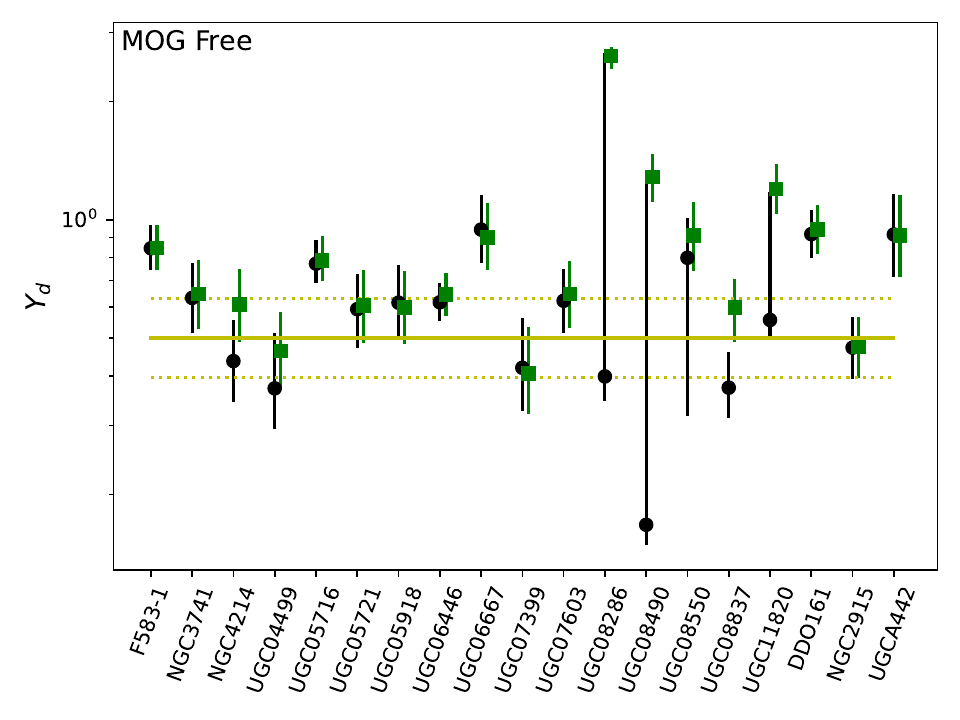}}
    \resizebox{0.49\hsize}{!}{\includegraphics{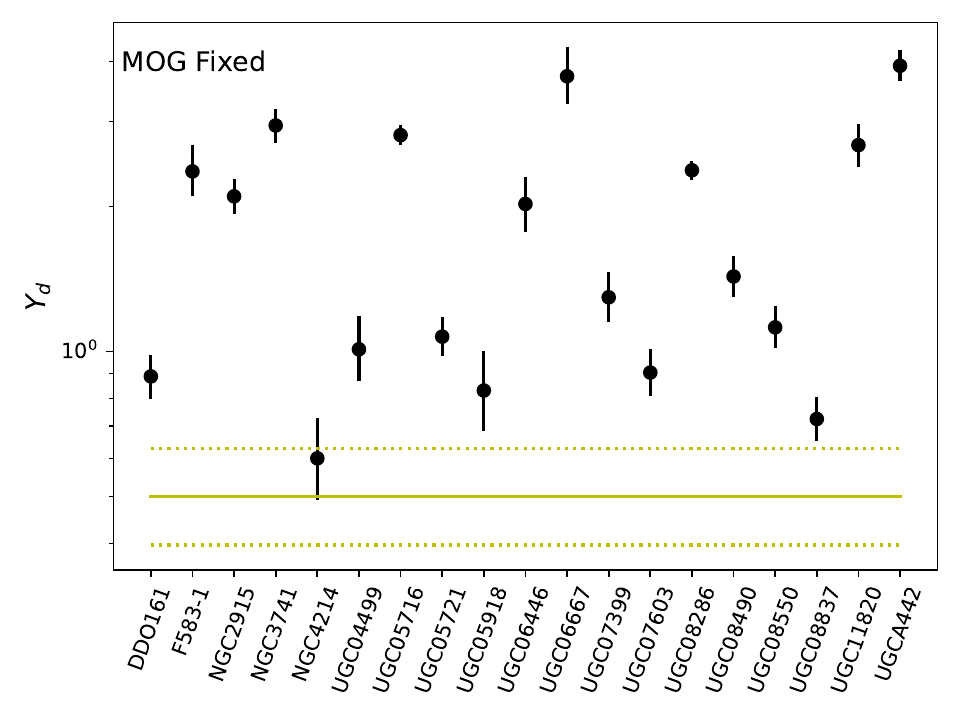}}
    \caption{MOG model values of $D$, $E$, and $Y_d$ for the various galaxies in the SPARC dwarf sample. Black circles display the flat prior result, green squares are that for a lognormal prior. For $Y_d$ the yellow lines reflect the prior from \citet{schombert_2018}. Whereas, for $D$ it reflects $6250$ M$_\odot^{1/2}$ kpc$^{-1}$ and for $E$ it is $25000$ M$_\odot^{1/2}$.}
    \label{fig:sparc-mog-prior}
\end{figure*}

The MOG fits for SPARC are displayed in Fig.~\ref{fig:sparc-mog-prior}. There are some rather significant deviations from the universal values of both $D$ and $E$ as well as a single rather large mass-to-light ratio. The mean value of $D$ is $\sim 2\times 10^4$ M$_\odot^{1/2}$ kpc$^{-1}$, around 3 times the universal value, with a standard deviation of $5.6\times 10^4$ M$_\odot^{1/2}$ kpc$^{-1}$. There are five highly problematic galaxies with a flat prior: UGC05918, UGC06446, UGC06667, UGC08286, and UGC08490. These favour $D \to \infty$. Note that these galaxies never find a convergent finite $D$ value, regardless of the upper limit on the $D$ prior. The rotation curves of these galaxies have a particular commonality: the MOG fit cannot accommodate the data at both small and large radii. Each fit substantially underestimates the inner regions of the rotation curve. The physical interpretation of such a large $D$ value suggests that the exponential factor is reduced to $0$, leaving only a mass-dependent boost to the gravitational strength. We can re-examine trends in $D$ by looking at a lognormal prior. Here there are 8 galaxies that diverge from the universal value by at least an order of magnitude at high significance (highest is $13\sigma$). These are FS583-1, UGC4499, UGC05716, UGC05721, UGC05918, UGC06446, UGC06667, and UGC07399. Notably, the lognormal prior largely tames the problems in UGC08286 and UGC08490. 

As for $E$, the mean is $\sim 1.4\times 10^4$ M$_\odot^{1/2}$, around half the universal value, with a standard deviation of $2.9\times 10^4$ M$_\odot^{1/2}$. Out of 19 galaxies, 8 deviate at $> 2\sigma$, with 4 of these exceeding $3\sigma$, and 1 of them exceeding $10\sigma$ (NGC3741). NGC3741 requires $E$ an order of magnitude or more below the universal value. NGC2915 and UGCA442 are similar, but at $3.9\sigma$. There is far less tension between the fitted $E$ values and the universal than existed for $D$. 

Thus, universal values for both parameters seem unlikely to avoid substantial disagreement with at least some dwarf galaxies. Importantly, the variance here is at the level of orders of magnitude. When $D$ and $E$ are set to their universal values we find a need for somewhat concerning mass-to-light ratios, 7 of which exceed 2. This seems somewhat unlikely following the population synthesis work in \citet{schombert_2018} or the study of the LITTLE THINGS sample in \citet{2016AJ....152..177H}. In addition, the model robustness degrades substantially, as seen in Fig.~\ref{fig:ts}.

Finally, we display the average deviation in the nuisance parameters in Table~\ref{tab:mog-nuisance}. We can see this is low for the free MOND model, but quite comparable to DM halos for MOND with universal $a_0$. For free MOG, we see low adjustments relative to DM halos. However, the nuisance adjustments are greatly enhanced for the universal MOG model, indicating it is very common to significantly adjust the nuisance parameters to achieve better fits. 
\begin{table}
    \caption{Average relative deviation in distance and inclination with MOND and MOG models in SPARC.}
    \label{tab:mog-nuisance}
    \begin{tabular}{|l|l|l|}
        \hline
        Model & $\bar{\sigma}_d$ & $\bar{\sigma}_\theta$ \\
        \hline
        MOND Free & 0.33 & 0.43 \\
        MOND Universal & 1.06 & 1.19 \\
        MOG Free & 0.54 & 0.55 \\
        MOG Universal & 2.33 & 1.41 \\
        \hline
    \end{tabular}
\end{table}

\subsection{LITTLE THINGS}
Now we turn to the LITTLE THINGS sample. Notably, these are not included in SPARC, due to a concern that the resolution of the observations forces the inclusion of non-circular motion effects~\citep{lelli_sparc_2016}. 

\begin{figure*}
    \resizebox{0.49\hsize}{!}{\includegraphics{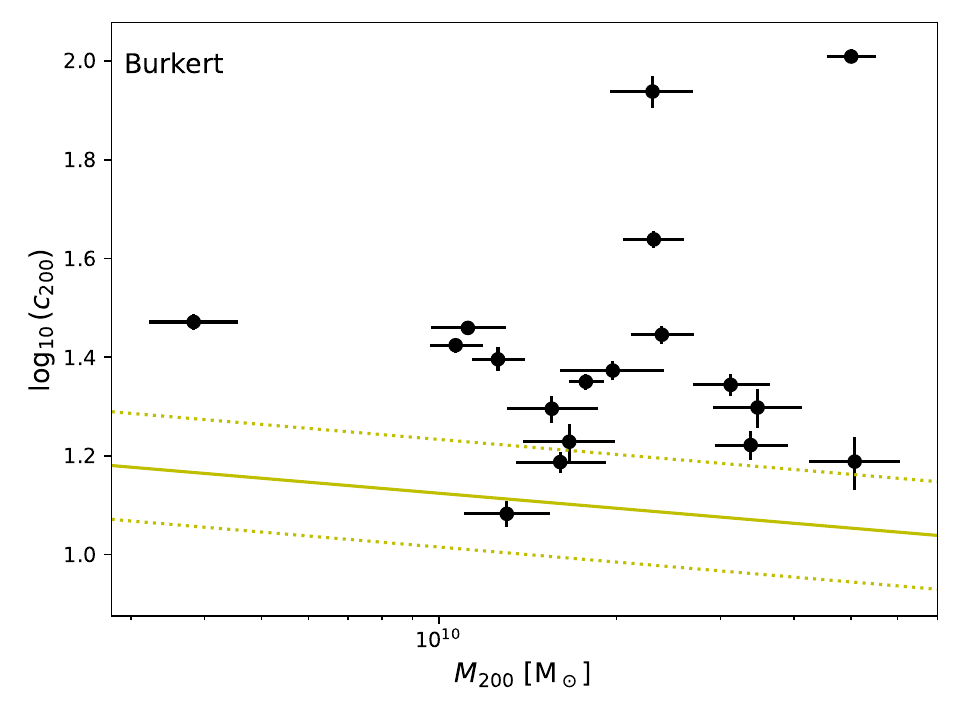}}
    \resizebox{0.49\hsize}{!}{\includegraphics{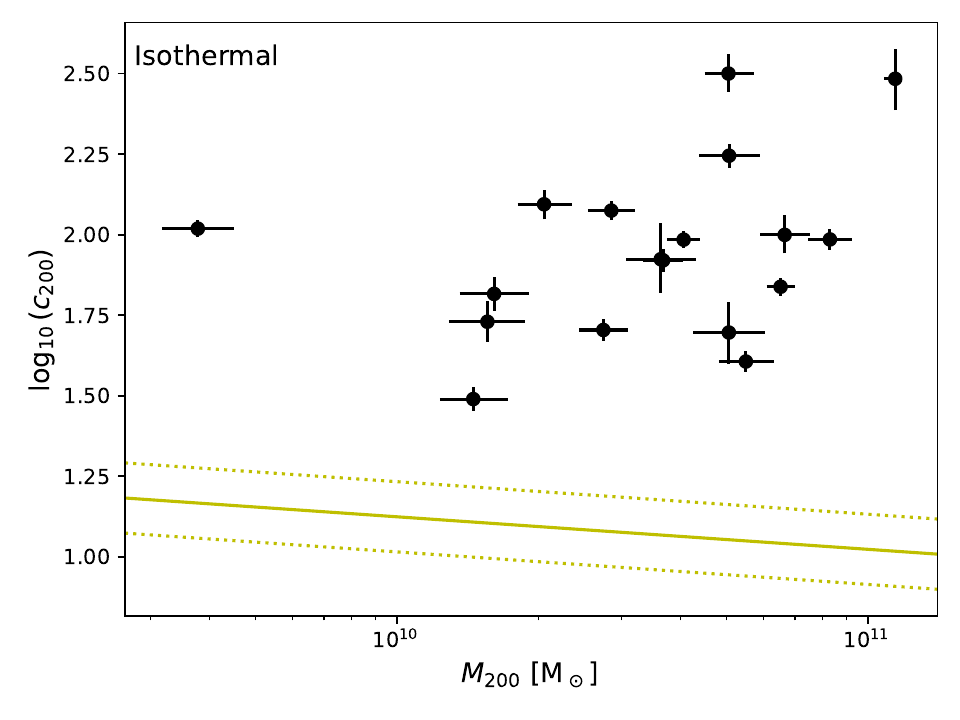}}

    \resizebox{0.49\hsize}{!}{\includegraphics{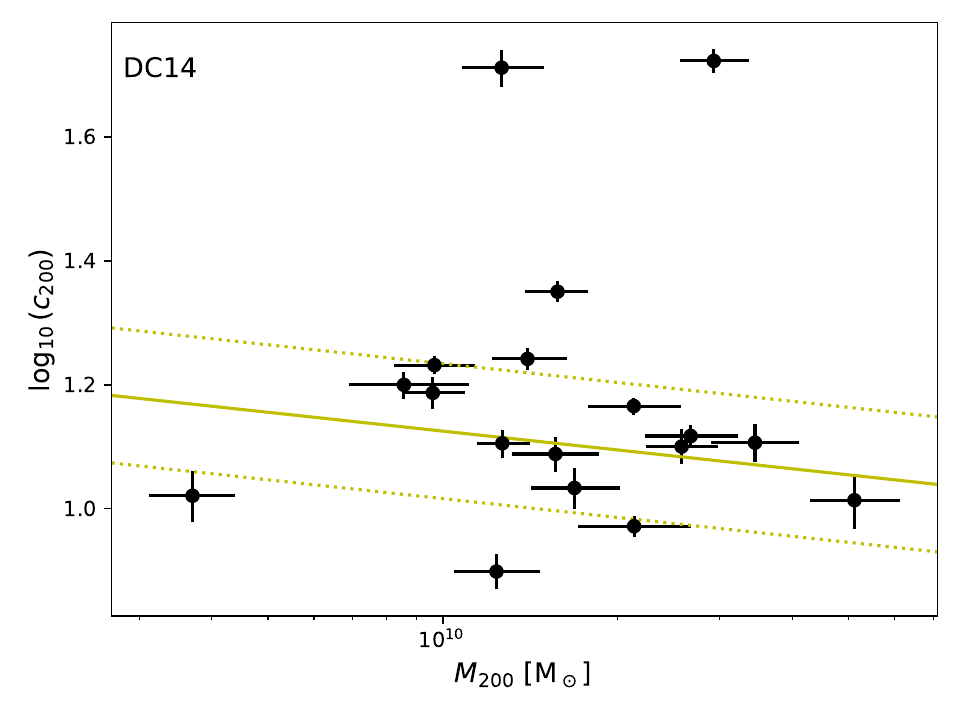}}
    \resizebox{0.49\hsize}{!}{\includegraphics{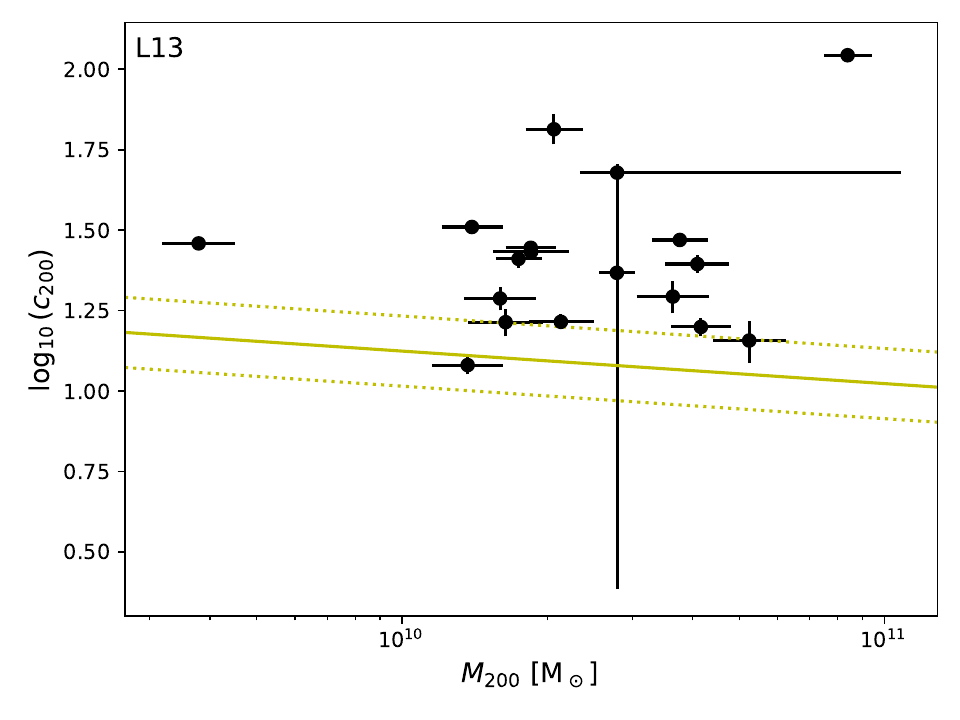}}

    \resizebox{0.49\hsize}{!}{\includegraphics{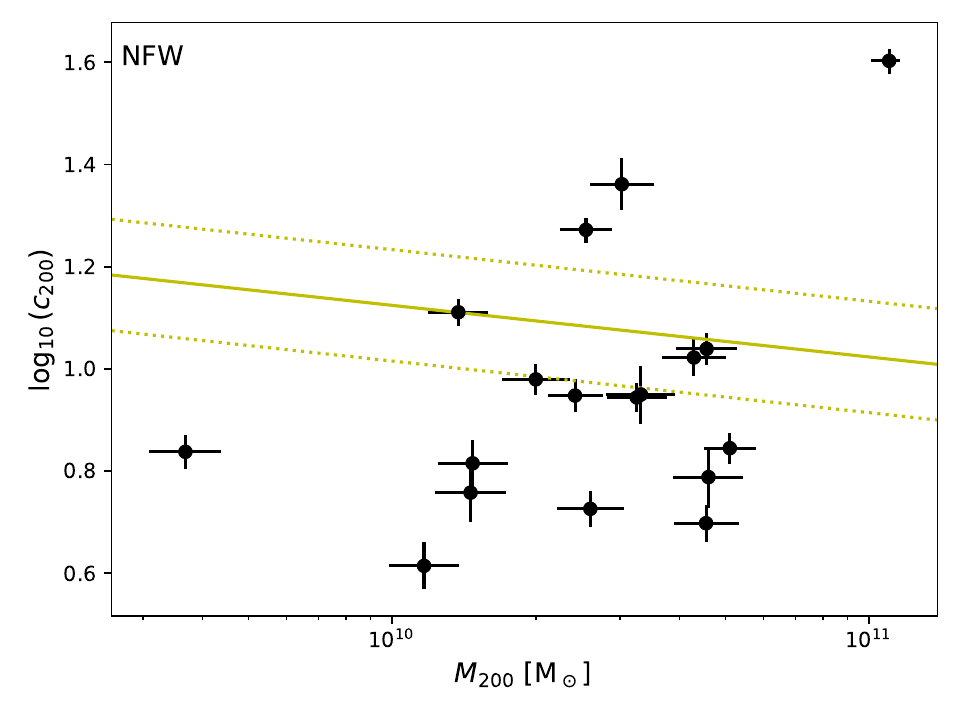}}
    \resizebox{0.49\hsize}{!}{\includegraphics{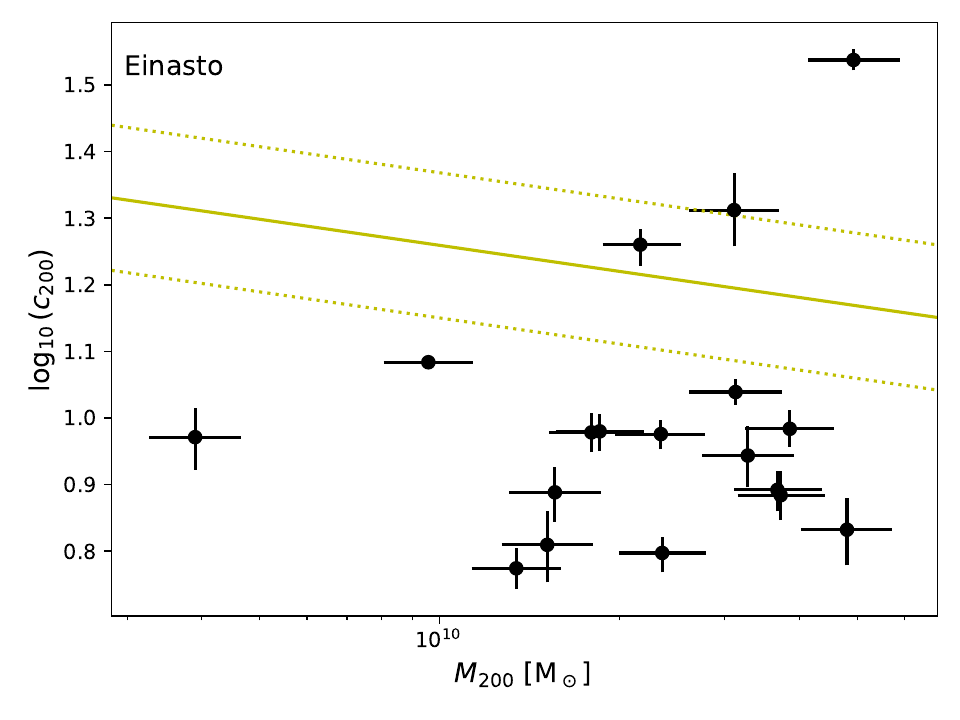}}
    \caption{$c_{200}$ vs $M_{200}$ for the various halo models fitted to the LITTLE THINGS dwarf sample. The yellow lines reflect the concentration model from \citet{dutton2014}. Note that this prior was not applied to the Burkert or isothermal halos.}
    \label{fig:lt-c200}
\end{figure*}

For the concentration we display the results in Fig.~\ref{fig:lt-c200}. This largely reiterates the trends from the SPARC sample. However, the NFW and Einasto cases scatter notably more than for the SPARC galaxies, with both of these halos preferring smaller than expected concentrations. DC14 has significant deviations in DDO101 ($4.3\sigma$) and NGC3738 ($4.8\sigma$). In Einasto we have CVnldwA ($3.3\sigma$), DDO126 ($3.1\sigma$), DDO53 ($2.1\sigma$), DDO216 ($2.3\sigma$), NGC2366 ($2.1\sigma$), and NGC3738 ($2.3\sigma$). NFW has 6 galaxies deviating at $> 2.1\sigma$ with NGC3738 reaching $4.5\sigma$. L13 has 11 significant deviations with NGC3738 reaching $8\sigma$. 
The $c_{200}$ fitting exercise has little success here, with L13 and Burkert requiring $\sim 3.6$ times larger $c_{200}$ values and little mass-dependence, without ameliorating the major outliers.

\begin{figure*}
    \resizebox{0.49\hsize}{!}{\includegraphics{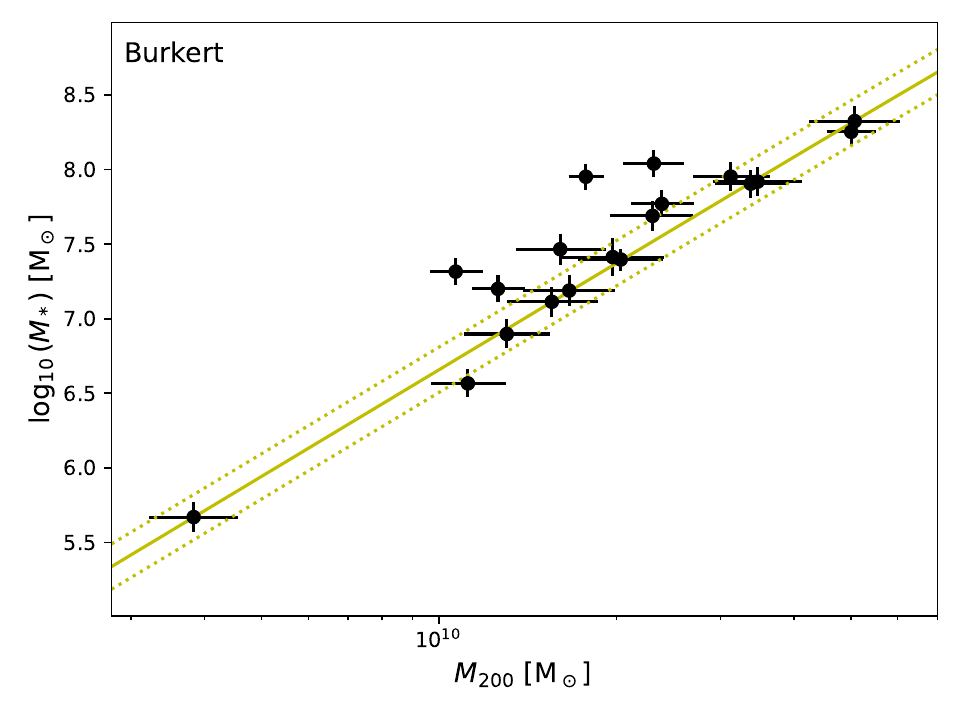}}
    \resizebox{0.49\hsize}{!}{\includegraphics{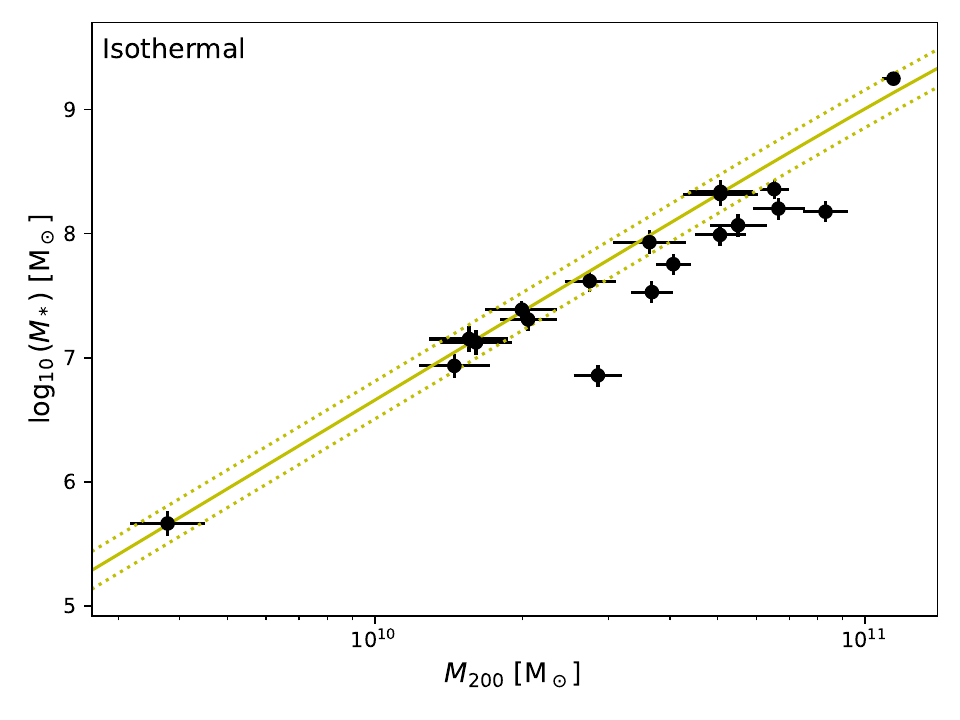}}

    \resizebox{0.49\hsize}{!}{\includegraphics{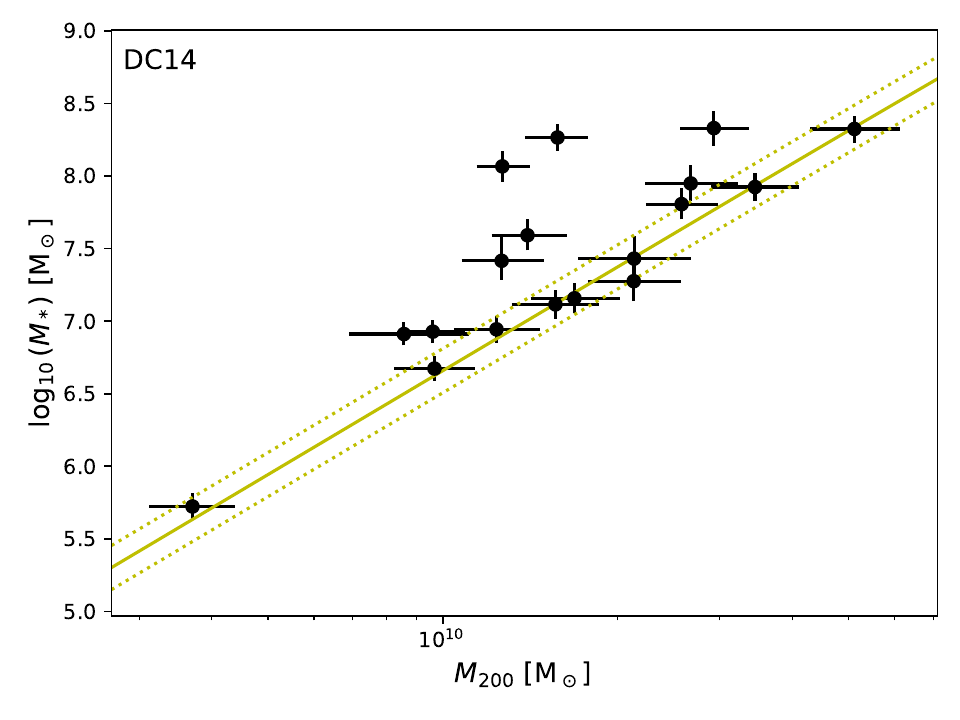}}
    \resizebox{0.49\hsize}{!}{\includegraphics{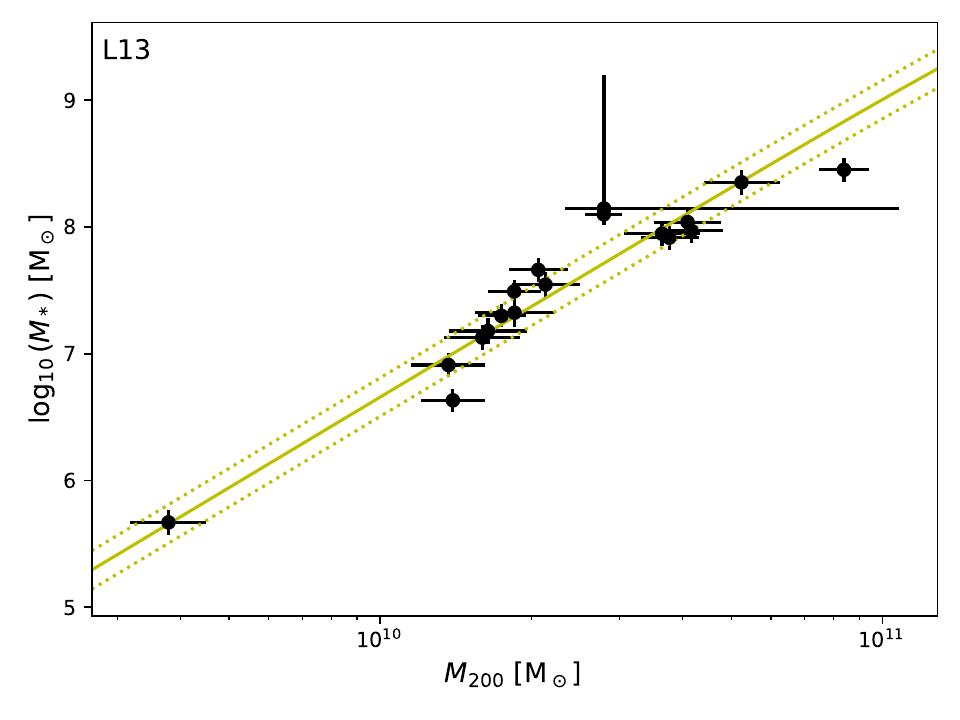}}

    \resizebox{0.49\hsize}{!}{\includegraphics{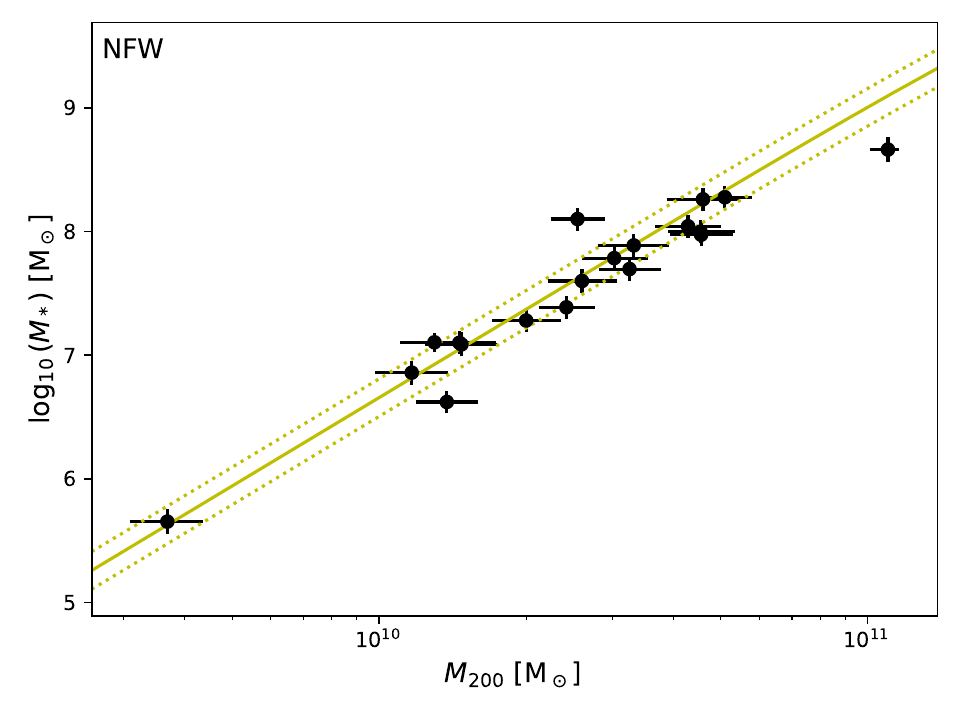}}
    \resizebox{0.49\hsize}{!}{\includegraphics{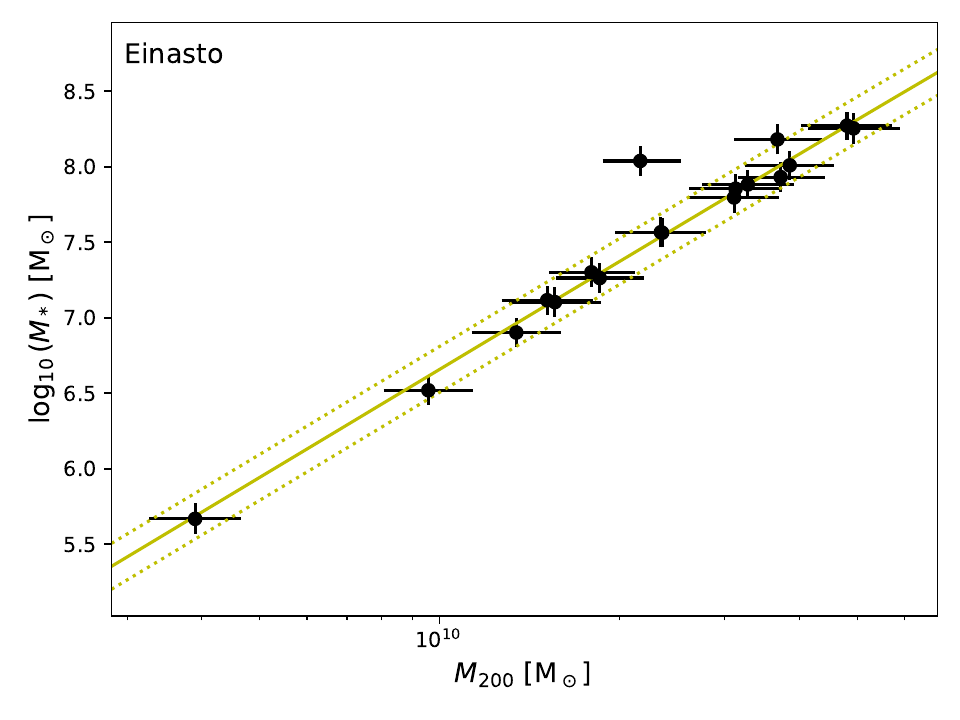}}
    \caption{$M_*$ vs $M_{200}$ for the various halo models fitted to the LITTLE THINGS dwarf sample. The yellow lines prior from \citet{moster_2013}.}
    \label{fig:lt-m200}
\end{figure*}

In the case of the halo mass, Fig.~\ref{fig:lt-m200} shows similar trends to the SPARC sample. Interestingly, the isothermal halo is far less divergent from the prior in these galaxies, with only DDO168 ($2.2\sigma$) and UGC8508 ($2.9\sigma$) significantly deviating. In Einasto we have only DDO50 at $2.2\sigma$. DC14 has significant deviations in DD50 and NGC2366, both at $4.5\sigma$. Burkert has DDO133 and NGC2336 deviating at $2.3\sigma$ and $2.9\sigma$ respectively. NFW and L13 show no significant deviations. 

\begin{figure*}
    \resizebox{0.49\hsize}{!}{\includegraphics{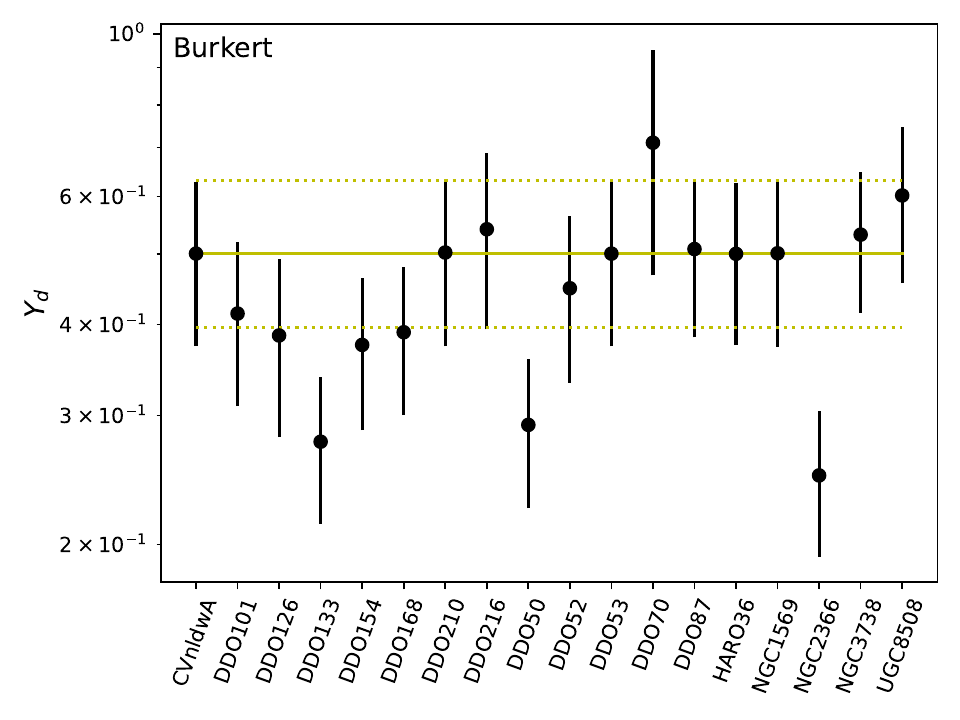}}
    \resizebox{0.49\hsize}{!}{\includegraphics{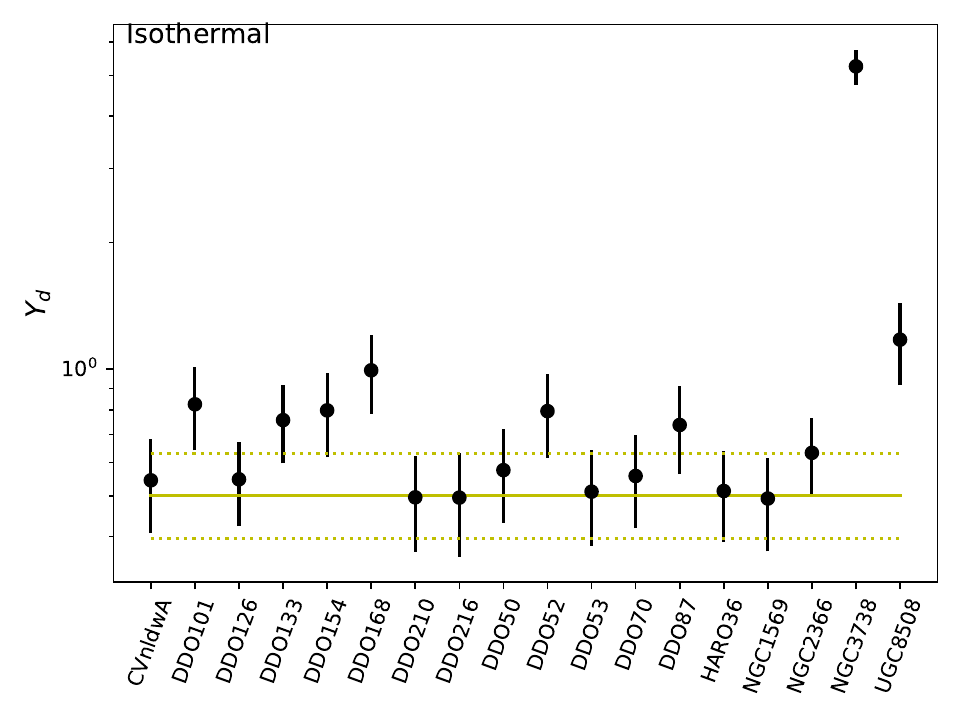}}

    \resizebox{0.49\hsize}{!}{\includegraphics{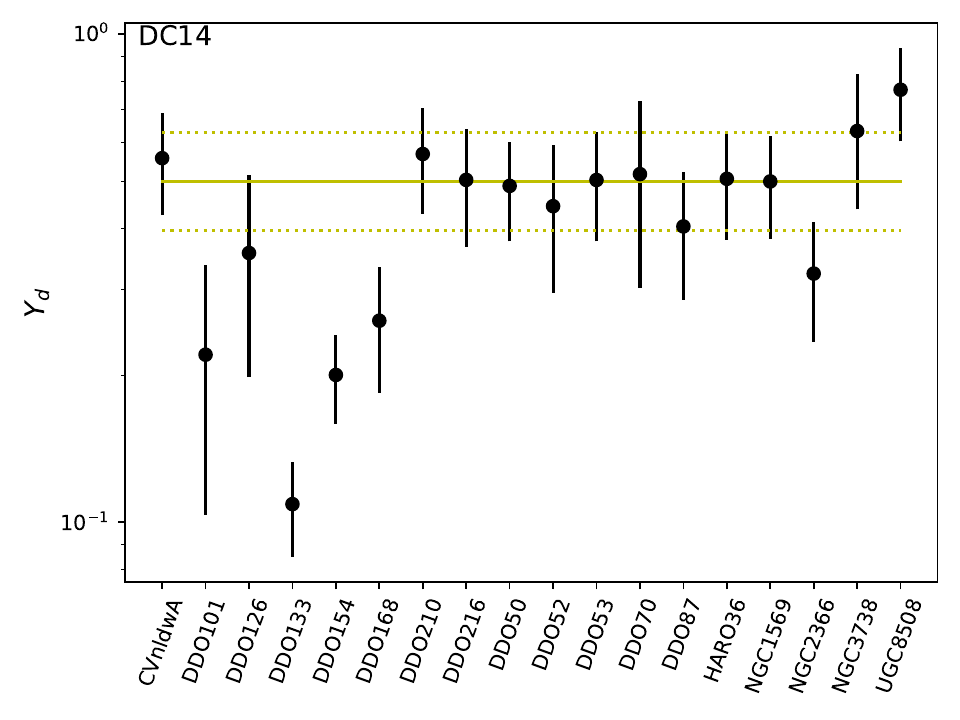}}
    \resizebox{0.49\hsize}{!}{\includegraphics{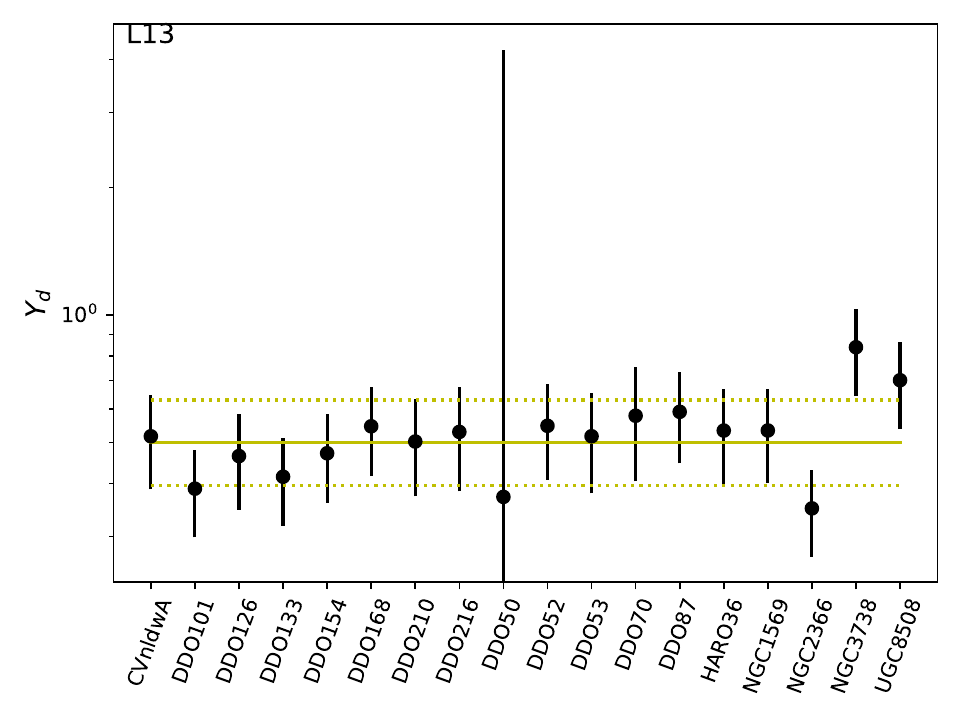}}

    \resizebox{0.49\hsize}{!}{\includegraphics{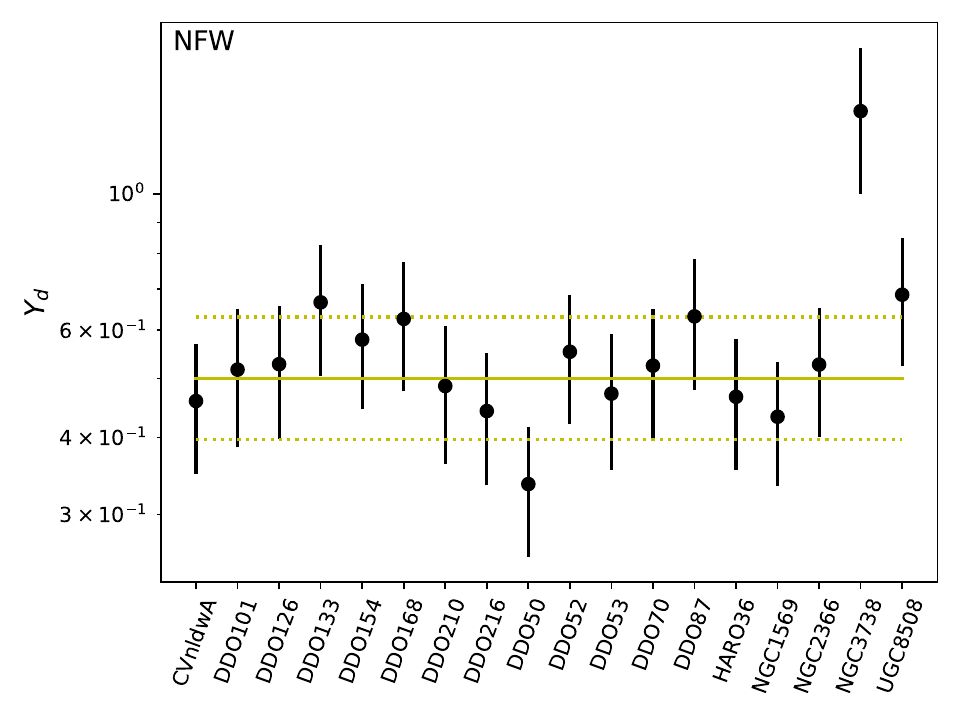}}
    \resizebox{0.49\hsize}{!}{\includegraphics{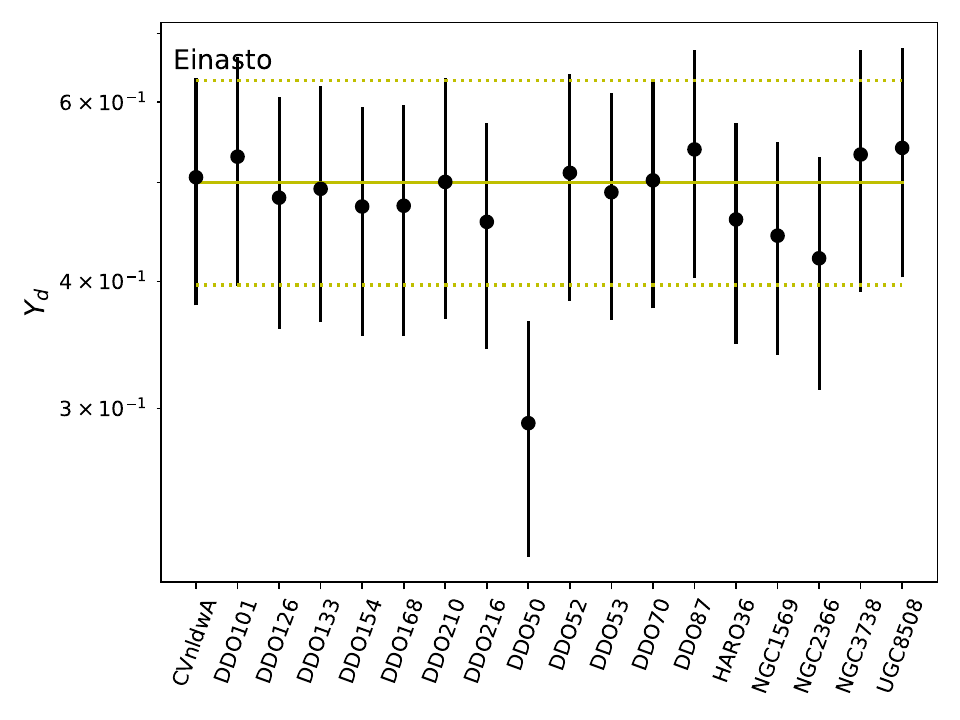}}
    \caption{$Y_d$ fitted for the various halo models fitted to the LITTLE THINGS dwarf sample. The yellow lines reflect the prior from \citet{schombert_2018}.}
    \label{fig:lt-yd}
\end{figure*}

For the mass-to-light ratios Fig~\ref{fig:lt-yd} shows that Einasto remains the best at recovering the prior. Burkert has DDO133 at $2.3\sigma$ and NGC2336 at $2.9\sigma$. Followed by L13 and then NFW and Burkert. Interestingly, the isothermal halo displays one very significant deviation here with NGC3728 reaching a value $\sim 5 \sigma$.  

Interestingly, the LITTLE THINGS halos show smaller average adjustments to the nuisance parameters across all models in Table~\ref{tab:dm-nuisance-lt}.
\begin{table}
    \caption{Average relative deviation in distance and inclination with halo model in LITTLE THINGS.}
    \label{tab:dm-nuisance-lt}
    \begin{tabular}{|l|l|l|}
        \hline
        Halo & $\bar{\sigma}_d$ & $\bar{\sigma}_\theta$ \\
        \hline
        Burkert & 0.18 & 0.72 \\
        DC14 & 0.20 & 0.89 \\
        L13 & 0.16 & 0.39 \\
        Einasto & 0.17 & 0.63 \\
        Isothermal & 0.41 & 0.59 \\
        NFW & 0.41 & 0.55 \\
        \hline
    \end{tabular}
\end{table}

\begin{figure*}
    \resizebox{0.49\hsize}{!}{\includegraphics{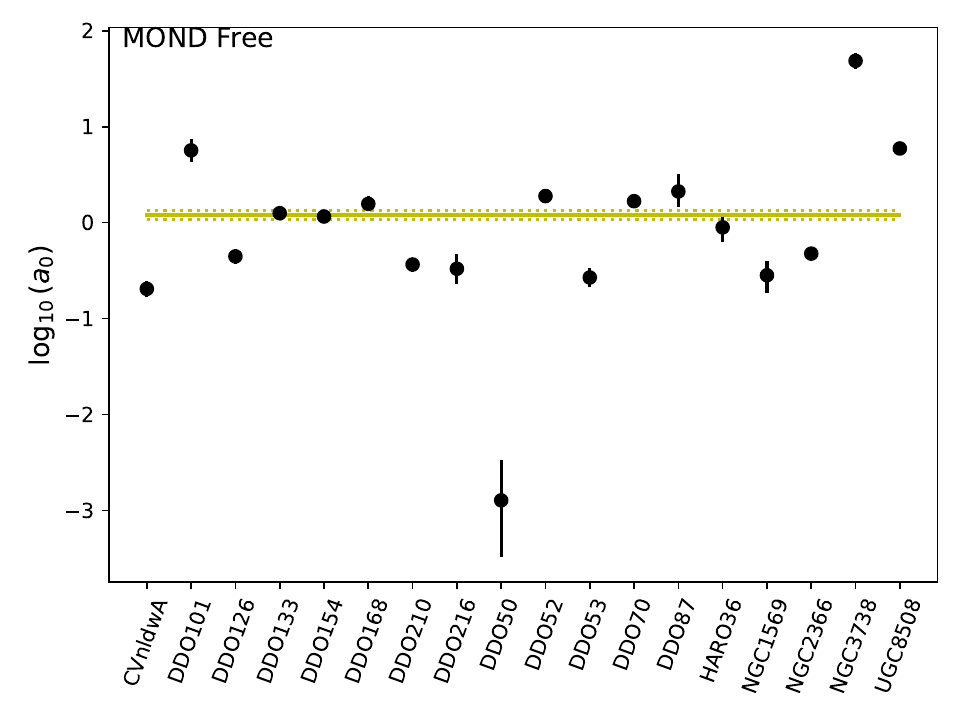}}
    \resizebox{0.49\hsize}{!}{\includegraphics{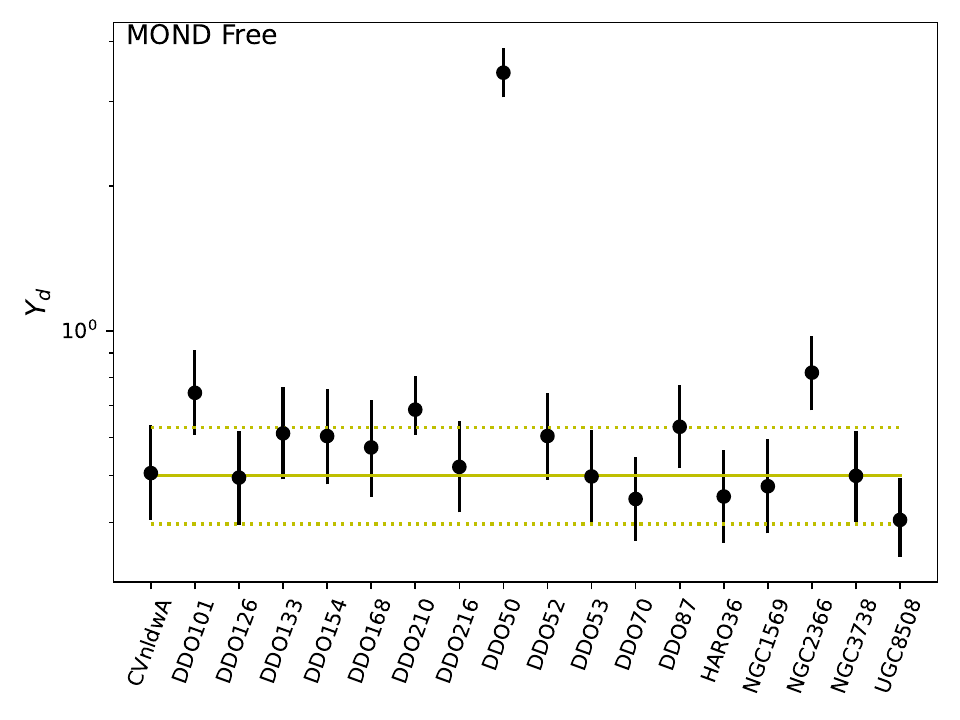}}

    \resizebox{0.49\hsize}{!}{\includegraphics{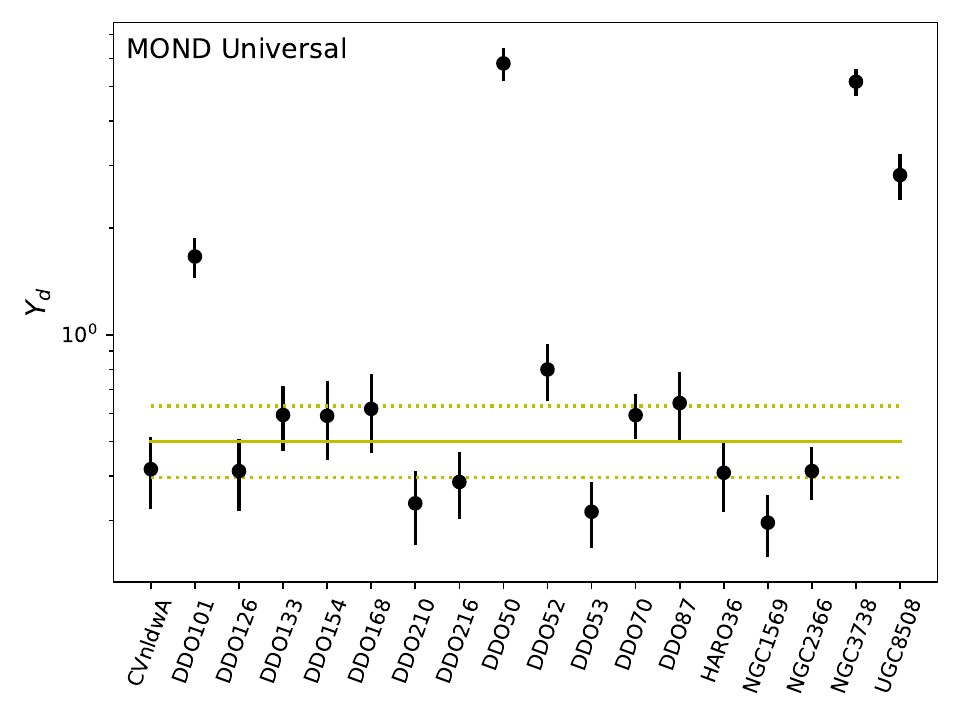}}
    \caption{MOND model values of $a_0$ and $Y_d$ for the various galaxies in the LITTLE THINGS dwarf sample. For $Y_d$ the yellow lines reflect the prior from \citet{schombert_2018}. Whereas, for $a_0$ it reflects $1.2\times 10^{-13}$ km s$^{-2}$ with a 12\% tolerance from \citet{desmond-rar-2023}.}
    \label{fig:lt-mond-prior}
\end{figure*}

For the MOND fits we have Fig~\ref{fig:lt-mond-prior}. Here we see that mass-to-light ratios are well-recovered for the free MOND model, with only DDO50 deviating by $> 3\sigma$. However, there are substantial discrepancies from the universal value for $a_0$, with 10 galaxies achieving significant deviations (even including the $a_0$ variance from \citet{desmond-rar-2023}). In particular, CVnldwA, DDO50, NGC3738, and UGC8508 all exceed a $5\sigma$ deviation. The other 6 all exceed $3\sigma$. In DDO50, MOND cannot replicate the low radius behaviour without a large mass-to-light ratio, which in turn forces $a_0$ lower as the gas and stellar contributions from \citet{oh_high-resolution_2015} already flatten at larger radii. For NGC3738, the lower radii are not a problem, but the magnitude of the rotation curve, compared to the baryonic contributions, requires a large $a_0$ to compensate. The mean value of $a_0 = 0.78$ with a relative error factor of $9.9$. In addition, when we fix $a_0$, we find three significant and substantial deviations for $Y_d$ in DDO50, NGC3738, and UGC8508. 

\begin{figure*}
    \resizebox{0.49\hsize}{!}{\includegraphics{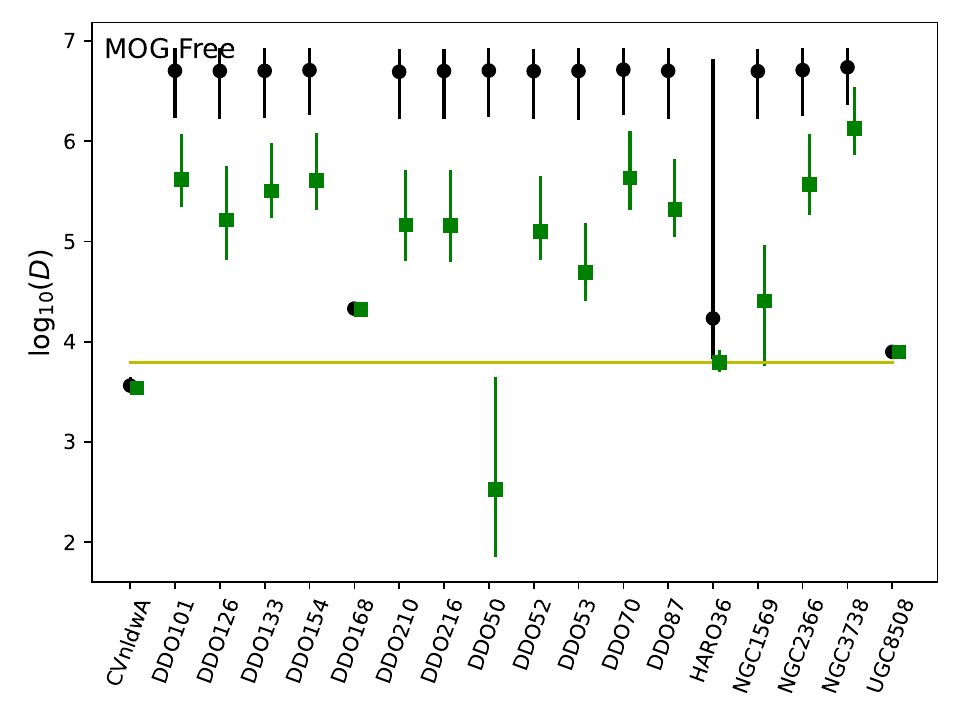}}
    \resizebox{0.49\hsize}{!}{\includegraphics{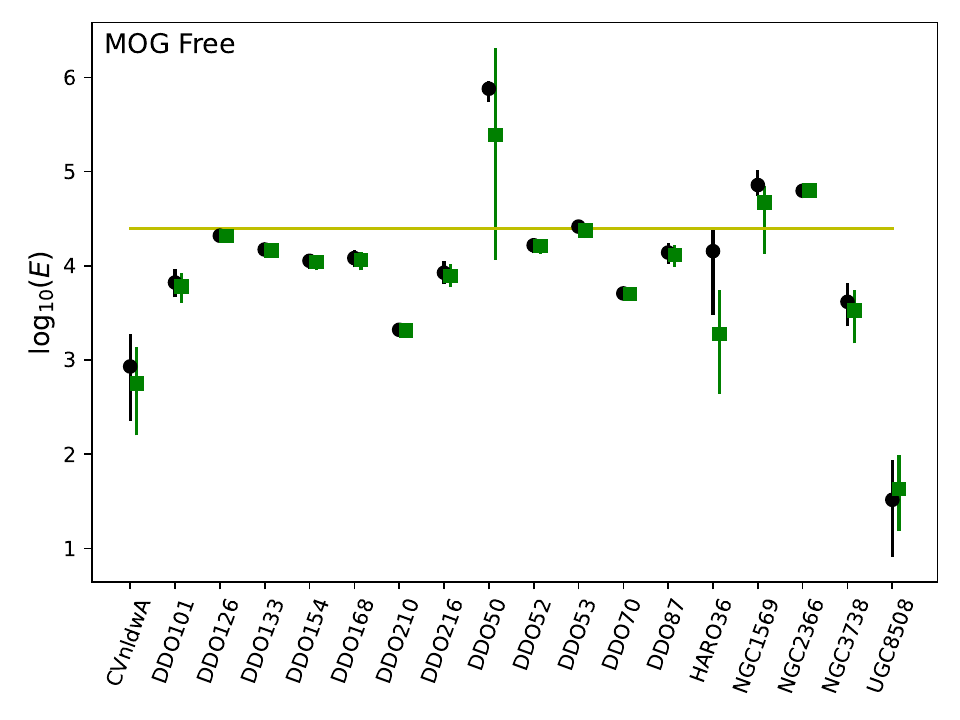}}

    \resizebox{0.49\hsize}{!}{\includegraphics{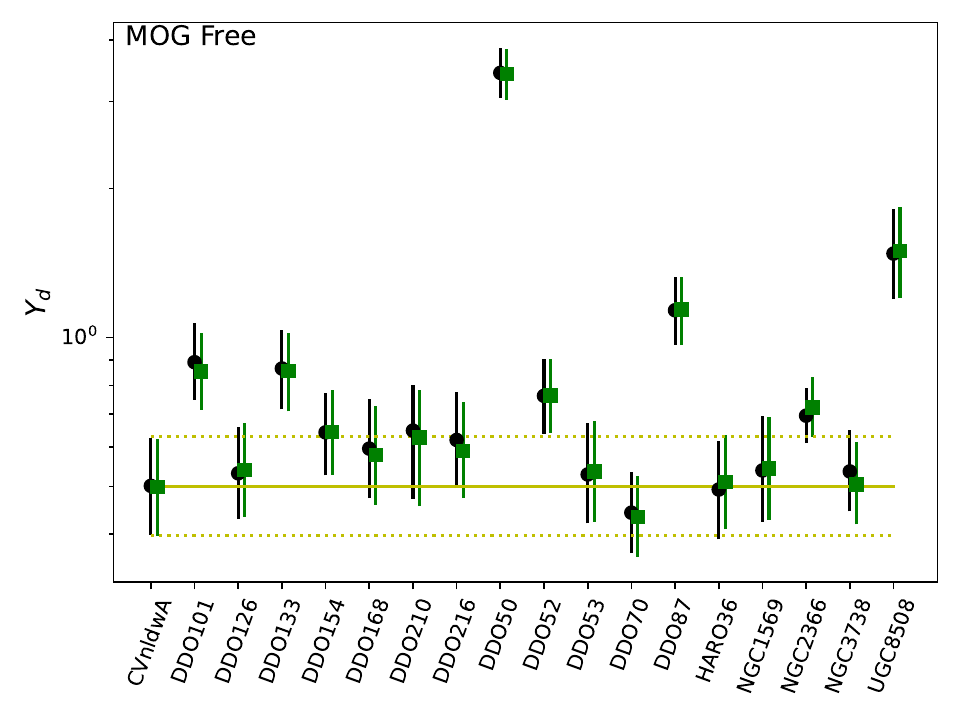}}
    \resizebox{0.49\hsize}{!}{\includegraphics{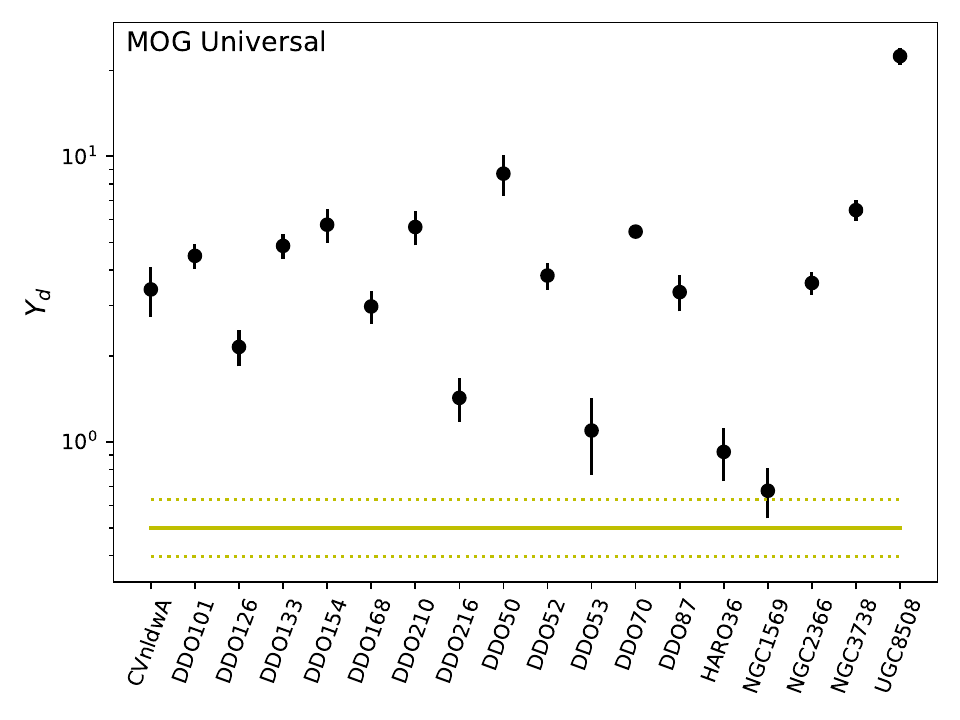}}
    \caption{MOG model values of $D$, $E$, and $Y_d$ for the various galaxies in the LITTLE THINGS dwarf sample. Black circles display the flat prior result, green squares are that for a lognormal prior. For $Y_d$ the yellow lines reflect the prior from \citet{schombert_2018}. Whereas, for $D$ it reflects $6250$ M$_\odot^{1/2}$ kpc$^{-1}$ and for $E$ it is $25000$ M$_\odot^{1/2}$.}
    \label{fig:lt-mog-prior}
\end{figure*}

For MOG fits we display Table~\ref{fig:lt-mog-prior}. Here we see the Free MOG model recovers $Y_d$ well, but again shows very significant divergences from the universal values of $D$ and $E$. Particularly, fourteen galaxies favour $D \to \infty$ with a flat prior. All of these objects' fits underestimate the low radius velocities, over-estimate those at very large radii, or both. 

Not all of these problem galaxies have a significant preference for DM halos over modified gravity, but a universal MOG model is generally significantly incompatible with these targets. Since we tested this up values of $10^{50}$, it is unclear if there is any finite lower limit. 

If we employ a lognormal prior on $D$ and $E$ we find that the $D\to \infty$ tendency is moderated. However, all but one of these galaxies still display significant disagreement with the universal value. The mean value of $D$ is $\sim 6.9\times 10^4$ M$_\odot^{1/2}$ kpc$^{-1}$, around triple the universal value, with a standard deviation of $4.8\times 10^5$ M$_\odot^{1/2}$ kpc$^{-1}$. While for $E$ we find $\sim 7.7\times 10^3$ M$_\odot^{1/2}$, around a third of the universal value, with a standard deviation of $4.2\times 10^4$ M$_\odot^{1/2}$.

Finally, if we fix $D$ and $E$ we find extremely large values of $Y_d$ are required, all but 4 being in excess of 2, and UGC8508 exceeding 20. In most cases, the model ELPD worsens considerably.

In Table~\ref{tab:mog-nuisance-lt} we see that the universal models again require substantial average shifts in the distance.
\begin{table}
    \caption{Average relative deviation in distance and inclination for MOND and MOG in LITTLE THINGS.}
    \label{tab:mog-nuisance-lt}
    \begin{tabular}{|l|l|l|}
        \hline
        Model & $\bar{\sigma}_d$ & $\bar{\sigma}_\theta$ \\
        \hline
        MOND Free & 0.14 & 0.30 \\
        MOND Universal & 2.34 & 0.89 \\
        MOG Free & 0.26 & 0.31 \\
        MOG Universal & 2.98 & 1.17 \\
        \hline
    \end{tabular}
\end{table}

\subsection{Einasto halos}
\begin{figure}
    \resizebox{0.99\hsize}{!}{\includegraphics{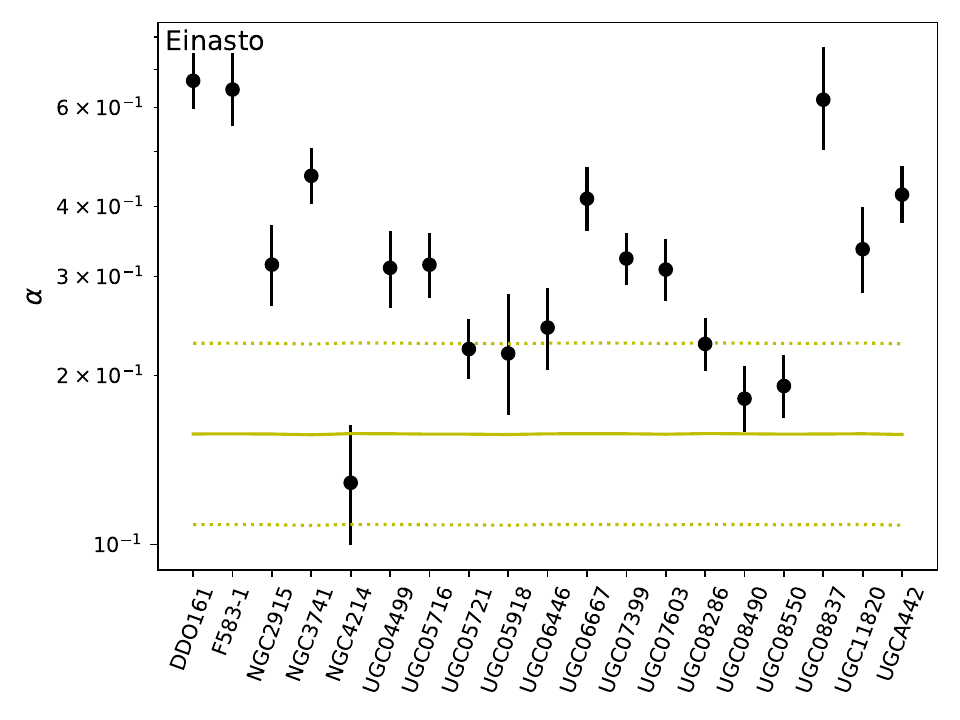}}
    \resizebox{0.99\hsize}{!}{\includegraphics{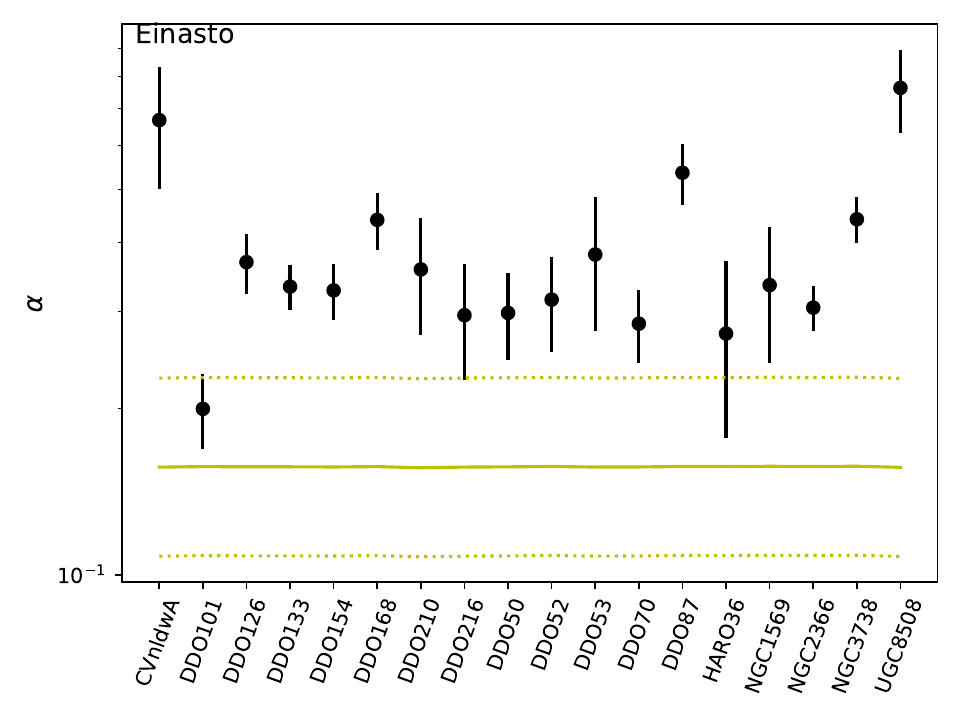}}
    \caption{Einasto $\alpha_e$ values for each galaxy in the two samples. The \citet{dutton2014} value is displayed in the yellow line. Upper: SPARC. Lower: LITTLE THINGS.}
    \label{fig:ein-prior}
\end{figure}

Here we display the fits to the Einasto index $\alpha_e$ for all our galaxies in Fig.~\ref{fig:ein-prior}. It is notable that most of these galaxies fall within $2\sigma$ of the values cited in \citet{dutton2014}. The values are generally somewhat larger than expected, especially for LITTLE THINGS, which leads to more cored halo behaviour for the inferred density profiles. This further demonstrates the preference for cored type halos in these targets.

\section{Comparison to the literature}
\label{sec:comp}
We find very similar DM fit results to \citet{li_comprehensive_2020} for our subsample of galaxies. This at least provides validation for our approach, which uses far fewer MCMC walkers or samples.

Our SPARC galaxies display no significant preference either way between free and universal MOND. This is in concordance with radial acceleration analyses that find a compatible universal value for the MOND scale~\citep{rar_lelli,rar_li,desmond-rar-2023}. However, for LITTLE THINGS, we find a strong preference for free MOND over the universal version.

In terms of the MOND external field effect, there are studies finding evidence for it~\citep{for_efe_1,for_efe_2,sep-zeta} and against it~\citep{against_efe_1,against_efe_2,against_efe_3}, while \citet{desmond-rar-2023} finds weak evidence in its favour. Here we only contribute weak evidence against it if MOND has a universal $a_0$.

The work \citet{khelashvili2024} argues the SPARC galaxies show strong preferences for DM halos over MOND. However, this is not fully replicated here. Cored halos do indeed achieve some preference over MOND, although only 6 prefer a Burkert halo to universal MOND, 5 pass $3\sigma$, and UGC05721 reaches $4.7\sigma$. Thus, less than $30$\% of our SPARC dwarfs achieve significant preferences for Burkert halos over MOND. A more robust preference only exists at the level of the combined sample. We should be cautious about drawing conclusions from this without more evidence from individual galaxies. An important difference in our analysis is that \citet{khelashvili2024} uses flat priors for all the DM parameters, and does not include distance and inclination as nuisance parameters. The former tilts the playing field by failing to make DM halos conform to expected limits on their properties, such as the mass relation from \citet{moster_2013}. For example, if we free isothermal from the \citet{moster_2013} relation it goes from a $4\sigma$ advantage over MOND to $\sim 8\sigma$ (at the cost of unrealistically massive halos). The lack of nuisance parameters can enhance the significance of other effects. Notably, our $a_0$ values compare well with those in \citet{khelashvili2024}, further evidencing that differences occur in the DM modelling. These authors also find a correlation between preference for MOND and inclination angle, i.e. more face-on galaxies tend to prefer MOND. We find a $0.5$ Spearman-$r$ value at $p=0.03$ (by permutation test) for Burkert or L13 vs MOND, weakly suggesting that more edge-on galaxies do favour Burkert/L13 over MOND (no significant relationship for DC14, isothermal, or Einasto). For NFW halos we find the opposite correlation, $r = -0.43$ at $p=0.07$. We note that \citet{khelashvili2024} also suggest that lower quality rotation curves favour MOND. Interestingly, we find no correlation between number of data points or average relative error and preference for MOND. However, we do find a $-0.5$ Spearman $r$ coefficient with a $p$-value of $0.06$ for a correlation between the standard deviation of the $\frac{\delta v_\mathrm{obs}}{v_\mathrm{obs}}$. Insignificantly indicating larger variations in relative error lead to a preference for Burkert halos, again in the opposite direction to \citet{khelashvili2024}. For the Einasto halo we find $r \approx 0.5$ at $p=0.03$, other halos showed no significant results.

In the original analysis of LITTLE THINGS~\citep{oh_high-resolution_2015}, the authors found a preference for isothermal halos over NFW. We find a $5.9\sigma$ preference for isothermal over NFW in this sample, but more significant preferences for L13, Einasto, and Burkert. In particular, L13 achieves a $3.1\sigma$ advantage over isothermal. This sample seems far more favourable to isothermal than SPARC. 

In \citet{obrien_alternative_2018} the authors find that a curated subsample of LITTLE THINGS is compatible with both conformal gravity and MOND. Particularly, they find agreement in CVnldwA, DDO43, DDO53, DDO126, DDO133, DDO154, DDO210, and DDO216. Of these we highlight CVnldwA, which shows some tension with the universal MOND scale in our fitting (around a factor of 5 different at high significance). A notable difference between our work and theirs is they use a fitting function for the baryonic contributions, rather than the rotation velocities found in \citet{oh_high-resolution_2015}. We also do not test conformal gravity, as it is unclear it actually predicts flat rotation curves~\citep{2021PhRvD.104f4014H}.

The authors in \citet{little-things-mond-mog} test various alternative gravity models, including MOND and MOG, on the LITTLE THINGS galaxies. In general, they find these models require larger than expected mass-to-light ratios. Notably, their treatment of MOG does not involve a mass dependence in $\mu$ or $\alpha$ (as suggested by rotation curve fits in \citet{moffat_fundamental_2009}). Thus, our work improves upon this previous analysis via inclusion of predicted parameter scaling. Additionally, the parameters of the models were not treated as free, so problematic galaxies that require, for instance, $D\to \infty$ were not noticed. Similarly, their treatment of MOND used only a fixed value for $a_0$, so tensions with the universal value would only manifest in the highly uncertain $Y_d$. This misses the opportunity to compare universal values of $a_0$ to the tolerances from studies of the radial acceleration relation, such as \citet{desmond-rar-2023}. Finally, there is no accounting for distance and inclination as nuisance parameters in \citet{little-things-mond-mog}. 

There are no sample overlaps between this work and \citet{brownstein2006,moffat_fundamental_2009}. Thus, there are no direct comparisons possible. However, we have presented several galaxies that strongly exclude the universal parameters found in \citet{moffat_fundamental_2009}. Notably, the sample level statistics also strongly disfavour universal MOG when contrasted with its free counterpart.

\section{Conclusions}
\label{sec:conc}

The dwarf galaxy samples from SPARC and LITTLE THINGS display some important common conclusions: cored DM halos are preferred to cusps, and cored halos are preferred to MOND or MOG. These results exceed $4\sigma$ and seem relatively robust, as the physically motivated priors are reasonably recovered for the cored halo models, although the concentration-mass relation had significant outliers in LITTLE THINGS. On the other hand, it must be noted that many significant discrepancies emerge for MOND and MOG between fitted parameters and universal model values. The latter model attaining order of magnitude level disagreement. The use of universal values strongly degraded the ELPD in all cases except MOND in the SPARC sample. Importantly, our model comparison was made on the basis of a leave-one-out analysis, more robust and less biased than the often employed $\chi^2$, or indeed metrics like the BIC, especially when fewer data points are involved.   

After assessing the performance of both DM and modified gravity, we can see that LITTLE THINGS rotation curves are definitely problematic compared to SPARC. There is a substantial incompatibility with a universal MOND $a_0$, which is not observed in SPARC, and we cannot recover physically motivated $c_{200}$ relationships in L13 and Burkert halos. The latter has to make one doubtful of the physical validity of the universal MOND problems here. The additional flexibility of the DM halos might be allowing inaccurate halo models to account for significant non-circular velocity effects. Despite this, the same general trend of preferring DM to MOG and MOND also reappears in the SPARC sample, albeit with reduced significance. 

Notably, our results disagree with \citet{moffat_fundamental_2009}, who concluded that MOG with universal parameters was compatible with galaxy rotation curves. Additionally, we also extend the exploration of \citet{li_comparing_2017}, by considering more DM profiles and including the SPARC dwarf galaxies. However, we conclude that, although MOND does provide an adequate description of these galaxies, there is still a sample-level $>4\sigma$ preference for cored DM halos instead. 

The fact that only a single galaxy significantly preferred MOND over a cored halo indicates that future studies with BIG SPARC~\citep{bigsparc} may be troubling for MOND. In particular, re-examination of the LITTLE THINGS targets, with updated infrared photometry, may pay dividends in the era of MeerKAT and SKA. 

Despite the preceding, it is a remarkable fact that a universal MOND scale is largely compatible with the SPARC data (although it strongly disfavoured by LITTLE THINGS). Such phenomena seem to indicate a need for more complex DM formulations to produce explanatory baryonic correlations.

\section*{Acknowledgements}

GB acknowledges funding from the National Research Foundation of South Africa under the incentive funding for rated researchers program. 

\section*{Data Availability}
This work is based on publicly available SPARC and LITTLE THINGS data. All corner plots and inference data from MCMC sampling is made available on Zenodo, along with example analysis scripts.



\bibliographystyle{mnras}
\bibliography{dwarfs} 








\bsp	
\label{lastpage}
\end{document}